\documentclass[aps, pra, reprint, tightenlines, letterpaper, amsmath, amssymb, preprintnumbers, floatfix, longbibliography, nofootinbib,superscriptaddress]{revtex4-2}
\usepackage{tabularx}
\usepackage{dsfont}
\usepackage{extarrows}
\usepackage{amsmath}
\usepackage{amssymb}
\usepackage{physics}
\usepackage{tabu}
\usepackage{tabularx}
\usepackage{bm}
\usepackage{graphicx}
\usepackage{placeins}
\usepackage{textgreek}
\usepackage{multirow}
\usepackage{makecell}
\usepackage{soul}
\usepackage{diagbox}
\usepackage{multirow}
\usepackage{listings}
\newcommand{\lmag}[1]{|{\bf l}_{#1}|}
\newtheorem{theorem}{Theorem}
\usepackage[normalem]{ulem}

\makeatletter

\usepackage[hypertexnames=false]{hyperref}
\hypersetup{
    colorlinks=true,       
    linkcolor=blue,          
    citecolor=blue,        
    filecolor=blue,      
    urlcolor=blue           
}

\newcolumntype{Y}{>{\centering\arraybackslash}X}

\makeatletter
\pretocmd\frontmatter@thefootnote{\color{black}}{}{}
\makeatother

\usepackage{orcidlink}
\newcommand{\sgn}{\text{sgn}}

\makeatletter  
\renewcommand\onecolumngrid{
\do@columngrid{one}{\@ne}%
\def\set@footnotewidth{\onecolumngrid}
\def\footnoterule{\kern-6pt\hrule width 1.5in\kern6pt}%
}
\makeatother    

\begin{document}

\title{Universal enhancement of the pairing gap in a dilute Fermi superfluid}

\author{Silas R.~Beane}
\email{silas@uw.edu}
\affiliation{Department of Physics, University of Washington, Seattle, WA 98195, USA}

\author{Roland C.~Farrell}
\email{rolandf@caltech.edu}
\affiliation{Institute for Quantum Information and Matter (IQIM) and Department of Physics,
California Institute of Technology, CA 91125, USA}

\author{Cameron P.~Shotwell}
\email{cshotwel@uw.edu}
\affiliation{Department of Physics, University of Washington, Seattle, WA 98195, USA}

\preprint{NT@UW-26-23}
\date{\today}
\begin{abstract}
\noindent
At zero temperature, a dilute Fermi gas with an attractive contact interaction forms a superfluid with gapped fermionic excitations.
We show that the gap at weak coupling can be organized as a trans-series expansion in the interaction strength and the asymptotic Gor’kov-Melik-Barkhudarov gap. Low-energy virtual quasiparticles generate corrections that are linear in the gap with a leading coefficient nearly 25$\times$ larger than the other known coefficients. This substantially increases the gap at intermediate densities, highlighting that systematic corrections are crucial for quantitative predictions.
\end{abstract}

\maketitle

\begingroup
\renewcommand*\addcontentsline[3]{}  
\section{Introduction}
\label{sec:intro}
\endgroup
\pagenumbering{arabic}
\setcounter{page}{1}
\noindent
The fermionic excitation gap is a defining energy scale in a superfluid that sets the pair-breaking threshold and suppresses fermionic excitations at low temperatures.
It shapes the low-energy properties of paired fermions across systems ranging from ultracold gases~\cite{RevModPhys.80.1215} and superconductors~\cite{Chen_2024} to neutron stars~\cite{Chamel:2017wwp,Haskell:2017lkl,10.1093/acprof:oso/9780198719267.003.0011,Ramanan:2020xrj} and dense quark matter~\cite{Schmitt:2025cqi}.
Despite its importance, systematically improvable calculations of the gap remain challenging because the pairing instability generates a non-perturbative energy scale that is exponentially small in the interaction strength.
As a result, even the structure of the weak coupling expansion of the gap has not been established.

A paradigmatic system that exhibits superfluidity is a spin-half Fermi gas with short-range attraction.
In the cold and dilute limit, the interaction is universally described by a contact force whose renormalized strength is proportional to the $s$-wave scattering length $a$~\cite{Braaten:2004rn}.
At zero temperature, the behavior is governed by two dimensionless quantities
\begin{align}
\lambda=\frac{2k_Fa}{\pi}<0 \ ,
\qquad
x=\frac{\Delta}{E_F} \ ,
\end{align}
where $k_F$ is the Fermi momentum, $E_F=k_F^2/(2M)$ is the Fermi energy and $M$ is the fermion mass.
The coupling constant can be tuned through $a$ or by varying $k_F$ through the number density $k_F= \sqrt[3]{3\pi^2 N/V}$.

Bardeen, Cooper, and Schrieffer (BCS) obtained the leading nonperturbative prediction for the gap~\cite{PhysRev.108.1175},
\begin{align}
x_{\mathrm{BCS}}
=
\frac{8}{e^2}e^{1/\lambda}\ .
\end{align}
Gor'kov and Melik-Barkhudarov (GMB) subsequently found that particle-hole
polarization screens the attractive interaction and suppresses the gap by a
universal factor~\cite{GorkovMelik},
\begin{align}
x_{\mathrm{GMB}}
\equiv x_0 = 
\frac{x_{\mathrm{BCS}}}{(4e)^{1/3}}\ .
\label{eq:xGMB}
\end{align}
The GMB suppression comes from a perturbative $O(\lambda^2)$ correction to the effective potential that is enhanced to an $O(1)$ multiplicative shift in the gap energy by the Cooper logarithm $\log x\sim1/\lambda$.

Numerous approaches based on diagrammatic expansions~\cite{Pisani_2018,Pisani_2018_2,Ruan2013EffectOT,Chen_2016,Urban:2026trm,kagan2013modern,Haussmann_2007,Fujimoto:2025zfa,Furnstahl:2006pa,BCF2024,Haussmann1993CrossoverFB} and functional methods~\cite{Diener_2008,Floerchinger_2010} have investigated pairing beyond the BCS mean field solution.
Much of this work seeks quantitative predictions across the BCS-BEC crossover, including at unitarity where $\lambda$ diverges~\cite{Veillette:2007largeN,Nishida_2006}.
These crossover calculations often employ diagrammatic or functional truncations followed by a numerical solution of coupled self-consistency equations.
Such approaches can resum important classes of contributions but ultimately produce a number at each coupling, and therefore cannot reveal the structure of the weak coupling expansion or isolate its coefficients.
Without an accompanying power counting, a diagrammatic truncation may retain only partial contributions at a given order, making its accuracy difficult to assess.

In this Letter, we show that the weak-coupling expansion of the gap is a trans-series in $(\lambda,x_0)$ that contains a sector linear in the gap whose leading coefficient is large.
The linear-in-gap correction originates from a kinematic enhancement in the self-energy and the effective interaction between Cooper pairs when two virtual quasiparticles have energies $E\sim\Delta$.
A similar term in the static spin susceptibility of superconductors was previously identified in Refs.~\cite{Aristov_1997,PhysRev.116.898}, but to the best of our knowledge its effect on the gap has not been determined.
The trans-series expansion systematically extends the GMB result and yields predictions for the gap away from asymptotically weak coupling.
These predictions are equipped with uncertainty bands coming from truncating higher order terms and can be meaningfully compared to Quantum Monte Carlo (QMC) and experiment. 

To organize the weak-coupling expansion, we start from the exact condition that the pairing vertex has a pole at zero total four-momentum corresponding to the particle-number breaking Goldstone mode. 
Under mild assumptions on the structure of diagrammatic perturbation theory, we show that the pole condition implies that the gap at weak coupling is a trans-series in $\lambda$ and $x_0$,
\begin{align}
\log{x} \ = \  \sum_{m\geq 0,n} c_{n,m} \, \lambda^n \, x_{0}^m \ .
\label{eq:gapTransSeries}
\end{align}
The sum over $n$ is bounded below for each $m$ sector.
Keeping only the $c_{-1,0}$ and $c_{0,0}$ coefficients recovers the GMB result.
For previous work on the perturbative expansion of the gap see, for example, Refs.~\cite{Efremov_2000,Mari_o_2019,BCF2024}.
The coefficients are obtained by expanding the pole condition in explicit powers of the interaction strength while holding the gap fixed, and then expanding each term in the gap.
The explicit order in the interaction does not fix the order in $\lambda$ because the loop integrals generate logarithms $\log{x}\sim1/\lambda$, and higher explicit orders can shift lower-order coefficients.
To track this, we develop a power counting that identifies the diagrams and momentum regions that contribute to the trans-series coefficients.

We obtain closed form expressions for the leading coefficient $c_{2,1}$ of the linear-in-gap sector, as well as for $c_{1,0}$.
The latter was previously calculated numerically in Ref.~\cite{BCF2024}, but contained errors that we correct. 
The $c_{2,1}$ coefficient is nearly $25\times$ larger than any of the other known coefficients, and is the dominant correction to the GMB gap already for $|\lambda|\gtrsim 0.3$.~\cite{scalenote}
We compare the trans-series prediction to QMC simulations~\cite{Gezerlis:2007fs} and ultracold-atom experiments~\cite{Biss_2022} and find tension at the lowest densities available.

\vspace{2em}

\begingroup
\renewcommand*\addcontentsline[3]{}  
\section{Goldstone-Pole Gap Equation}
\endgroup
\noindent
Consider non-relativistic spin-1/2 fermions in three dimensions with a contact interaction ${\cal H}_{\text{int}} = g_0\psi^{\dagger}_{\uparrow}\psi^{\dagger}_{\downarrow}\psi_{\downarrow}\psi_{\uparrow}$ 
where the field $\psi_\sigma^{\dagger}$ creates a fermion of spin $\sigma$ and $g_0$ is the bare coupling strength that gets renormalized to $g=4\pi a/M$.
This interaction universally describes fermions with finite-range interactions at low energies and its strength can be tuned in ultracold atoms using Feshbach resonance~\cite{FESHBACH1962287,Chin_2010}.
An arbitrarily weak attractive interaction destabilizes the free Fermi gas and the system is a superfluid at zero temperature.
We will work in units with $M=1$.
\begin{figure}
    \centering
    \includegraphics[width=\linewidth]{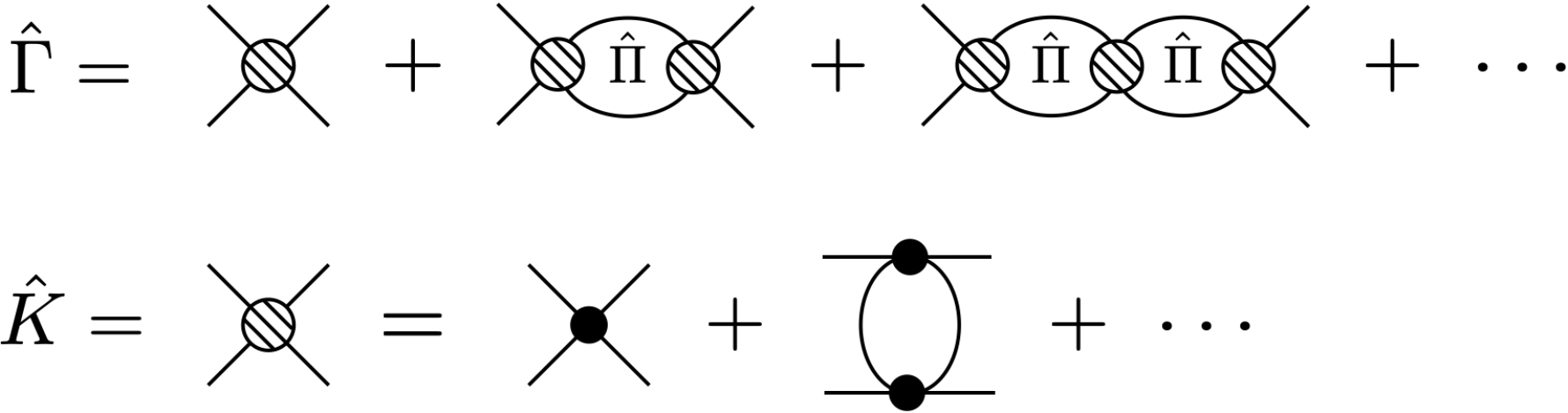}
    \caption{The BSE for the pairing vertex $\hat{\Gamma}$. The shaded circle is the pairing-irreducible potential $\hat{K}$ and the black circles are insertions of the contact interaction $g_0$.}
    \label{fig:MatrixFigs}
\end{figure}

There is a gapless Goldstone mode in the superfluid phase due to spontaneous breaking of $U(1)$ particle number symmetry.
This gapless mode manifests as a pole at zero total four-momentum in the pairing correlator
$\langle  \Phi(k) \Phi^{\dagger}(k')\rangle$. where $\Phi(k) = [\psi_{\uparrow}(-k) \psi_{\downarrow}(k), \, \psi^{\dagger}_{\downarrow}(k) \psi^{\dagger}_{\uparrow}(-k)]$ is the two-component pairing field~\cite{kagan2013modern,Pisani_2018,Haussmann1993CrossoverFB}.
The pole is inherited by the amputated vertex function $\hat{\Gamma}$ that is exactly given by a matrix Bethe-Salpeter equation (BSE),
\[
\hat\Gamma
\ = \ 
\hat K+\hat K\,\hat\Pi\,\hat\Gamma
\ = \
\left(\hat{\mathds 1}-\hat K\,\hat\Pi\right)^{-1}\hat K ,
\]
where $\hat K$ is the exact pairing-irreducible potential and $\hat\Pi$ is built from the full Nambu–Gor'kov propagators. 
Products imply matrix multiplication
in two-dimensional pairing space and integration over intermediate relative momenta.
The BSE is shown diagrammatically in Fig.~\ref{fig:MatrixFigs}.

The poles of the vertex function correspond to the zero eigenvalues of $(\hat{\mathds{1}}-\hat{K}\, \hat{\Pi})$.
Symmetries decouple the BSE into separate channels in two-dimensional pairing space and the Goldstone mode couples to the linear combination $\Gamma_- = \Gamma_{11}-\Gamma_{12}$, with $K_-$ and $\Pi_-$ defined analogously. The Goldstone pole condition is
$(1-K_-\, \Pi_-)\phi_-=0$ where $\phi_-$ is the zero-eigenvector.
Restoring the dependence on relative momentum and re-arranging gives,
\begin{align}
\int_m \! K_-(k,m)\,\Pi_-(m)\phi_{-}(m)
\ = \ \phi_{-}(k) \ ,
\label{eq:exact_gap}
\end{align}
where $\int_m\equiv\int\!d^4m/(2\pi)^4$.
The gap is defined as the lowest positive-energy pole in the Nambu–Gor'kov propagator and enters the Goldstone pole condition through internal propagators in both $K_-$ and $\Pi_-$.

Eq.~\eqref{eq:exact_gap} is homogeneous in $\phi_-$ and can be projected onto a single scalar condition by multiplying by $\Pi_-(k)$ and integrating over $k$. 
In SI~\ref{sec:power_counting} it is shown that a suitable normalization can be chosen to re-arrange the gap equation as,
\begin{align}
1 =  -g_0\int_k\Pi_-(k)  +  \frac{\int_{k,m}\Pi_-(k)\delta K_-(k,m)\Pi_-(m)\phi_-(m)}{\int_k\Pi_-(k)} ,
\label{eq:exact_scalar_gap}
\end{align}
where $K_- = -g_0 + \delta K_-$.
Up to this point, no approximations have been made, and Eq.~\eqref{eq:exact_scalar_gap} is an exact relation that the gap must satisfy.

The trans-series coefficients in Eq.~\eqref{eq:gapTransSeries} are extracted by expanding Eq.~\eqref{eq:exact_scalar_gap} in explicit powers of $g=(2\pi^2) \lambda/k_F$, and then expanding the resulting terms in $\Delta$.
In SI~\ref{sec:power_counting} it is shown that $\phi_-(k)$ can be iteratively solved in terms  of $\Pi_-$ and $K_-$, which in turn can be expanded as,
\begin{align}
K_- = -g_0  +  K_-^{[2]} + K_-^{[3]} + \cdots \ , \quad
\Pi_- = \Pi_-^{[0]} + \Pi_-^{[2]} + \cdots 
\label{eq:explicit_expansion}
\end{align}
where the superscript $[b]$ denotes the explicit order in $g$ at fixed $\Delta$.
The leading self-energy is absorbed into the chemical potential and gap giving $\Pi_-^{[1]}=0$.
Collecting terms with the same explicit order, the gap equation becomes 
\begin{align}
    1 \ = \   \sum_{b\geq 1}R^{[b]} \ , 
    \label{eq:Rb_gap}
\end{align}
where the first three terms are
\begin{align}
R^{[1]} &= -g_0\int_k\!\Pi_-^{[0]}(k)\ , \nonumber \\
R^{[2]}  &= \frac{\int_{k,m}\Pi_-^{[0]}(k)\,K_-^{[2]}(k,m)\,\Pi_-^{[0]}(m)}{\int_k\Pi_-^{[0]}
(k)}\ ,\nonumber \\
R^{[3]}  &= -g_0\int_k\!\Pi_-^{[2]}(k)+\frac{\int_{k,m}\Pi_-^{[0]}(k)\,K_-^{[3]}(k,m)\,\Pi_-^{[0]}(m)}{\int_k\Pi_-^{[0]}(k)} \ .
\label{eq:expanded_gap}
\end{align}
\begin{table}[t]
\caption{Contributions to the trans-series coefficients through $b=3$.
A term of explicit order $g^b$ carrying $p$ net logarithms first contributes to $c_{b-p-1,m}$.
In the hard region the internal energies $\xi$, defined in Eq.~\eqref{eq:E_def}, satisfy $|\xi|\gg\Delta$. In the soft region $|\xi|\sim\Delta$. All quantities in the Value column are known in closed form.}
\label{tab:power_counting}
\setlength{\tabcolsep}{4pt}
\renewcommand{\arraystretch}{1.3}
\begin{tabular}{llcccl}
\hline\hline
Term & Origin & $p$ & Region & Coeff. & Value \\
\hline
$R^{[1]}$ & Cooper log              & $1$  & hard & $c_{-1,0}$ & $1$ \\
$R^{[1]}$ & BCS equations                & $0$  & hard & $c_{0,0}$  & $\log(8/e^2)$ \\
$R^{[2]}$ & static $K^{[2]}_-$    & $1$  & hard & $c_{0,0}$  & $-\log(4e)/3$ \\
$R^{[2]}$ & dynamic $K^{[2]}_-$   & $0$  & hard & $c_{1,0}$  & $1.2067$ \\
$R^{[3]}$ & $K_-^{[3]}$, $Z$, $M^*$ & $1$  & hard & $c_{1,0}$  & $-1.3593$ \\
$R^{[2]}$ & $K^{[2]}_-$             & $-1$ & soft & $c_{2,1}$  & $17.6388$ \\
$R^{[3]}$ & on-shell self-energy             & $0$  & soft & $c_{2,1}$  & $6.6847$ \\
\hline\hline
\end{tabular}
\end{table}
Each $R^{[b]}$ is then expanded in $\Delta$.
Under mild assumptions on the form of this expansion, we show in the End Matter that solving Eq.~\eqref{eq:Rb_gap} for $\log x$ yields the trans-series in Eq.~\eqref{eq:gapTransSeries}.
The logarithms $\log x\sim1/\lambda$ generated in this expansion lower the effective order of a term, while the normalization $1/\int_k\Pi_-^{[0]}(k)$ raises it by one.
Hence, a term of explicit order $g^b$ carrying $p$ net logarithms first contributes to $c_{b-p-1,m}$.
Table~\ref{tab:power_counting} lists the components that contribute to the trans-series coefficients considered in this work. Extracting $c_{1,0}$ and $c_{2,1}$ requires evaluating up to $b=3$ and we next present the results from the $b=1,2,3$ sectors, with details provided in the SI.

\vspace{2em}

\begingroup
\renewcommand*\addcontentsline[3]{}  
\section{Systematic expansion of the Gap}
\endgroup
\noindent
\paragraph{Leading order}
\noindent
After integrating over energy, the leading order $b=1$ gap equation is
\begin{align}
&R^{[1]}  =  -g_0\int\!\frac{d^3l}{(2\pi)^3}\frac{1}{2E_l}\ = \ 1  \ , 
\label{eq:BCSgap}
\end{align}
where the quasiparticle energy is 
\begin{align}
    E_l=\sqrt{\xi_l^2+\Delta^2}, \quad \xi_l = |{\bf l}|^2/2- \mu
    \label{eq:E_def}
\end{align}
with $\mu$ the chemical potential.
As explained in SI~\ref{sec:LO}, it suffices to take $\mu=E_F$  for the trans-series coefficients considered here.
Eq.~\eqref{eq:BCSgap} is the BCS gap equation and the ultraviolet divergence is removed by renormalizing $g_0\to g$.
Expanding $R^{[1]}$ for small $\Delta$ generates terms of the form $\Delta^{m} \log^p{\Delta}$ with even $m$. Relevant to this work are the $m=0$ coefficients,
\begin{align}
c_{-1,0}^{[1]}=1 \ ,
\qquad 
c_{0,0}^{[1]}=\log{\left (8/e^2 \right )} \  ,
\label{eq:BCS_coeffs}
\end{align}
where the superscript $[1]$ denotes that they come from leading order.
While $c_{-1,0}$ is completely determined at leading order, next-to-leading order shifts $c_{0,0}$.
The BCS equation generates only even powers of $\Delta$, so $c^{[1]}_{n,m} = 0$ for odd $m$, and any linear-in-gap contribution must come from higher orders.

\paragraph{Next-to-leading order}
\noindent
The new element in $R^{[2]}$ of Eq.~\eqref{eq:expanded_gap} is the one-loop correction to the pairing-irreducible potential $K_-^{[2]}$ shown in Fig.~\ref{fig:MatrixFigs}.
The potential is no longer momentum independent and after integrating over energy one finds,
\begin{align}
K^{[2]}_{-} = g^2 \int \! \frac{d^3 l}{(2\pi)^3} \frac{(\Delta^2+\xi_{\bf l}\xi_{{\bf l}+{\bf q}} - E_{\bf l}E_{{\bf l}+{\bf q}})(E_{\bf l}+E_{{\bf l}+{\bf q}})}{2E_{\bf l}E_{{\bf l}+{\bf q}}[(E_{\bf l}+E_{{\bf l}+{\bf q}})^2-q_0^2-i\epsilon]}  ,
\label{eq:K2minus}
\end{align}
where $(q_0, {\bf q})$ is the four-momentum transfer and we have taken $g_0\to g$ as there are no ultraviolet divergences to renormalize at this order.

The integrals in $R^{[2]}$ can be split into regions in which the energies $\xi$ of the inner and outer lines of $K_-^{[2]}$ are either hard, $|\xi|\gg\Delta$, or soft, $|\xi|\sim\Delta$.
The outer integration over $\Pi_-^{[0]}$ produces a Cooper logarithm from the interval $\Delta\ll|\xi|\ll E_F$ if the rest of the integrand tends to a constant near the Fermi surface.
This occurs when the internal lines are hard as the leading dependence can be extracted by placing the outer lines on the Fermi surface, and $K_-^{[2]}$ reduces to the $s$-wave average of the static Lindhard function.
This gives a contribution to $c_{0,0}^{[2]}$ which, together with $c_{0,0}^{[1]}$, fully determines the GMB value, $c_{0,0}=\tfrac{7}{3}\log(2/e)$~\cite{GorkovMelik}. 
Non-static pieces with outer lines away from the Fermi surface carry one fewer Cooper logarithm and contribute to $c_{1,0}$, which is fully determined at next-to-next-to-leading order.

\begin{figure}
    \centering
    \includegraphics[width=\linewidth]{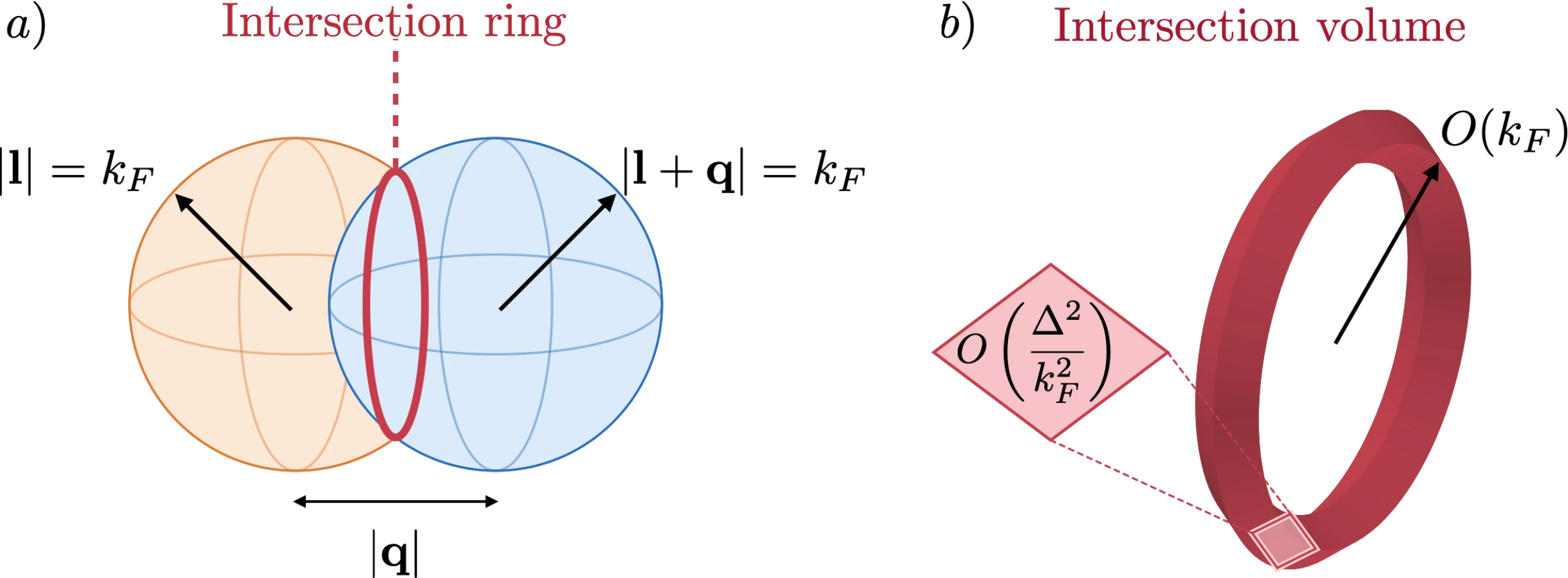}
    \caption{The integration region that generates the linear-in-$\Delta$ term in the irreducible pairing potential is the intersection of two spherical shells, each with radius $k_F$ and thickness proportional to $\Delta/k_F$.
    a) For zero thickness and sphere separation $0<|{\bf q}|< 2k_F$, the intersection is a ring with $O(k_F)$ radius.
    b) Adding in the thickness, the intersection volume has a toroidal shape with cross-sectional area $O(\Delta^2/k_F^2)$ and volume $O(\Delta^2/k_F)$.}
    \label{fig:ring_intersection}
\end{figure}
The situation changes in the soft region where
$\{|q_0|,|\xi_{\bf l}|,|\xi_{{\bf l}+{\bf q}}|\}\sim \Delta$.
For these kinematics, the integrand in Eq.~\eqref{eq:K2minus} scales as $1/\Delta$ and the overall contribution to $K_-^{[2]}$ depends on the scaling of $d^3 l$, i.e. the available phase space.
In the soft regime, the two internal momenta must satisfy
$\big||{\bf l}|-k_F\big|\lesssim\Delta/k_F$ and
$\big||{\bf l}+{\bf q}|-k_F\big|\lesssim\Delta /k_F$.
Geometrically, these constraints describe the intersection of two
spherical shells, each of radius $k_F$ and thickness proportional to
$\Delta/k_F$, displaced by ${\bf q}$.
For $|{\bf q}|\sim k_F$, the intersection volume scales as $\Delta^2/k_F$, as illustrated in Fig.~\ref{fig:ring_intersection}.
Consequently,
\[
\left .\frac{K_-^{[2]}}{g^2}\right|_{S}
\sim
\underbrace{O(\Delta^2/k_F)}_{\text{intersection volume}}
\ \times \ 
\underbrace{O(1/\Delta)}_{\text{integrand}}
\ = \ 
O(\Delta/k_F) \ .
\]
Compared with the leading contribution from the hard region $K_-^{[2]}\big |_H\sim g^2k_F$, the soft contribution is suppressed by one power of $x$.
In contrast, expanding the hard region in produces only even powers of $x$.

A linear-in-gap term in the irreducible potential can only contribute to a $c_{n,1}$ coefficient if it persists when integrated against the outer $\Pi_-$ in $R^{[2]}$.
In SI~\ref{sec:NLO}, we show that this selects the region where the outer energies are also soft and $R^{[2]}$ reduces to
\begin{align}
\!\!
   R^{[2]}\big |_S  =  -\frac{g^3k_F}{128\pi^6}\int_{-\infty}^{\infty}\!d^4\xi \frac{\Delta^2+\xi_1\xi_2-E_1E_2}{E_0E_1E_2E_3(E_0+E_1+E_2+E_3)}  ,
    \label{eq:soft_R2}
\end{align}
where $E_{0,3}$ are the outer energies and $E_{1,2}$ are the inner energies.
The extra factor of $g$ has come from replacing the Cooper logarithm in the denominator by its leading value $\int_k\Pi^{[0]}_-(k)\to-1/g$.
The outer and inner integrals are coupled by the summed energy in the denominator, so the outer integrands scale as $1/|\xi_{0,3}|^2$ for $\Delta \ll |\xi_{0,3}|$, and no Cooper logarithms are produced.
Hence, this term has $b=2$, $p=-1$ and the leading coefficient in the linear-in-gap sector is $c_{2,1}$.
In SI~\ref{sec:NLO} we show that, after removing the overlap with the hard region, Eq.~\eqref{eq:soft_R2} reduces to moments of modified Bessel functions of the second kind evaluated in Ref.~\cite{Bailey:2008ib}.
The result is,
\begin{align}
c_{2,1}^{[2]}  =\frac{\pi^2}{2}+\frac{\pi^4}{8}{}_4F_3\!\left(\tfrac12,\tfrac12,\tfrac12,\tfrac12;\,2,1,1;\,1\right) \ = \ 17.63884\ldots 
\label{eq:cNLO}
\end{align}
where ${}_4F_3$ is a generalized hypergeometric function.

\paragraph{Next-to-next-to-leading order}
\noindent
Table~\ref{tab:power_counting} shows that the $b=3$ terms in the gap equation contribute to both $c_{1,0}$ and $c_{2,1}$.
In the $c_{n,0}$ sector, $\Delta$ only enters through the Cooper logarithm and $c_{1,0}$ can also be obtained from the pairing instability of the free Fermi gas~\cite{THOULESS1960553}.
This approach is simpler than the present method and was taken in Ref.~\cite{BCF2024}, see also Refs.~\cite{Maiti_2013,Baranov,Efremov_2000} for applications to $p$-wave pairing.
We follow this strategy, correcting errors in Ref.~\cite{BCF2024}, and find that
\begin{align}
&c_{1,0}  =  \tfrac{79}{180}+\tfrac{5\pi^2}{36}-\tfrac{3\pi^3}{16}-\tfrac{44}{15}\log2+(\tfrac{4}{9}-\tfrac{\pi}{4})\log^22\nonumber \\
&+4{\rm G}\log2+8{\rm Im}\,{\rm Li}_3\!\big(\tfrac{1+i}{2}\big)-\tfrac{7}{8}\zeta(3) = -0.15263\ldots
\label{eq:c10}
\end{align}
where $\zeta$ is the Riemann zeta function, ${\rm G}$ is the Catalan's constant and ${\rm Li}_3$ is the trilogarithm function.
See SI~\ref{sec:NNLO} for details.

Relevant to $c_{2,1}$ is the self-energy correction to $\Pi_-^{[2]}$ in $R^{[3]}$.
Surprisingly, there is an exact cancellation in the soft region and $
\int_k\Pi_-^{[2]}(k) \big |_S =0$.
However, this is not the only place that the self-energy enters.
The gap is defined as the lowest positive-energy pole of the Nambu–Gor'kov propagator, and the shift of this pole by the self-energy must also be included.
This shift is evaluated at the pole itself and there is no cancellation in the soft region.
This contributes to $c_{2,1}$ and the requisite integrals again reduce to moments of modified Bessel functions of the second kind.
We were unable to find these integrals in the literature but the following closed form is identified with PSLQ~\cite{Ferguson1999AnalysisOP} and verified to 500 digits,
\begin{align}
    c_{2,1}^{[3]}  =  \frac{1}{\sqrt{2}}\left[\tfrac{47}{1536}\big[\Gamma(\tfrac18)\Gamma(\tfrac38)\big]^2-\frac{\pi^4}{\big[\Gamma(\tfrac18)\Gamma(\tfrac38)\big]^2}\right] =  6.684\ldots 
\end{align}
Combining with Eq.~\eqref{eq:cNLO} gives
\begin{align}
    c_{2,1} \ = \ c_{2,1}^{[2]}+c_{2,1}^{[3]} \ = \ 24.32356\ldots 
\end{align}
We have assumed that, as for $K_-^{[2]}$, soft contributions from $K_-^{[3]}$ and beyond pick up no Cooper logarithms from the outer lines and first enter at $c_{3,1}$, so that $c_{2,1}$ is complete.

\begin{figure}
    \centering
    \includegraphics[width=\linewidth]{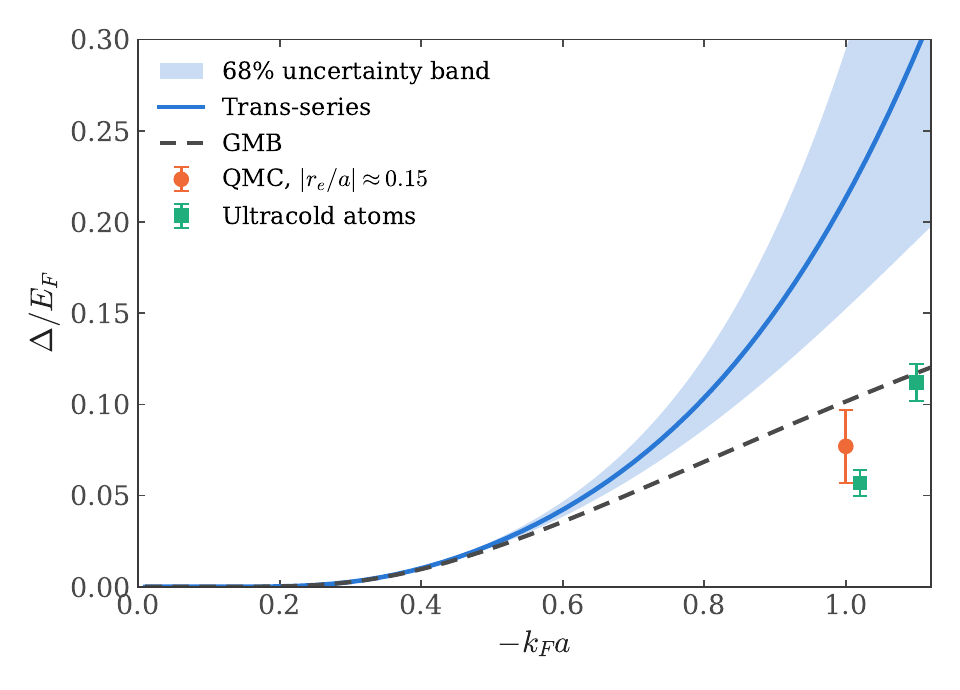}
    \caption{The solid blue line is the superfluid gap predicted from the weak coupling expansion including the $c_{1,0}$ and $c_{2,1}$ corrections to GMB.
    It is compared to GMB (dashed gray line), QMC with an effective range $|r_e/a|\approx0.15$~\cite{Gezerlis:2007fs} (orange) and ultracold atom experiment~\cite{Biss_2022} (green).
    The shaded blue region is a $68\%$ uncertainty band on the trans-series prediction.
    }
    \label{fig:gap_prediction}
\end{figure}

\begingroup
\renewcommand*\addcontentsline[3]{}  
\section{Gap prediction and comparison with data}
\endgroup
\noindent
To ensure no effects are included beyond working order, we solve for $x$ in Eq.~\eqref{eq:gapTransSeries} and expand to obtain,
\begin{align}
    x \ = \ x_0\left (1+c_{1,0}\lambda \ + \ c_{2,1}\lambda^2 x_0 \right ) \ .
\end{align}
Fig.~\ref{fig:gap_prediction} compares this prediction to GMB as well as QMC simulations~\cite{Gezerlis:2007fs} and ultracold atom experiments~\cite{Biss_2022}.
The QMC data models neutron matter and has an effective range $|r_e/a|\approx0.15$.
The leading effective-range correction can be absorbed into an effective scattering length, $1/a_{\text{eff}}= 1/a-r_e k_F^2/2$~\cite{Schwenk:2005ka,Beane:2021bpr}. 
Using the GMB solution to estimate its size, removing the effective range would raise the QMC gap by $\approx 12\%$.
The blue $68\%$ uncertainty band is obtained by treating the next-order coefficients $\{c_{2,0},c_{3,1},c_{-2,2}\}$ as independent normal variables whose standard deviations are set by the magnitudes of the known coefficients in their respective trans-series sectors.
This is explained in SI~\ref{sec:uncertainty}.
The leading source of uncertainty at low densities comes from $c_{2,0}$ but transitions to $c_{3,1}$ for $-k_F a >0.76$. 
The contribution from $c_{-2,2}$ is negligible.

There is about a $2\sigma$ disagreement between the trans-series prediction and the points near $-k_F a\approx 1$.
At this value of the coupling, the trans-series corrections are larger than the GMB gap itself and the expansion is likely no longer reliable.
However, both known corrections to GMB favor an increase of the gap, while QMC and experiment either agree with GMB or are lower.
Additional data at weak coupling, $-k_F a\in[0.6,0.9]$, are needed to validate the weak coupling expansion developed here.

\vspace{2em}

\begingroup
\renewcommand*\addcontentsline[3]{}  
\section{Outlook}
\endgroup
\noindent
We have shown that the weak-coupling expansion of the gap is a trans-series in $(\lambda,x_0)$ that includes a previously unidentified sector of corrections that are linear in the gap.
This sector originates from virtual quasiparticles with energies $E\sim\Delta$ in the self-energy and the irreducible potential between Cooper pairs.
The leading coefficient is obtained in closed form, $c_{2,1}=24.323\ldots$, which is large compared to the other known coefficients $|c_{n,0}|\leq1$.
This is not necessarily inconsistent with naturalness as $c_{2,1}$ is generated by soft kinematics that are distinct from the hard processes that fix the $c_{n,0}$ coefficients.
The dominant correction to GMB is $c_{2,1}$ for $-k_F a>0.49$, and the trans-series structure predicts a gap with an uncertainty band based on estimated higher-order terms.
The prediction lies about $2\sigma$ above QMC and experimental measurements for $-k_F a\approx 1$. 
However, this is beyond where the trans-series is reliable and data at $-k_F a\in[0.6,0.9]$ are needed to properly test the weak coupling expansion.

The phase-space argument of Fig.~\ref{fig:ring_intersection} carries over to two dimensions and the two-dimensional Fermi gas should also have linear-in-gap corrections whose leading trans-series coefficients can be computed using the methods developed here.
At the same time, the computational cost is reduced for QMC in two dimensions as fewer particles are needed to obtain accurate results at low density.
New QMC simulations, building on Refs.~\cite{PhysRevA.93.023602,PhysRevA.92.033603}, could test the weak coupling expansion in two dimensions.
An accurate gap would also fix the leading non-perturbative contribution to the energy density, which complements its perturbative expansion~\cite{Wellenhofer:2021eis,Beane:2022wcn}.

In tandem, the error bands on the trans-series prediction could be tightened by  going beyond the leading corrections to the GMB asymptotics.
Many of the diagrams that are needed to compute $c_{2,0}$ were evaluated in Ref.~\cite{Efremov_2000} to determine the induced $p$-wave gap~\cite{Fujimoto:2025zfa,Kumamoto:2024muw,Kohn:1965zz,Maiti_2013}.
An important extension will be to replace the contact interaction used in this work with a finite-range potential that models the forces between neutrons.
This would enable gap predictions in dilute neutron matter with quantified truncation uncertainties, and may help clarify the differences among many-body treatments and QMC~\cite{Gandolfi_2022,Allard:2024dtl}.
An accurate determination of the neutron gap would provide valuable microscopic input for pairing models of neutron-rich nuclei~\cite{Grams:2024kzz,Margueron:2025ugs} and for the evolution of neutron-star crusts~\cite{Allard:2024dtl}.

\vspace{2em}

\begingroup
\renewcommand*\addcontentsline[3]{}  
\begin{acknowledgements}
\noindent
We acknowledge Yuki Fujimoto and Sanjay Reddy for helpful comments.
S.B. and C.S. acknowledge support of
the U.~S.~Department of Energy (DOE) grant \textbf{DE-FG02-97ER-41014} (UW Nuclear Theory).
R.F.~acknowledges support from the U.S.~DOE QuantISED program through the theory consortium “Intersections of QIS and Theoretical Particle Physics” at Fermilab, from the U.S.~DOE, Office of Science, Accelerated Research in Quantum Computing, Quantum Utility through Advanced Computational Quantum Algorithms (QUACQ), and from the Institute for Quantum Information and Matter, an NSF Physics Frontiers Center
(PHY-2317110). 
R.F.~additionally acknowledges support from a Burke Institute prize fellowship.
This work was performed in part at Aspen Center for Physics, which is supported by NSF grant PHY-2210452.

\textit{Disclaimer of AI use}: The Claude Opus-5.5, Claude Fable-5 and Claude Fable-5.1 models from Anthropic and the GPT-5.6 Sol and GPT-6 Astra models from OpenAI were used in the preparation of this manuscript.
The trans-series structure and linear-in-gap sector were first identified by the authors and then refined by AI.
Many of the strategies for evaluating the Feynman diagrams in closed form were first developed by AI and then cross checked by the authors.
The authors take full responsibility for the content of this manuscript.

\end{acknowledgements}
\endgroup

\clearpage
\let\savedaddcontentsline\addcontentsline
\renewcommand{\addcontentsline}[3]{}
\bibliography{bibi}
\let\addcontentsline\savedaddcontentsline
\clearpage

\onecolumngrid
\begingroup
\renewcommand*\addcontentsline[3]{}  %
\section*{End Matter}
\label{app:transseries}
\noindent
Here we show that the weak-coupling expansion of the gap is the trans-series in $\lambda$ and $x_0$ given in Eq.~\eqref{eq:gapTransSeries}.
We make three assumptions:
\begin{enumerate}
    \item[1.] At fixed $\mu$, the pole equation in Eq.~\eqref{eq:Rb_gap} admits weak-coupling expansions in powers of $\lambda$, $x$ and $L\equiv\log x$.
    The number density equation in SI Eq.~\eqref{eq:Number_Density} determines the corrections to $\mu$ as an expansion of the same form, whose $x^0$ part is the standard Fermi-liquid power series in $\lambda$.
    \item[2.] The expansion only has non-negative powers of $x$.
    \item[3.]  In the $x^m$ sector with $m\geq1$, the power of $L$ is at most some finite $Q_m$ at all orders in $\lambda$. In the $x^0$ sector, every term other than the leading constant is at least of order $\lambda$ and $Q_0=1$.
\end{enumerate}
Assumption 1 holds for all contributions that have been evaluated. 
If assumption 2 were false, negative powers of $x$ would dominate exponentially as $x\to0$ and alter the established GMB asymptotics.
Assumption 3 states that, in the $x^0$ sector, each $R^{[b]}$ carries at most one net Cooper logarithm, $p\leq1$ in Table~\ref{tab:power_counting}.
The breakdown of this would represent a new type of infrared singularity in the normal Fermi liquid, distinct from the Cooper logarithm, and would be at odds with the renormalization group arguments in Refs.~\cite{Shankar,Polchinski:1992ed}.
Higher powers of $L$ occur for $m\geq1$, for example, the mean-field BCS gap and number equations give a term $\lambda x^2L^2/8$, see SI Eq.~\eqref{eq:LOgapExpand}.
The bound on the maximum power of $L$ is needed as otherwise there could be a tower of terms like $\lambda^nL^n$ that would contribute at the same order since $L\sim 1/\lambda$.

After using the number density equation SI Eq.~\eqref{eq:Number_Density} to solve for $\mu/E_F$ order by order in $x$, $\lambda$ and $L$, the assumptions imply that the pole condition in Eq.~\eqref{eq:Rb_gap} can be organized as
\begin{align}
1\ + \ \sum_{m\geq0}\sum_{n\geq1}\sum_{q\leq Q_m}A_{n,m,q}\,\lambda^n x^m L^q \ = \ 0 \ ,
\label{eq:diagram_expansion}
\end{align}
Powers of $\lambda$ come from vertices in Feynman diagrams and therefore $n\geq1$. 
The leading Cooper logarithm gives $A_{1,0,1}=-1$.
To put this into the desired trans-series form we will repeatedly apply the implicit function theorem (IFT):
\begin{theorem}[Implicit function theorem]
\label{theorem:IFT}
Let $F(u,v)=\sum_{j,k\geq0}F_{jk}\,(u-u^*)^j\,v^k$, where each coefficient $F_{jk}$ is a Laurent series in $w$ with finitely many negative powers. If
\begin{enumerate}
    \item $F(u^*,0)=0$, and
    \item $\partial_uF(u^*,0)\neq0$,
\end{enumerate}
then there is a unique power series $u(v)=u^*+\sum_{k\geq1}u_k\,v^k$ that satisfies $F(u(v),v)=0$, and whose coefficients $u_k$ are Laurent series in $w$.
\end{theorem}
This theorem does not require any of the series to converge.

It is convenient to define $y\equiv\lambda L$ and treat $\lambda$, $x$ and $y$ as independent variables.
At the end, we set $x=e^{y/\lambda}$.
Eq.~\eqref{eq:diagram_expansion} becomes the constraint
\begin{align}
F(\lambda,x,y) \ \equiv \ 1+\sum_{m\geq0}\sum_{n\geq1}\sum_{q\leq Q_m}A_{n,m,q}\,\lambda^{n-q}x^m y^q \ = \ 0 \ .
\label{eq:rearrange_implicit_gap_expansion}
\end{align}
By assumption 3, a given power $\lambda^kx^m$ receives contributions only from $1\leq n\leq k+Q_m$, so the coefficient of each $x^m$ is a Laurent series in $\lambda$ with a pole of order at most $Q_m-1$.
Negative powers of $y$ can be expanded around any nonzero $y^*$, so $F$ has the form required by the IFT.
We will first apply the IFT to obtain
\begin{align}
    y(\lambda,x) \ = \ \sum_{m\geq0}x^m y_m(\lambda) \ ,
    \label{eq:ym}
\end{align}
where $y_m(\lambda)$ is a Laurent series in $\lambda$.
To obtain $y_0(\lambda)$ apply the IFT to $F(\lambda,0,y)$ with $u=y$ and $v=\lambda$.
At $\lambda=0$, $F(0,0,y)=1-y$ and therefore IFT condition 1 is satisfied with $y^*=1$. Condition 2 is satisfied because
\[
\partial_y F(\lambda,0,y)\vert_{y=1,\lambda=0} \ = \ \left ( \sum_{n\geq1}\sum_{q\leq 1} A_{n,0,q}\lambda^{n-q}y^{q-1}q\right )\Bigg |_{y=1,\lambda=0} \ = \ A_{1,0,1} \ = \ -1 \  \neq  \ 0\ ,
\]
where we used $Q_0=1$ by assumption 3.
The IFT gives $y_0(\lambda)=1+\sum_{n\geq1}b_n\lambda^n$.
Next, we apply the IFT to Eq.~\eqref{eq:rearrange_implicit_gap_expansion} with $u=y$ and $v=x$ and coefficients that are Laurent series in $w=\lambda$.
Condition 1 is satisfied with $u^*=y_0(\lambda)$, since $F\big(\lambda,0,y_0(\lambda)\big)=0$ by construction.
Condition 2 holds because
\[
\partial_yF\big(\lambda,x,y\big)\bigg \vert_{x=0,y=y_0(\lambda)} \ = \ \left ( \sum_{n\geq1}\sum_{q\leq 1} A_{n,0,q}\lambda^{n-q}y_0(\lambda)^{q-1}q\right ) \ = \ A_{1,0,1}+O(\lambda) \ \neq \  0 \ .
\]
Applying the IFT gives the desired form in Eq.~\eqref{eq:ym}.
Dividing Eq.~\eqref{eq:ym} by $\lambda$ and setting $y= \lambda L$ gives
\begin{align}
L \ = \ \sum_{m\geq0}x^m\ell_m(\lambda) \ = \ \sum_{m\geq0}\sum_{n\geq-N_m}a_{n,m}\lambda^nx^m \ ,
\label{eq:implicit_lambda_x_expansion}
\end{align}
with $\ell_m(\lambda)=y_m(\lambda)/\lambda$, each $N_m$ finite and $a_{n,m}$ coefficients of the Laurent series.
The Cooper logarithm manifests in ${\ell }_0=y_0/\lambda = 1/\lambda+O(1)$.

The remaining step in deriving the trans-series expansion is to remove the $x$ dependence on the right-hand side.
First, factor out the asymptotic (GMB) gap,
\[
x_0 \ = \ e^{1/\lambda + c_{0,0}} \ , \qquad c_{0,0} \ = \ a_{0,0} \ ,
\]
and define
\begin{align}
L \ = \ \log{x_0} +u \ , \qquad u \ = \ A(\lambda) + \sum_{m\geq1} x^m\ell_m(\lambda) \ ,
\end{align}
with $A(\lambda)=\sum_{n\geq1}a_{n,0}\lambda^n$.
Eq.~\eqref{eq:implicit_lambda_x_expansion} becomes
\begin{align}
u\ = \ A(\lambda)\ + \ \sum_{m\geq1}x_0^m e^{m u }\ell_m(\lambda) \ ,
\label{eq:u_expansion}
\end{align}
where we used $x = x_0e^u$.
Next, apply the IFT with $u=u$, $v=x_0$ and and coefficients that are Laurent series in $w=\lambda$ to
\begin{align}
F(u,x_0) \ = \ A(\lambda) \ - \ u \ + \ \sum_{m\geq1}x_0^m e^{m u }\ell_m(\lambda) \ = \ 0 \ .
\end{align}
At $x_0=0$, condition 1 is satisfied by $u^*=A(\lambda)$, and condition 2 follows from
\begin{align}
\partial_uF(u,x_0)\Big|_{u=A(\lambda),\,x_0=0} \ = \ -1 \ \neq \ 0 \ .
\end{align}
The IFT then gives
\begin{align}
u(x_0,\lambda) \ = \ A(\lambda) \ + \ \sum_{m\geq 1}x_0^mC_m(\lambda) \ .
\end{align}
As a result,
\begin{align}
u \ &= \ A(\lambda)\ + \ \sum_{m\geq1}x_0^mC_m(\lambda) \ = \   A(\lambda)\ + \ \sum_{m\geq1}\sum_{n\geq-N'_m}c_{n,m}\lambda^nx_0^m \ ,
\end{align}
and therefore the gap has the desired weak-coupling trans-series expansion in $(\lambda, x_0)$,
\begin{align}
\log{x} \ = \ \log{x_0} \ + \ \sum_{n\geq 1}c_{n,0}\lambda^n \ + \ \sum_{m\geq 1}\sum_{n\geq-N'_m} c_{n,m}\lambda^n x_0^m \ ,
\label{eq:transseries_EM}
\end{align}
with $c_{n,0}=a_{n,0}$ and each $N'_m$ finite.

\endgroup
\clearpage
\begingroup
\hypersetup{linkcolor=black}
\tableofcontents
\endgroup

\clearpage

\clearpage


\section*{Supplementary information}
\label{sec:methods}
\setcounter{section}{0}
\renewcommand\thesection{}
\renewcommand\thesubsection{\Alph{subsection}}
\renewcommand{\theHsection}{supp.\arabic{section}}
\noindent
Throughout the supplementary information we will adopt the notation that three-vectors are denoted as ${\bf k}$ and non-relativistic four-vector are $k=(k_0,{\bf k})$. 
Integration over four-momentum will be compressed as $\int\frac{d^4l}{(2\pi)^4} \to \int_l$.
The Fermi momentum and energy is $k_F$ and $E_F=k_F^2/(2M)$ where $M$ is the fermion mass.
The renormalized interaction strength is $g=4\pi a /M$ where $a$ is the $s$-wave scattering length.
The dimensionless expansion quantities are $\lambda = k_FgM/(2\pi^2) = 2 k_F a/\pi$ and $x=\Delta/E_F$.
The pairing amplitude is in general complex, and we use the global $U(1)$ phase freedom to take it real and positive.
Matrices in pairing space will be denoted by a hat, while those in the Nambu-Gor’kov basis will be denoted by a tilde.

The supplementary information is organized as follows:
SI~\ref{sec:preliminaries} provides a brief background on superfluidity and SI~\ref{sec:GoldstonePole} derives the Goldstone pole condition that must be satisfied by the gap.
The Feynman rules for the irreducible potential and self-energy are given in SI~\ref{sec:FeynmanRules}.
In SI~\ref{sec:LO}, the leading-order solution is derived.
SI~\ref{sec:power_counting} uses the knowledge of the leading-order solution to rewrite the Goldstone pole equation in a way that can be expanded in explicit powers of $g$.
This establishes the power counting that is used in the rest of the SI.
The next-to-leading order and next-to-next-to-leading order solutions to the gap are derived in SI~\ref{sec:NLO} and SI~\ref{sec:NNLO} respectively.
The method for generating the error bands on the trans-series prediction of the gap is described in SI~\ref{sec:uncertainty}.
The application of the Feynman rules to the irreducible potential is given in Appendix~\ref{app:FeynmanLONLO} and the remaining Appendices provide detailed derivations of the many-body integrals.

\subsection{Preliminaries}
\label{sec:preliminaries}
\noindent
This work considers a system of non-relativistic spin-1/2 fermions with a Hamiltonian density
\[
{\cal H} \ = \ \sum_{\sigma=\uparrow,\downarrow}\psi_\sigma^{\dagger}\left (-\frac{\vec{\nabla}^2}{2M} \right )\psi_\sigma \ + \ g_0\psi^{\dagger}_\uparrow\psi^{\dagger}_\downarrow\psi_\downarrow\psi_\uparrow \ , 
\]
where the field $\psi_\sigma^{\dagger}$ creates a fermion of spin $\sigma$ and $g_0$ is the coupling in dimensional regularization with minimal subtraction (equivalently PDS~\cite{Kaplan:1998tg} at $\mu_{\rm PDS}=0$) where power divergences vanish.
The ground state of  the free Fermi gas ($g=0$) has all momentum modes up to $k_F$ filled, schematically
\[
|\Omega\rangle_{\text{FG}} \ = \ \bigotimes_{\sigma=\downarrow,\uparrow}\bigotimes_{{\bf k}=0}^{k_F}|{\bf k},\sigma\rangle \ .
\]
In a free Fermi gas, the chemical potential is the energy at the Fermi surface $\mu = E_F$ which is related to the number density through $N/V = k_F^3/(3\pi^2)$.
A system with repulsive interactions ($a>0$) can be studied using Fermi liquid theory which perturbatively treats the effects of interactions around this ground state.
The free Fermi gas is unstable in the presence of an arbitrarily weak attractive interaction.
This can be diagnosed by examining the poles in the Cooper channel, $\langle\psi_{\uparrow}^{\dagger}(-{{\bf k}}'+P/2) \psi_{\downarrow}^{\dagger}({{\bf k}'}+P/2)\psi_{\uparrow}(-{{\bf k}}+P/2)\psi_{\downarrow}({{\bf k}}+P/2)\rangle$ at total four-momentum $P=(i\Delta,{ \bf 0})$~\cite{THOULESS1960553}.
Such a pole signals the breakdown of the Fermi liquid description as the excitation grows with time $e^{-i t \Omega}\to e^{t \Delta}$.
This is the normal-state Cooper instability and has been used to compute the $c_{n,0}$ coefficients in the trans-series expansion for the s-wave~\cite{GorkovMelik,BCF2024} and p-wave~\cite{Efremov_2000,PhysRevB.48.1097,Maiti_2013} gaps.

The proper starting point for perturbation theory with attractive interactions is not the free-fermion fixed point and its associated ground state $|\Omega\rangle_{\text{FG}}$. 
The correct description is a superfluid, and a Hubbard-Stratonovich transformation to pairing fields shows that there is a saddle point of the effective action that corresponds to the mean-field BCS ground state,
\begin{align}
|\Omega\rangle_{\text{BCS}} \ = \ \bigotimes_{{\bf k}=0}^{\infty} \left (u_k|0\rangle_{-{\bf k}\uparrow,{\bf k}\downarrow} \ + \ v_ke^{i\theta}|-{\bf k},\uparrow\rangle|{\bf k},\downarrow\rangle \right ) \ ,
\label{eq:BCS_GS}
\end{align}
where 
\begin{align}
u_k^2, v_k^2 = (1\pm\xi_k/E_k)/2 \quad , \quad E_k^2 = \Delta^2+\xi_k^2 \quad , \quad \xi_k \ = \ |{\bf k}|^2/(2M) - \mu\ .
\label{eq:BCS_occ_energy}
\end{align}
Due to the interactions, the chemical potential gets shifted, i.e. $\mu\neq E_F$, but the number density is still related to the Fermi momentum by $N/V = k_F^3/(3\pi^2)$~\cite{Luttinger:1960zz}.
The single-fermion excitations are Bogoliubov quasiparticles with energy $E_k\geq\Delta$.
Additionally, there is a gapped Higgs mode corresponding to amplitude fluctuations $v\to v+\delta v$, as well as a gapless Goldstone mode corresponding to phase fluctuations $\theta \to \theta +\delta\theta$.
The BCS wavefunction is mean-field because it factorizes over pairs of momentum modes, whereas the exact superfluid ground state has correlations between the different momentum modes.
The form of the BCS wavefunction shows that, for each momentum mode, there is a superposition over the unoccupied $|0\rangle$ and doubly occupied Cooper pair states.
For weak interactions, the superfluid gap energy $\Delta$ is very small and, for momentum away from the Fermi surface, $u_k^2,v_k^2\to (1\pm\text{sgn}(\xi_k))/2$.
This recovers the free Fermi gas distribution where all momentum modes well below the Fermi surface are occupied and the modes well above the Fermi surface are unoccupied.

\subsection{The Goldstone Pole Condition}
\label{sec:GoldstonePole}
\noindent
It is straightforward to show that the energy of $|\Omega\rangle_{\text{BCS}}$ in Eq.~\eqref{eq:BCS_GS} is independent of the phase $\theta$.
A uniform fluctuation of $\theta$ therefore costs no energy.
This collective excitation of all the Cooper pairs is the Goldstone mode that arises due to the spontaneous breakdown of $U(1)$ particle number symmetry in a superfluid.
The masslessness of the Goldstone mode is protected by the associated Ward identity and always corresponds to fluctuations in the $U(1)$ phase.

In this work, we determine the gap via a self-consistent relation that is satisfied if there is a massless $U(1)$ phase mode.
The Goldstone mode can be excited by coupling to Cooper pairs in the superfluid, and manifests as a pole in the pairing correlation function at zero total four-momentum.
Specifically, define the two component pairing field at zero total four-momentum,
\[
\Phi(k) =\ 
\frac{1}{\sqrt{2}}\begin{pmatrix}
\epsilon_{\alpha \beta} \psi_{\beta}(-k)\psi_{\alpha}(k)  \\  \epsilon_{\alpha \beta}\psi^{\dagger}_{\alpha}(k) \psi^{\dagger}_{\beta}(-k)
\end{pmatrix}
  \ .
\]
The pairing correlator is,
\begin{align}
\hat{D}(k, k') \ &= \ \langle \Phi(k)\,\Phi^{\dagger}(k') \rangle \ = \ 
\frac{1}{2}(\delta_{\alpha \gamma}\delta_{\beta \delta}-\delta_{\alpha \delta}\delta_{\beta \gamma})
\begin{pmatrix}
\langle\psi_\beta(-k)\psi_\alpha( k)  \psi^{\dagger}_\gamma(k')\psi^{\dagger}_\delta(-k')  \rangle & \langle \psi_\beta(-k)\psi_\alpha(k)\psi_\delta(- k')\psi_\gamma(k')   \rangle\\
\langle \psi^{\dagger}_\alpha(k)\psi^{\dagger}_\beta(-k) \psi^{\dagger}_\gamma(k')\psi^{\dagger}_\delta(-k')\rangle & \langle \psi^{\dagger}_\alpha(k)\psi^{\dagger}_\beta(-k) \psi_\delta(-k')\psi_\gamma(k')   \rangle 
\end{pmatrix} \nonumber \\[4pt]
&=\ D_{11}\hat{\mathds{1}} \ + \ D_{12}\hat{\tau}_1 
\end{align}
where the $\hat{\tau}_i$ are the three Pauli matrices acting in the $2\times 2$ pairing space and all correlators are connected and time-ordered. 
Matrices in this space will be denoted with a hat.
The second line has assumed time reversal symmetry and a real gap to equate the two diagonal entries and, separately, the two off-diagonal entries.
Changing the relative four-momentum $k,k'$ only effects the residue of the pole.

\begin{figure}
    \centering
    \includegraphics[width=0.5\linewidth]{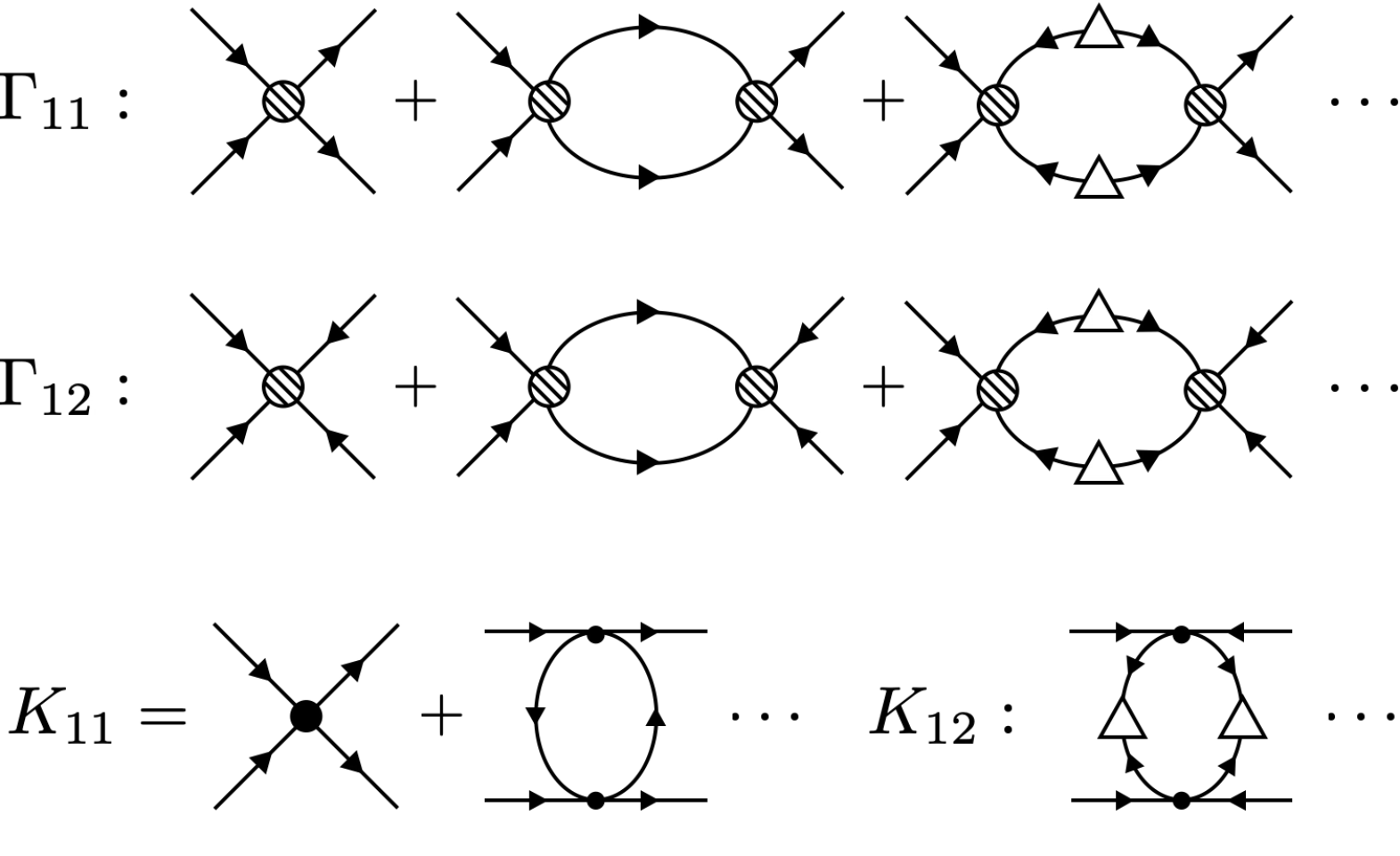}
    \caption{Upper: Diagrammatic expansion for the entries of the pairing vertex function $\hat{\Gamma}$. 
    The shaded circle is the irreducible potential $\hat{K}$ and the white triangle denotes an anomalous propagator.
    Lower: Diagrams contributing to the entries of $\hat{K}$.}
    \label{fig:FullMatrixFigs}
\end{figure}
Instead of $\hat{D}$, it is more convenient to work with the amputated vertex function $\hat{\Gamma}$ with the spin-singlet projector factored out.
Each entry of $\hat{\Gamma}$ has a diagrammatic expansion, with examples shown in Fig.~\ref{fig:FullMatrixFigs}.
It is clear that there is mixing between the different pairing sectors,\footnote{There is additionally mixing between the pairing and particle-hole sector and one could consider a four-dimensional space spanned by all the fermionic bilinears $\{\psi\psi, \psi^{\dagger}\psi,\psi\psi^{\dagger},\psi^{\dagger}\psi^{\dagger}\}$.
However, for a real gap and external legs at zero total four-momentum, the Goldstone pole is unaffected by coupling to the particle-hole sector.
As a result, the self-consistency equation for the gap decouples from the particle-hole sector and it is sufficient to only consider the pairing basis.} and the complete, i.e. non-perturbative, pairing vertex is the solution to a matrix Bethe-Salpeter equation (BSE)
\begin{align}
\hat{\Gamma}(k,k') \ &= \ \hat{K}(k,k') \ + \ \int_l \hat{K}(k,l)\,\hat{\Pi}(l)\,\hat{\Gamma}(l,k') \ ,
\label{eq:BSeqn}
\end{align}
where $\hat{K}(l,l')$ is the pairing-irreducible potential with external incoming (outgoing) fermion momentum $\pm l$ ($\pm l'$), and $\hat{\Pi}(l)$ is the pairing propagator. 
The simplified notation $\hat{\Pi}(l)$ uses the fact that the pairing propagator is diagonal in relative momentum, $\Pi(l,l')\propto\delta^4(l-l')$.
Each operator in the BSE is a $2\times 2$ matrix in pairing space and an infinite-dimensional operator in relative-momentum space.
The BSE can be formally solved through inversion,
\begin{align}
    \hat{\Gamma}(k,k') \ &= \  \int_l \Big[\big(\hat{\mathds{1}} - \hat{K}\hat{\Pi}\big)^{-1}\Big](k,l)\;\hat{K}(l,k')
    \label{eq:BSeqn_formal}
\end{align}
where the inverse operator is defined through,
\begin{align}
\int_m \Big[\hat{\mathds{1}}\,(2\pi)^4\delta^4(k-m) - \hat{K}(k,m)\,\hat{\Pi}(m)\Big]
\Big[\big(\hat{\mathds{1}} - \hat{K}\hat{\Pi}\big)^{-1}\Big](m,l)
\ = \ \hat{\mathds{1}}\,(2\pi)^4\delta^4(k-l) \ .
\end{align}
There is no zero total four-momentum pole in either $\hat{K}$ or $\hat{\Pi}$, and a pole in Eq.~\eqref{eq:BSeqn_formal} is equivalent to the following zero-eigenvalue condition,
\begin{align}
\int_m \Big[\hat{\mathds{1}}\,(2\pi)^4\delta^4(k-m) - \hat{K}(k,m)\,\hat{\Pi}(m)\Big]\vec{v}(m)
\ = \ 0 \ .
\end{align}
The eigenvector $\vec{v}(m)$ is a two-component vector in pairing space, with each component a function of the relative momentum $m$.

Time-reversal symmetry and a real gap constrain the matrices in pairing space to have the form,
\begin{align}
\hat{K} \ = \ 
\begin{pmatrix}
K_{11} & K_{12} \\
K_{12} & K_{11}
\end{pmatrix} \quad , \quad  \hat{\Pi} \ = \ \begin{pmatrix}
\Pi_{11} & \Pi_{12} \\
\Pi_{12} & \Pi_{11}
\end{pmatrix} \ .
\end{align}
This structure implies that the eigenvectors split into two classes,
\begin{align}
    \vec v_\pm(m) = \frac{1}{\sqrt2}\begin{pmatrix}1\\ \pm1\end{pmatrix}\phi_\pm(m) \ ,
\end{align}
that satisfy,
\begin{align}
   \int_m\Big[(2\pi)^4\delta^4(k-m)-K_\pm(k,m)\,\Pi_\pm(m)\Big]\phi_\pm(m)  \ =  \  \eta_\pm\,\phi_\pm(k)  \ ,
\end{align}
where $\eta_{\pm}$ are the eigenvalues and, e.g., $K_-=(K_{11}-K_{12})$.
The Goldstone mode corresponds to $\vec{v}_-$ as can be seen by considering the action of an infinitesimal phase rotation $e^{i\delta \theta}=1+i\delta \theta$ on vectors in the pairing basis,\footnote{
The other mode, $v_+$, corresponds to a collective amplitude fluctuation. 
In the BCS state, this is a fluctuation $u_k+\delta u_k$ and $v_k + \delta v_k$ in Eq.~\eqref{eq:BCS_GS}.
This mode has a nonzero excitation energy, and therefore does not show up as a pole at the kinematics being probed.}
\[
\delta
\begin{pmatrix}
\psi \psi\\
\psi^{\dagger} \psi^{\dagger}
\end{pmatrix} \ = \  \begin{pmatrix}
2i\delta \theta \, \psi \psi\\
-2i\delta \theta \, \psi^{\dagger} \psi^{\dagger}
\end{pmatrix} 
\ = \  2i\delta\theta\begin{pmatrix}
 \, \,\psi \psi\\
- \, \psi^{\dagger} \psi^{\dagger} 
\end{pmatrix} \ .
\]
The existence of a massless Goldstone mode requires that one of the eigenvalues vanishes, i.e. $\eta_-=0$ for one eigenfunction $\phi_-(m)$.
Re-arranging the eigenvector equation gives the self-consistent gap equation,
\begin{align}
\int_m K_-(k,m)\,\Pi_-(m)\phi_{-}(m)
\ = \ \phi_{-}(k) \ . 
\label{eq:BS_gapEqn}
\end{align}

The internal propagators in $K_-$ and $\Pi_-$ depend on the gap $\Delta$ and the chemical potential $\mu$.
There is an additional constraint that comes from fixing the number density,
\begin{align} 
N/V  \ &= \  \frac{k_F^3}{3\pi^2} \ =\ -2i\int_l  G(l) e^{i l_0  0^+}
\label{eq:Number_Density}
\end{align}
where $G(l)$ is a gapped propagator that will be defined below.
The two constraints in Eq.~\eqref{eq:BS_gapEqn} and Eq.~\eqref{eq:Number_Density} are sufficient to determine both $\Delta$ and $\mu$.
This is easiest to see in the case of discretized relative four-momentum, e.g. on a finite space-time lattice.
Defining $M=K_- \Pi_-$, the gap equation becomes the eigenvector equation 
\[
\sum_{j=1}^N M_{ij} (\phi_-)_j \ = \ (\phi_-)_i\ ,
\]
where the subscripts $(i,j)$ correspond to the $N$ discrete four-momentum modes. 
This equation is homogeneous so the normalization of $\phi_-$ is arbitrary and the eigenvector has $N-1$ unknowns.
The matrix $M_{ij}$ depends on the two additional unknowns, $\Delta$ and $\mu$, and in total this linear system gives $N$ equations for $N+1$ unknowns.
The number density provides the remaining constraint.
This carries through in the continuum limit and $\Delta$, $\mu$ and $\phi_-(m)$ are fixed by solving Eq.~\eqref{eq:BS_gapEqn} and Eq.~\eqref{eq:Number_Density}.

\subsection{Diagrammatic Organization and Feynman rules}
\label{sec:FeynmanRules}
\noindent
Determining the coefficients of the trans-series in Eq.~\eqref{eq:gapTransSeries} requires reducing the pole condition in Eq.~\eqref{eq:BS_gapEqn}, together with the number equation Eq.~\eqref{eq:Number_Density}, to a scalar relation that can be solved as $\log{x} = \ldots$, where the right-hand side is an expansion in powers of $\lambda$ and $x_0$.
In diagrammatic perturbation theory, explicit powers of $\lambda$ come from the interaction vertices in diagrams that contribute to $K_-$ and $\Pi_-$.
The correspondence between the number of vertices and powers of $\lambda$ is modified by logarithms, $\log x = 1/\lambda+\ldots$, that are generated from certain loop integrals.
Expanding these same loop integrals for a small gap also generates higher powers of $x$ possibly with logarithms, i.e. terms of the form $x^m\log^px$.
A power counting scheme should give a prescription for associating a diagram and a kinematical regime with a trans-series coefficient.
Each $c_{n,m}$ would then be computable from a finite set of known Feynman diagrams.
The primary focus of the present work is to extract the leading coefficient in the linear-in-gap sector, which does not require a complete power counting scheme. 
To organize perturbation theory, we compute $K_-$ and $\Pi_-$ up to an explicit order in $g=\lambda(2\pi^2)/(M k_F)$.
The coefficients we extract are complete as long as there are no unexpected logarithms that enhance nominally higher-order contributions.

\begin{figure}
    \centering
\includegraphics[width=0.4\linewidth]{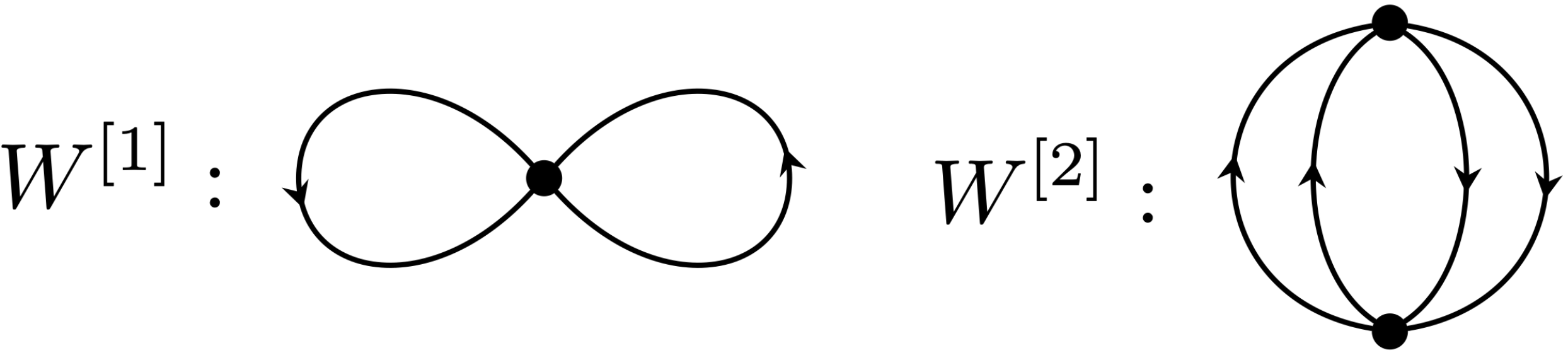}
    \caption{Hugenholtz diagrams that contribute to the functional $iW$ up to $O(g^2)$.
    Cutting these diagrams via functional differentiation is used to obtain the self-energy $\tilde{\Sigma}$ (one cut) and the pairing-irreducible potential $\tilde{K}$ (two cuts).
    The symmetry factors are ${\cal S}=\{1/2,1/8\}$ for $\{W^{[1]},W^{[2]} \}$.}
    \label{fig:hugenholtz}
\end{figure}

We now give the Feynman rules for computing $K_-$ and $\Pi_-$ in perturbation theory. 
Closed, connected and two-particle-irreducible Hugenholtz diagrams define the functional $W$, with the relevant ones shown in Fig.~\ref{fig:hugenholtz}. 
Taking one functional derivative with respect to the internal propagators opens an internal line and generates contributions to the self-energy, $\Sigma$.
Taking two functional derivatives with respect to the internal propagators opens two internal lines and generates the pairing-irreducible kernel. 
The Feynman rules are an extension of Ref.~\cite{Furnstahl:2006pa}:
\begin{enumerate}
    \item[1.] Assign non-relativistic four-momenta to all lines and enforce four-momentum conservation at each vertex. Arrows on internal legs are used to denote momentum flow, not particle number.
    \item[2.] Each vertex gets a factor of $(-ig_0)\left [(\tilde{\tau}_3)_{\alpha \gamma}(\tilde{\tau}_3)_{\beta \delta} + (\tilde{\tau}_3)_{\alpha \delta}(\tilde{\tau}_3)_{\beta \gamma}  \right ]$. The incoming Nambu indices are $\alpha,\beta$ and the outgoing indices are $\gamma,\delta$. 
    The Nambu basis is distinct from the pairing basis, hence the tilde, see rule 7.
    Each line gets a Nambu-Gor’kov propagator $i\tilde{G}_0(l)$ defined as,
    \begin{align}
    \tilde{G}_0(l_0,{\bf l}) \ = \  \begin{pmatrix}
        G_0(l_0,{\bf l}) & F_0(l_0,{\bf l}) \\
        F_0(l_0,{\bf l}) & -G_0(-l_0,{\bf l})
    \end{pmatrix}
    \label{eq:GNG}
    \end{align}
    where $G_0$ and $F_0$ are,
    \begin{align}
    G_0(l_0,{\bf l}) \ = \ \frac{l_0 + \xi_l}{l_0^2 - E_l^2+i\epsilon} \quad , \quad F_0(l_0,{\bf l}) \ = \ \frac{-\bar{\Delta}}{l_0^2 - E_l^2+i\epsilon} \ .
    \label{eq:G0F0}
    \end{align}
    \item[3.] Sum over all Nambu indices. Each trace gets a factor of $(-1)$.
    \item[4.] Integrate over all distinct four momentum $\int d^4 l/(2\pi)^4$. 
    \item[5.] Multiply by a symmetry factor ${\cal S}$. The relevant ones to this work are given in the caption of Fig.~\ref{fig:hugenholtz}. Define the resulting expression $iW$.
    \item[6a.] The $\tilde{\Sigma}$ matrix in the Nambu-Gor’kov basis is obtained by cutting one internal line by taking a functional derivative with respect to the propagators,
    \[
        i\tilde{\Sigma}_{ba}(k) \ = \ \frac{\delta\,[iW]}{\delta\,[i\tilde{G}_0(k)_{ab}]}  \ .
    \]
    where only the connected components are kept.
    \item[6b.]
    The $\tilde{K}$ matrix in the Nambu-Gor’kov basis is obtained by cutting two internal lines by taking two functional derivatives with respect to the propagators,
    \[
    i\tilde{K}_{ba;dc} \ = \ \frac{\delta^2\,[iW]}{\delta\,[i\tilde{G}_0(k)_{ab}]\delta\,[i\tilde{G}_0(k')_{cd}]}  \ .
    \]
    The incoming (outgoing) four-momenta are $k$ and $k'$ respectively.
    \item[7.] The irreducible potential carries Nambu indices associated with the spinor $\Psi(k)=[
    \psi_\uparrow(k), \,
    \psi^\dagger_\downarrow(-k)]$.
    Contracting the opened Nambu indices with $\tilde{\tau}_-$ and $\tilde{\tau}_+$ selects the pair-annihilation and pair-creation components, respectively. 
    Projection onto the pairing basis is therefore achieved by contracting the external Nambu indices with $\tilde{\tau}_{\pm}$,
    \[
    \hat{K} = 
    \begin{pmatrix}(\tilde{\tau}_-)_{ab}\tilde{K}_{ba;dc}(\tilde{\tau}_+)_{cd} & (\tilde{\tau}_-)_{ab}\tilde{K}_{ba;dc}(\tilde{\tau}_-)_{cd} \\
    (\tilde{\tau}_+)_{ab}\tilde{K}_{ba;dc}(\tilde{\tau}_+)_{cd} & (\tilde{\tau}_+)_{ab}\tilde{K}_{ba;dc}(\tilde{\tau}_-)_{cd}
    \end{pmatrix} \ ,
    \]
    where $\tilde{\tau}_{\pm} = (\tilde{\tau}_1\pm i\tilde{\tau}_2)/2$ are raising and lowering operators.
\end{enumerate}
Note that these Feynman rules are only valid for extracting spin-singlet components.\footnote{Because of this, the pairing irreducible potential is only guaranteed to be correct when projecting onto even partial waves.} 
As discussed below, the parameter $\bar{\Delta}$ in the propagators defined in Eq.~\eqref{eq:G0F0} differs from the fermionic excitation gap $\Delta$ at $O(g^2)$.

The gap equation also requires the pairing propagator $\Pi_-$.
Using Wick's theorem, the matrix elements of $\hat{\Pi}$ take the form,
\begin{align}
&\langle \psi_{\beta} \psi_{\alpha}\psi^\dagger_{\gamma}\psi_{\delta}^{\dagger}\rangle \ \stackrel{\text{two propagator contribution}}{\xrightarrow{\hspace{3cm}}} \ \langle\psi_{\alpha}\psi_{\gamma}^{\dagger}\rangle\langle\psi_{\beta}\psi_{\delta}^{\dagger}\rangle \ - \  \langle\psi_{\alpha}\psi_{\delta}^{\dagger}\rangle\langle\psi_{\beta}\psi_{\gamma}^{\dagger}\rangle \nonumber \\
&\to \Pi_{11} = -i G(l)G(-l)
\label{eq:Pi11}
\end{align}
and
\begin{align}
&\langle \psi_{\beta} \psi_{\alpha}\psi_{\delta}\psi_{\gamma}\rangle \ \stackrel{\text{two propagator contribution}}{\xrightarrow{\hspace{3cm}}} \ -\langle\psi_{\alpha}\psi_{\gamma}\rangle\langle\psi_{\beta}\psi_{\delta}\rangle \ + \  \langle\psi_{\alpha}\psi_{\delta}\rangle\langle\psi_{\beta}\psi_{\gamma}\rangle \nonumber \\
&\to \Pi_{12} =  i F(l)F(-l)
\label{eq:Pi12}
\end{align}
where the full gapped propagators are defined through
\[
\langle \psi_{\alpha}(l)\psi^{\dagger}_\beta(l)\rangle = i\delta_{\alpha\beta}\, G(l) \quad , \quad \langle \psi_{\alpha}(-l)\psi_\beta(l)\rangle = i\epsilon_{\alpha\beta}\, F(l) \ .
\]

The Wick contractions in Eqs.~\eqref{eq:Pi11} and~\eqref{eq:Pi12} involve the full propagators $G(l)$ and $F(l)$, which are the entries of the full Nambu-Gor’kov propagator,
\begin{align}
i\tilde{G}(l) \ \equiv \ \langle \Psi(l)\,\Psi^{\dagger}(l)\rangle \ = \ i\begin{pmatrix} G(l) & F(l) \\ F(l) & -G(-l) \end{pmatrix} \ ,
\end{align}
with the same structure as $\tilde{G}_0$ in Eq.~\eqref{eq:GNG}.
The full propagator is determined by Dyson's equation,
\begin{align}
\tilde{G}^{-1}(l) \ = \ \tilde{G}_0^{-1}(l) \ - \ \tilde{\Sigma}(l) \ ,
\label{eq:Dyson}
\end{align}
where the inverse of the leading gapped propagator is,
\[
\tilde{G}_0^{-1}(l) \ = \ \begin{pmatrix}
    l_0-\xi_l & \bar{\Delta}\\
    \bar{\Delta} & l_0+\xi_l
\end{pmatrix} \ = \ l_0\tilde{\mathds{1}} \  - \ \xi_l\tilde{\tau}_3 \ + \  \bar{\Delta} \tilde{\tau}_1\ .
\]
The self-energy can be decomposed in the same basis,
\begin{align}
    \tilde{\Sigma}(l) \ = \ \sigma_o(l) \tilde{\mathds{1}} \ + \ \sigma_e(l) \tilde{\tau}_3 \ + \ \Sigma_1(l) \tilde{\tau}_1 \ .
    \label{eq:Sigmadef}
\end{align}
The self-energy was computed up to $O(g^3)$ in Ref.~\cite{Urban:2026trm} for a momentum-dependent potential regulated to reproduce the contact limit.
The $O(g)$ self-energy is independent of four-momentum, and we absorb
its $\tilde{\tau}_3$ component into the chemical potential $\mu$ and its $\tilde{\tau}_1$ component is the leading gap $\bar{\Delta}$ in $\tilde{G}_0$.
With these conventions, $\tilde{\Sigma}$ in Eq.~\eqref{eq:Dyson} begins at $O(g^2)$ and $\Delta = \bar{\Delta}[1+O(g^2)]$.
Note that because $\tilde{\Sigma}$ and $\tilde{K}$ are both generated from the symmetric functional $W$, this construction preserves the Ward identity that guarantees the Goldstone pole at every order in $g$.

\subsection{Leading-order: the BCS equations}
\label{sec:LO}
\noindent
The power counting used in this work is introduced in Sec.~\ref{sec:power_counting}.
This is motivated by properties of the leading order gap solution which is detailed here.
As discussed in the previous section, we can set $\bar{\Delta}=\Delta$ as corrections first enter at next-to-next-to-leading order.

The leading contribution to Eq.~\eqref{eq:BS_gapEqn} comes from $K_-$ evaluated at tree level and $\Pi_-$ evaluated with the leading gapped propagators in Eq.~\eqref{eq:GNG}.
Here and below, a superscript $[n]$ denotes a contribution with $n$ interaction vertices. 
The relevant Hugenholtz diagram for $K_-^{[1]}$ is $W^{[1]}$ in Fig.~\ref{fig:hugenholtz}.
The application of the Feynman rules is detailed in Appendix~\ref{app:FeynmanLONLO} and gives 
\[
K_-^{[1]} \ = \ -g_0 \ .
\] 
The gap equation in Eq.~\eqref{eq:BS_gapEqn} is
\begin{align}
-g_0\int_m \Pi_-^{[0]}(m)\phi_{-}(m)
\ = \ \phi_{-}(k) \ .
\end{align}
The LHS is independent of $k$ and therefore solved exactly by $\phi_-(m) = c$ where $c$ is a constant.
We take $c=1$ without loss of generality because the gap equation is homogeneous in $\phi_-$.
Inserting the leading pairing propagators from Eqs.~\eqref{eq:Pi11} and~\eqref{eq:Pi12} gives
\begin{align}
    1\ &= \ ig_0\int_l\left [ G_0(l)G_0(-l) +F_0(l)F_0(-l)\right ] \nonumber \\
    &= -g_0\int\! \frac{d^3l}{(2\pi)^3} \frac{1}{2E_l} \ . 
    \label{eq:LOgapEqn}
\end{align}
Before going further, notice that this equation implies the leading scaling $K_-^{[1]} \Pi_-^{[0]}\sim O(1)$.
Thus, each additional insertion of $K_-\Pi_-$ in the ladder series for $\Gamma$ starts at $O(1)$ rather than $O(g)$. 
All such terms are of the same order which justifies the resummation implicit in the solution of the BSE in Eq.~\eqref{eq:BSeqn}.

The second constraint on the gap comes from the number equation in Eq.~\eqref{eq:Number_Density} which, at leading order, is 
\begin{align}
    N/V \ = \ \frac{k_F^3}{3\pi^2}
\ = \
\int\frac{d^3l}{(2\pi)^3}
\left(1-\frac{\xi_l}{E_l}\right) \ .
\label{eq:BCS_number}
\end{align}
Together, Eq.~\eqref{eq:LOgapEqn} and Eq.~\eqref{eq:BCS_number} are the coupled BCS gap and number equations. 
Eq.~\eqref{eq:LOgapEqn} has a linear ultraviolet divergence that is removed by the renormalization of $g_0$. 
The renormalization and exact evaluation are well known, see for example Refs.~\cite{Papenbrock:1998wb,BCF2024}.
Using the number equation to solve for $\mu/E_F$ and then expanding the gap equation up to $O(x^2)$ gives,
\begin{align}
    1 \ = \ \lambda \left [\log{x}\ + \ \log{\left (\frac{e^2}{8} \right )}  \ + \ \frac{x^2}{16}\left (2\log^2{x} \ - \ 12\log{2}\log{x} \ + \ 1 \ + 18\log^2{2} \right ) \right ] \ \ .
    \label{eq:LOgapExpand}
\end{align}
Solving for $\log{x}$ gives,
\begin{align}
\log{\left (x \right )} \ = \ \frac{1}{\lambda} +\log{\left (\frac{8}{e^2} \right )}  \ - \ \frac{x^2}{16}\left [2\log^2{x} \ - \  12\log{2}\log{x}+1+18\log^2{2}\right ]\ .
\label{eq:LOgapd0}
\end{align}
This recovers the Cooper log scaling $\log{x}=1/\lambda + \ldots$ that is responsible for the $O(1/g)$ component of $\Pi_-^{[0]}$.
Matching onto the leading-order transmonomials is done by substituting $x\to x_0$ and $\log x\to1/\lambda$ on the right-hand side giving,
\begin{align}
\log{\left (x \right )} \ = \ \frac{1}{\lambda} \ + \ \log{\left (\frac{8}{e^2} \right )} \ - \ \frac{x_0^2} {8}\frac{1}{\lambda^2} \ + \ \ldots \ .
\label{eq:LOgapd0expand}
\end{align}
The contribution to the trans-series coefficients can be read off,
\begin{align}
&c_{-1,0}^{[1]} \ = \ 1  , \quad c_{0,0}^{[1]} \ = \ \log{\left (\frac{8}{e^2} \right )} , \quad  c_{-2,2}^{[1]} \  = \ -1/8 \ .\
\label{eq:LO_coeffs}
\end{align}
The superscript $[1]$ indicates that these coefficients arise from the leading-order gap equation.
Only $c_{-1,0}$ is fully determined at this order as $c_{0,0}$ and $c_{-2,2}$ get contribution from $K_-^{[2]}$.
We emphasize that the mean-field BCS gap and number equations contain only sectors with even powers of $x_0$, and in particular do not generate any terms linear in the gap.

\subsection{Power counting}
\label{sec:power_counting}
\noindent
The leading-order solution reveals that the eigenvector is independent of momentum at lowest order, i.e. $\phi_-(k) = 1 + \delta\phi(k)$.
We will use this to rewrite the gap equation in Eq.~\eqref{eq:BS_gapEqn} in a way that makes powers of $g$ coming from $\delta \phi(k)$ manifest.

Separating the leading-order irreducible potential, $K_-^{[1]}=-g_0$, and writing $\phi_-(m)=1+\delta \phi(m)$ gives a gap equation, 
\begin{align}
-g_0\int_m \,\Pi_-(m)(1+\delta \phi(m)) \ + \ \int_m \delta K_-(k,m)\,\Pi_-(m)(1+\delta \phi(m))
\ = \ 1+\delta\phi(k) \ ,
\end{align}
where $K_-(k,m)=-g_0+\delta K_-$.
The eigenvector can be rescaled due to homogeneity
\[
1+\delta \phi(k) \ \to \ \frac{1+\delta \phi(k)}{1+\langle\delta \phi\rangle} \ = \ 1 \ + \ \underbrace{\frac{\delta \phi(k) - \langle \delta\phi \rangle}{1+\langle \delta\phi \rangle}}_{\delta\phi'(k)} 
\]
where
\[
\langle \delta\phi\rangle \ \equiv \ \frac{\int_l \Pi_-(l) \delta \phi(l)}{\int_l\Pi_-(l)} \ .
\]
It is straightforward to show that the rescaled eigenfunction satisfies $\langle \delta \phi'\rangle = 0$.
We will assume the eigenfunction has been rescaled in this way and drop the prime.
Using this property, the gap equation becomes,
\begin{align}
-g_0\int_m \,\Pi_-(m) \ + \ \left [ \int_m \delta K_-(k,m)\,\Pi_-(m)(1+\delta \phi(m)) \ - \ \delta \phi(k)\right ]
\ = \ 1 \ . 
\end{align}
Since the remaining terms are independent of $k$, the bracket must equal a constant, $C$. Thus,
\begin{align}
     \int_m \delta K_-(k,m)\,\Pi_-(m)(1+\delta \phi(m)) \ - \ \delta \phi(k) \ &= \ C \nonumber \\
    \int_{k,m} \Pi_-(k)\,\delta K_-(k,m)\,\Pi_-(m)(1+\delta \phi(m)) \ &= \ C\int_k \Pi_-(k) \nonumber \\
    \frac{\int_{k,m} \Pi_-(k)\,\delta K_-(k,m)\,\Pi_-(m)(1+\delta \phi(m))}{\int_k \Pi_-(k)} \ &= \ C \ .
    \label{eq:C_rearrange}
\end{align}
In the second line we have multiplied both sides by $\Pi_-(k)$, integrated over $k$ and used $\langle \delta \phi\rangle = 0 $.
Replacing the $[.]$ by $C$ gives an exact rewrite of the gap equation
\begin{align}
-g_0\int_m \Pi_-(m) \ + \ \frac{\int_{k,m}\Pi_-(k) \delta K_-(k,m)\Pi_-(m)}{\int_k\Pi_-(k)} \ + \ \frac{\int_{k,m}\Pi_-(k)\delta K_-(k,m)\Pi_-(m)\,\delta\phi(m)}{\int_k\Pi_-(k)} \ = \ 1 \ .
\label{eq:rearrange_gapEq}
\end{align}
The first line of Eq.~\eqref{eq:C_rearrange} determines the momentum dependence of the eigenfunction,
\begin{align}
\delta\phi(k) \ = \ \int_m \delta K_-(k,m)\,\Pi_-(m)\big(1+\delta\phi(m)\big) \ - \ \frac{\int_{k,m} \Pi_-(k)\,\delta K_-(k,m)\,\Pi_-(m)(1+\delta \phi(m))}{\int_k \Pi_-(k)}\ .
\label{eq:dphik}
\end{align}
The leading scaling is $\delta \phi(k)=O(g)$ as can be obtained by taking the leading terms for $\Pi_-$ and $K_-$, dropping the $\delta\phi(m)$ on the right-hand side, and assuming the integrals pick up the maximum number of Cooper logs.

To organize the gap equation in Eq.~\eqref{eq:rearrange_gapEq}, we expand $K_-=\sum_{r\geq1}K_-^{[r]}$ and $\Pi_-=\sum_{s\geq0}\Pi_-^{[s]}$ in explicit powers of the coupling coming from interaction vertices, and collect all terms with $b$ explicit powers into $R^{[b]}$.
The gap equation is
\begin{align}
\sum_{b\geq1}R^{[b]} \ = \ 1 \ ,
\label{eq:GapEqR}
\end{align}
where the first three terms are
\begin{align}
R^{[1]} &= -g_0\int_m\Pi_-^{[0]}(m) \ , \nonumber\\
R^{[2]} &= \frac{\int_{k,m}\Pi_-^{[0]}(k)\,K_-^{[2]}(k,m)\,\Pi_-^{[0]}(m)}{\int_k\Pi_-^{[0]}(k)} \ , \nonumber\\
R^{[3]} &= -g_0\int_m\Pi_-^{[2]}(m) \ + \ \frac{\int_{k,m}\Pi_-^{[0]}(k)\,K_-^{[3]}(k,m)\,\Pi_-^{[0]}(m)}{\int_k\Pi_-^{[0]}(k)} \ .
\label{eq:Rb_def}
\end{align}
We have used $\Pi_-^{[1]}=0$ as discussed at the end of Sec.~\ref{sec:FeynmanRules}.
The term containing $\delta\phi$ in Eq.~\eqref{eq:rearrange_gapEq} first enters $R^{[4]}$, as can be seen from Eq.~\eqref{eq:dphik}.
The $\delta\phi$ dependence can always be traded for $K_-$ and $\Pi_-$ by repeatedly iterating Eq.~\eqref{eq:dphik}.
Leading order, $R^{[1]}=1$, is the BCS gap equation that is treated in the previous section.

The explicit order $b$ does not by itself fix the power of $\lambda$, because $\int\Pi_-^{[0]}=O(1/g)$ through the Cooper logarithm.
In the $O(\Delta^0)$ sector, both integrals in the numerator of $R^{[2]}$ acquire Cooper logarithms and $R^{[2]}=O(\lambda)$, which fixes $c_{0,0}$.
In the $O(\Delta)$ sector, we will show that the leading contributions come from soft kinematics that carry no Cooper logarithms.
This term is $O(\lambda^3x_0)$ due to an extra power of $\lambda$ coming from the denominator of $R^{[2]}$ and contributes to $c_{2,1}$.

\subsection{Next-to-leading order: a linear-in-gap contribution}
\label{sec:NLO}
\noindent
Next-to-leading order comes from $R^{[2]}$ in Eq.~\eqref{eq:Rb_def}.
The relevant Hugenholtz diagram for $K_-^{[2]}$ is $W^{[2]}$ in Fig.~\ref{fig:hugenholtz}.
The application of the Feynman rules is detailed in Appendix~\ref{app:FeynmanLONLO} and gives
\begin{align}
K^{[2]}_{-} \ &=  -ig^2\int_r\left [F_0(r)F_0(r+\bar{Q}) \ + \ G_0(r)G_0(r+\bar{Q}) \right ] \ ,
\end{align}
where $\bar{Q} = k+k'$ with $k$ and $k'$ the four-momentum of the incoming and outgoing momentum respectively.
The bare coupling has been replaced by the renormalized coupling as there are no UV divergences at this order. We set $\bar{\Delta}=\Delta$ as corrections first enter at next-to-next-to-leading order.

Inserting $K^{[2]}_{-}$ into $R^{[2]}$ gives
\begin{align}
    \left ( \int_l\Pi_-^{[0]}(l)\right)R^{[2]} \ = \ ig^2\int_{k,r,k'} \frac{F_0(k)}{\Delta}\left [F_0(r)F_0(r+\bar{Q})+G_0(r)G_0(r+\bar{Q}) \right] \frac{F_0(k')}{\Delta} \ ,
    \label{eq:R2d4r}
\end{align}
where we have used the identity $\Pi_-^{[0]} = -i F_0(l)/\Delta$.
Performing the energy integrals and extracting the leading dependence by replacing the left hand side with $\left ( \int_l\Pi_-^{[0]}(l)\right)\to-1/g$ gives~\footnote{Corrections to this impact $c_{1,0}$.}
\begin{align}
    R^{[2]}  \ = \ -\frac{g^3}{8}\int\! \frac{d^3l_0}{(2\pi)^3}\int\! \frac{d^3l_1}{(2\pi)^3}\int\! \frac{d^3l_3}{(2\pi)^3}\frac{\Delta^2+\xi_1\xi_2-E_1E_2}{E_0E_1E_2E_3(E_0+E_1+E_2+E_3)}\ ,
    \label{eq:R2}
\end{align}
where we have defined ${\bf l}_0={\bf k}$, ${\bf l}_1={\bf r}$, ${\bf l}_2 = {\bf r}+\bar{{\bf Q}}$ and  ${\bf l}_3={\bf k}'$ and, for example, $\xi_i = \xi_{l_i}$.

Different regions of the internal momentum in
$R^{[2]}$ contribute to different $c_{n,m}$ coefficients.
There is a soft scale $S:\xi_i 
\sim \Delta$ and a hard scale $H:\xi_i\sim E_F$.
Depending on the kinematics, there can also be a marginal scale that generates a logarithm $M:\int_{\Delta}^{\Lambda} d\xi/\xi \sim \log{(\Lambda/\Delta)}$ where $\Delta<\Lambda < E_F$ is an arbitrary energy scale.
The outer integrals in Eq.~\eqref{eq:R2} are coupled to the inner ones by the $(E_0+E_1+E_2+E_3)$ term in the denominator. 
If a $c_{n,m}$ coefficient comes from the region where $E_{0,3}< E_{1,2}$ then the denominator can be expanded in $E_{0,3}/E_{1,2}$ and the integrals decouple.
This is the static approximation, and simplifies the calculation because the irreducible potential can be treated independent of the integration over $\Pi_-$.\footnote{The irreducible potential must be projected onto the s-wave as discussed in Appendix~\ref{app:contactph}.}

The breakdown of the regions that contribute to the trans-series coefficients considered in this work are provided in Table~\ref{tab:regions}.
\begin{table}
\renewcommand{\arraystretch}{1.4}
    \begin{tabular}{|c|c|c|}
        \hline
        coefficient & region of $(\xi_0,\xi_1,\xi_2,\xi_3)$ & static approximation? \\
        \hline
        $c_{0,0}$ & $(M,H,H,M)$ & yes\\
        \hline
        $c_{1,0}$ & $(H,H,H,M)$ and $(M,H,H,H)$& no\\
        \hline
        $c_{2,1}$ & $(S,S,S,S)$ & no \\
        \hline
        $c_{-2,2}$ & $(M,M,M,M)$ & no \\
        \hline
    \end{tabular}
\caption{The regions of $R^{[2]}$ in Eq.~\eqref{eq:R2} that generate the $c_{n,m}$ terms in the trans-series expansion of the gap. 
For each coefficient (left column), the integration region of the $\xi_i$ is given in the center column. 
The soft region is $S:|\xi|\sim \Delta$, the marginal region is $M:\Delta<|\xi|< E_F$ and the hard region is $H:|\xi|\sim E_F$.
The right column states whether the static approximation is valid.}
\label{tab:regions}
\end{table}
The Gor'kov screening contribution to $c_{0,0}$ comes from the $(M,H,H,M)$ region of the integrand where the static approximation is valid.
The $(M,H,H,H)$ and $(H,H,H,M)$ regions pick up one fewer Cooper logarithm and contribute one order higher, to $c_{1,0}$.
We will show that a linear-in-$\Delta$ contribution, $c_{2,1}$, comes from the $(S,S,S,S)$ soft region.
Note that if, for example, a $(M,S,S,M)$ region existed, then there could be a non-zero contribution to $c_{0,1}$.
However, there is no logarithm generated when $\xi_{1,2}\sim \Delta$ since the outer integrations become $\sim\int d\xi_{0,3}/\xi_{0,3}^2$ which does not generate a Cooper logarithm. 
Lastly, there is a contribution to the $c_{-2,2}$ coefficient that comes from a marginal $(M,M,M,M)$ region.
This corrects the $c_{-2,2}^{[1]}=-1/8$ coefficient computed in Sec.~\ref{sec:LO}, and its determination is left to future work.

The static approximation can be used to determine $c_{0,0}$ but not $c_{2,1}$.
In Appendix~\ref{app:contactph}, we provide a closed form expression for the $s$-wave projected $\hat{K}^{[2]}$ in the static approximation expanded to $O(\Delta^2\log^2{\Delta})$.
The $O(\Delta)$ term was previously derived in Ref.~\cite{PhysRev.116.898} but cannot be used to determine $c_{2,1}$ since it comes from the $(S,S,S,S)$ region where the static approximation is invalid.
Only the $O(\Delta^0)$ piece is needed for $c_{0,0}$ but the higher order terms may be useful in other contexts.

To extract $c_{2,1}$ we take $\mu=E_F$ as the relevant corrections from Eq.~\eqref{eq:BCS_number} are higher order in powers of $x$.
To make the separation into regions transparent, it is convenient to change integration variables to $\xi_i$ and set $k_F=M=1$.
The integral in Eq.~\eqref{eq:R2} becomes
\begin{align}
   R^{[2]} \ = \ -\frac{g^3}{256\pi^6}\int_{-1/2}^{\infty}\!d\xi_0\,d\xi_1\,d\xi_2\,d\xi_3\ {\cal L}(\xi_0,\xi_1,\xi_2,\xi_3)\,\frac{\Delta^2+\xi_1\xi_2-E_1E_2}{E_0E_1E_2E_3(E_0+E_1+E_2+E_3)}
    \label{eq:IKPi_Simplified}
\end{align}
where ${\cal L}(\xi_0,\xi_1,\xi_2,\xi_3)$ is the length of two overlapping intervals.
The evaluation of this integral is explained in Appendix~\ref{app:NLO}.
The soft $(S,S,S,S)$ contribution, $R^{[2]}\big \vert_{S}$, can be isolated by changing variables to $\xi_i=\Delta t_i$ and taking the $\Delta \to 0$ limit everywhere.
This gives
\begin{align}
R^{[2]}\big \vert_{S} \ = \ -\frac{g^3\Delta}{128\pi^6}\int_{\mathbb{R}^4}\!d^4t\ \frac{1+t_1t_2-e_1e_2}{e_0e_1e_2e_3(e_0+e_1+e_2+e_3)} \ , \qquad e_i=\sqrt{1+t_i^2} \ ,
\end{align}
where we have used ${\cal L}(\xi_0,\xi_1,\xi_2,\xi_3)=2+O(\Delta)$.
The large-$t_i$ region of this integral overlaps with the $O(\Delta^0)$ sector and produces a divergence that is removed by taking the finite part.
The finite part can be expressed using the closed-form expressions for moments of Bessel functions in Ref.~\cite{Bailey:2008ib}.
The result is,
\begin{align}
    R^{[2]}\big \vert_{S} \ = \ -\frac{g^3\Delta}{8\pi^6}\,J_A \ , \qquad
    J_A \ = \ \pi^2+\frac{\pi^4}{4}\;{}_4F_3\!\left(\tfrac12,\tfrac12,\tfrac12,\tfrac12;\,2,1,1;\,1\right) \ = \ 35.27767\ldots
\end{align}
where ${}_4F_3$ is a generalized hypergeometric function.
Adding this to the leading-order gap equation, $R^{[1]}$, in Eq.~\eqref{eq:LOgapExpand} and re-arranging gives,
\begin{align}
    c_{2,1}^{[2]} \ = \ \frac{J_A}{2} \ = \ \frac{\pi^2}{2}+\frac{\pi^4}{8}\;{}_4F_3\!\left(\tfrac12,\tfrac12,\tfrac12,\tfrac12;\,2,1,1;\,1\right) \ = \ 17.63884\ldots 
    \label{eq:jAdefinition}
\end{align}
%

\subsection{Next-to-next-to-leading order}
\label{sec:NNLO}
\noindent
The next-to-next-to-leading order contribution to the gap equation is $R^{[3]}$ in Eq.~\eqref{eq:Rb_def}.
This contains the self-energy correction to $\Pi_-^{[2]}$ and the two-loop irreducible potential $K_-^{[3]}$.
The gap equation at $O(g^3)$ also contains effects coming from $R^{[2]}$ beyond the leading-in-$g$ dependence.
The self-energy can contribute to both $c_{1,0}^{[3]}$ and $c_{2,1}^{[3]}$ and the two-loop irreducible potential contributes to $c_{1,0}^{[3]}$. 
We assume that, like for $K^{[2]}$, the linear-in-$\Delta$ component of $K^{[3]}$ comes from the soft region that does not pick up any Cooper logarithms and therefore affects $c_{3,1}^{[3]}$.
This is beyond the order considered in this work.

The calculation of $c_{1,0}^{[3]}$ is simpler in the framework set up in Ref.~\cite{BCF2024} where the gap is determined by identifying an instability in the free Fermi gas~\cite{THOULESS1960553}.
Although the setup was correct, Ref.~\cite{BCF2024} mistakenly double counted a momentum region in the static part of the irreducible potential, and also dropped a term when inverting the correlation function to obtain the pole. In Appendix~\ref{app:NNLO_c10}, we correct these mistakes and evaluate the integrals analytically which were previously computed numerically.
The $c_{1,0}$ coefficient is fully determined at this order and is given by
\begin{align}
    c_{1,0}\ = \ \frac{79}{180}+\frac{5\pi^2}{36}-\frac{3\pi^3}{16}-\frac{44}{15}\log2+\Big(\frac{4}{9}-\frac{\pi}{4}\Big)\log^22+4{\rm G}\log2+8\,{\rm Im}\,{\rm Li}_3\!\Big(\frac{1+i}{2}\Big)-\frac{7}{8}\zeta(3) \ = \ -0.1526\ldots
\end{align}
where ${\rm G}$ is the Catalan's constant, $\zeta$ is the Riemman zeta function and ${\rm Li}_3$ is the trilogratihm function.
This reverses the sign and decreases the magnitude relative to the $c_{1,0}=0.47619(20)$ reported in Ref.~\cite{BCF2024}. The correct version now increases the gap instead of suppressing it.

Next is the determination of $c^{[3]}_{2,1}$ from the self-energy.
At this order the fermionic excitation gap $\Delta$ and leading off-diagonal term in the Nambu-Gor'kov propagator $\bar{\Delta}$ differ.
The self-energy is parameterized by $(\sigma_o,\sigma_e,\Sigma_1)$ in Eq.~\eqref{eq:Sigmadef}.
Inserting the dressed propagators into $\Pi_-$ and expanding to first order in the self-energy gives,
\[\Pi_-^{[2]}(k) \ = \ \frac{i\big(2k_0\,\sigma_o(k)+2\xi_k\,\sigma_e(k)-2\bar{\Delta}\,\Sigma_1(k)\big)}{(k_0^2-E_k^2+i\epsilon)^2}\  .
\]
The $(\sigma_o,\sigma_e,\Sigma_1)$ are two-loop integrals obtained from cutting a single leg of $W^{[2]}$ in Fig.~\ref{fig:hugenholtz}.
The term that directly affects $c_{2,1}$ is generated when $k_0,\xi_k\sim\bar{\Delta}$ and the $\xi_i$ inside the self-energy are all soft.
In Appendix~\ref{app:NNLO_Sigma} it is shown that, at this order, the soft component $\sigma_e\big |_S =0$ and the contributions from $\sigma_o\big |_S$ and $\Sigma_1\big |_S$ exactly cancel upon integration giving,
\begin{align}
    \int_k\Pi_-^{[2]}(k) \Big |_S \ = \ 0 \ .
\end{align}
Thus, this term does not contribute to the gap equation.

However, the self-energy also changes the identification between the $\bar{\Delta}$ and $\Delta$.
The latter is defined as the lowest energy pole in the dressed propagator and represents the fermionic excitation gap.
This is measured in QMC through differences in the energy at fixed volume $\Delta = E(N+1) - (E(N)+E(N+2))/2$~\cite{Gezerlis:2007fs} and in the ultracold atom experiment of Ref.~\cite{Biss_2022} using Bragg spectroscopy to determine the energy threshold of the two-quasiparticle continuum, $2\Delta$.
The Nambu-Gor’kov propagator $\tilde{G}$ is a $2\times 2$ matrix and a pole in any one of its elements can be diagnosed from a zero of the determinant of its inverse,
\begin{align}
\det\tilde G^{-1}(k) \ = \ \big(k_0-\sigma_o(k)\big)^2-\big(\xi_k+\sigma_e(k)\big)^2-\big(\bar{\Delta}-\Sigma_1(k)\big)^2 \ .
\end{align}
The fermionic excitation gap $\Delta$ is defined as the positive $k_0$ that solves $\det\tilde G^{-1}(k)=0$, minimized over all momentum, 
\begin{align}
    \Delta = \min_{{\bf k}}[|k_0({\bf k})|] \ .
\end{align}
Here, the dispersion relation is defined implicitly through $\det\tilde G^{-1}(k)=0$, i.e.
\begin{align}
    k_0\ = \ \sigma_o(k)\pm\sqrt{\big(\xi_k+\sigma_e(k)\big)^2 + \big(\bar{\Delta}-\Sigma_1(k)\big)^2} \ .
\end{align}
Because $\sigma_o$ is odd and $\Sigma_1,\sigma_e$ are even in $k_0$, the roots come in $\pm$ pairs.
We therefore consider the positive root without loss of generality.
As shown in Appendix~\ref{app:NNLO_Sigma}, the soft self-energy corrections only depend on the energy $k_0$ and are independent of ${\bf k}$.
Hence, the three-momentum that minimizes $k_0$ has $|{\bf k}| =k_F$ such that $\xi_k=0$ and,
\begin{align}
    \Delta \ &= \ \sigma_o(\Delta)\big |_S  \ + \  \big(\bar{\Delta}-\Sigma_1(\Delta)\big |_S\big) \nonumber \\
     &=\ \bar{\Delta} \ + \ \sigma_o(\bar{\Delta})\big |_S  \  -\ \Sigma_1(\bar{\Delta})\big |_S \ + \ O(\lambda^4)\ ,
     \label{eq:Delta_Deltaqp}
\end{align}
where we used $\sigma_e(k)\big|_S=0$.
In the second line we replaced $\Delta \to \bar{\Delta}$ on the RHS as the differences are higher order. 
This is evaluated in Appendix~\ref{app:NNLO_Sigma} and found to be,
\[
\sigma_o(\bar{\Delta})\big |_S \ - \ \Sigma_1(\bar{\Delta})\big |_S \ = \ J_{os}\,\lambda^2\,\frac{\bar{\Delta}^2}{2E_F} \ ,
\]
where $J_{os}$ is given in terms of moments of modified Bessel functions of the second kind.
We were not able to find closed form expressions for these moments in the literature. However, the following relation was identified with PSLQ~\cite{Ferguson1999AnalysisOP} and verified to 500 digits,
\[
J_{os} \ = \ \sqrt{2}\,\left [\frac{47}{1536}[\Gamma(1/8)\Gamma(3/8)]^2-\frac{\pi^4}{[\Gamma(1/8)\Gamma(3/8)]^2}\right ] \ = \ 13.36945503\ldots 
\]
Inserting this into Eq.~\eqref{eq:Delta_Deltaqp} gives,
\begin{align}
\frac{\Delta}{\bar{\Delta}}  \ = \ 1+\frac{1}{2}J_{os}\,\lambda^2x \qquad \rightarrow \qquad
\log x \ = \ \log \bar{x}+\frac{1}{2}J_{os}\,\lambda^2x_0 \ ,
\end{align}
where $\bar{x} = \bar{\Delta}/E_F$.
This is an additive shift to the RHS of the trans-series expansion and the coefficient relevant to the fermionic excitation gap is obtained by adding $c_{2,1}^{[2]}$ from Eq.~\eqref{eq:jAdefinition},
\begin{align}
c_{2,1} \ = \ c_{2,1}^{[2]} \ + \ \frac{J_{os}}{2} \ = \ \frac{J_A+J_{os}}{2} \ = \ 24.32356479433\ldots 
\label{eq:c21qp}
\end{align}
The self-energy in the hard region has already been captured in $c_{1,0}$ through the effective mass $M^*$ and wavefunction renormalization $Z$~\cite{landau1980course}.

\subsection{Uncertainty bands}
\label{sec:uncertainty}
\noindent
The uncertainty band in Fig.~\ref{fig:gap_prediction} comes from modeling the omitted coefficients as independent normal variables,
\begin{align}
c_{2,0}&\sim\mathcal N(0,\sigma_{2,0}^2),&
\sigma_{2,0}&=\operatorname{median}
\left(|c_{1,0}|,|c_{0,0}|,|c_{-1,0}|\right),\nonumber\\
c_{3,1}&\sim\mathcal N(0,\sigma_{3,1}^2),&
\sigma_{3,1}&=|c_{2,1}|,\nonumber\\
c_{-2,2}&\sim\mathcal N(0,\sigma_{-2,2}^2),&
\sigma_{-2,2}&=|c_{-2,2}^{[1]}|.
\end{align}
These standard deviations are sector-dependent estimates of naturalness. 
The BCS contribution $c_{-2,2}^{[1]}=-1/8$ is used as it is our best estimate for the magnitude of $c_{-2,2}$. 
The gap is computed by expanding $x=\exp{\sum_{n,m}c_{n,m}\lambda^nx_0^m}$ up to the order included in the uncertainty band,
\begin{align}
    x \ = \   x_{\text{NNLO}} \ + \  x_0\Big[
\left(c_{2,0}+\tfrac12c_{1,0}^{\,2}\right)\lambda^2
+\left(c_{3,1}+c_{1,0}c_{2,1}\right)\lambda^3x_0
+c_{-2,2}\frac{x_0^2}{\lambda^2}
\Big] \ ,
\end{align}
where $x_{\text{NNLO}} = x_0(1+c_{1,0}\lambda + c_{2,1}\lambda^2 x_0)$.
The variance of $x$ is exactly given by,
\begin{align}
    \text{Var}(x) \ = \ x_0^2\left (\sigma_{2,0}^2 \lambda^4 \ + \  \sigma_{3,1}^2 \lambda^6 x_0^2 \ + \ \sigma_{-2,2}^2x_0^4/\lambda^4\right )
\end{align}
and the shaded band in Fig.~\ref{fig:gap_prediction} covers $x_{\text{central}}\pm\sqrt{\text{Var}(x)}$.
The uncertainty band is not centered on $x_{\text{NNLO}}$ as some of the higher order contributions are fixed by the lower order terms, 
\[
x_{\text{central}}\ = \ x_{\text{NNLO}}\  + \ x_0\Big[
\tfrac12c_{1,0}^{2}\lambda^2
\ + \ c_{1,0}c_{2,1}\lambda^3x_0
\Big] \ . 
\]

\begin{figure}
    \centering
    \includegraphics[width=\linewidth]{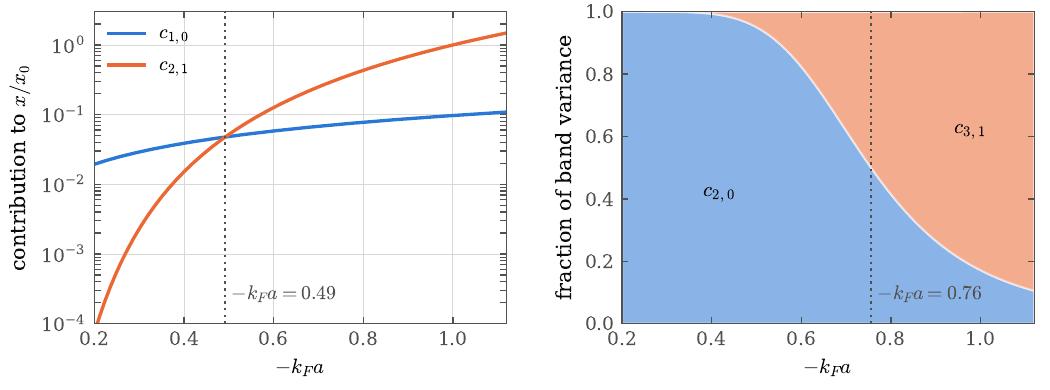}
    \caption{Left: the size of the correction to the GMB gap $x/x_0 =1+c_{1,0}\lambda + c_{2,1}\lambda^2 x_0$.
    The linear-in-gap correction is dominant for $-k_F a>0.49$.
    Right: the fraction of the uncertainty band variance from each unknown coefficient. 
    The uncertainty from the linear-in-gap sector is dominant for $-k_F a>0.76$.
    The fraction of the variance coming from $c_{-2,2}$ is $<10^{-4}$ over this interval.}
    \label{fig:error_band_breakdown}
\end{figure}
The contribution of $c_{1,0}$ and $c_{2,1}$ to the gap is shown in the left plot of Fig.~\ref{fig:error_band_breakdown}.
The dominant correction to GMB is $c_{1,0}$ at low densities and $c_{2,1}$ for $-k_F a>0.49$.
The contribution of each coefficient to the uncertainty band, measured by its fraction of $\text{Var}(x)$, is shown in in the right plot of Fig.~\ref{fig:error_band_breakdown}.
The uncertainty band is primarily driven by $c_{2,0}$ at low density and $c_{3,1}$ is the dominant term for $-k_F a >0.76$.
The fraction of the uncertainty band carried by $c_{-2,2}$ is negligible under our assumed prior.

\clearpage
\section*{Appendices}
\label{sec:appendix}
\setcounter{subsection}{0}
\renewcommand\thesubsection{\Alph{subsection}}

\subsection{Application of Feynman rules}
\label{app:FeynmanLONLO}
\subsubsection{Leading order}
\noindent
The irreducible potential at leading order is $K_-^{[1]}$ which is obtained by applying the Feynman rules in Sec.~\ref{sec:FeynmanRules} to $W^{[1]}$ in Fig.~\ref{fig:hugenholtz}.
Applying Feynman rules 1-5 gives
\begin{align}
iW^{[1]} \ &= \ i\frac{g_0}{2}\int_{l,l'}\left [ \text{Tr}\left ( \tilde{\tau}_3 \tilde{G}_0(l)\right)\text{Tr}\left ( \tilde{\tau}_3 \tilde{G}_0(l')\right) - \text{Tr}\left(\tilde{\tau}_3 \tilde{G}_0(l) \tilde{\tau}_3 \tilde{G}_0(l') \right )\right] \ .
\end{align}
Taking the functional derivatives gives
\begin{align}
i\tilde{K}_{ba;dc} = -\frac{\delta^2[iW^{[1]}]}{\delta \tilde{G}_{0}(k')_{cd} \delta \tilde{G}_{0}(k)_{ab}} \ = \ -ig \left [(\tilde{\tau}_3)_{ba}(\tilde{\tau}_3)_{dc}-(\tilde{\tau}_3)_{bc}(\tilde{\tau}_3)_{da} \right ] \ ,
\end{align}
where a couple of useful identities are
\[
\frac{\delta}{\delta \tilde{A}_{ab}} \text{Tr}(\tilde{A} \tilde{B}) = \tilde{B}_{ba} \quad , \quad \tilde{\tau}_3 = \tilde{\tau}_3^T \quad , \quad \tilde{G}_0 \ = \ \tilde{G}_0^T \ .
\]
Lastly, changing from the Nambu-Gor’kov to the pairing basis by contracting with $\tilde{\tau}_{\pm}$ gives
\[
\hat{K}^{[1]} \ = \ -g_0\mathds{1}  \ .
\]
This reproduces the expectation that the tree vertex is proportional to the identity in the pairing basis.

\subsubsection{Next-to-leading order}
\noindent
The irreducible potential at next-to-leading order is $K_-^{[2]}$ which is obtained by applying the Feynman rules in Sec.~\ref{sec:FeynmanRules} to $W^{[2]}$ in Fig.~\ref{fig:hugenholtz}.
Applying steps 1-5 of the Feynman rules gives
\begin{align}
iW^{[2]} \ = \ -\frac{g_0^2}{4}\int_{l,l',l''} &\bigg [\text{Tr}\left (\tilde{\tau}_3\tilde{G}_0(l'')\tilde{\tau}_3 \tilde{G}_0(l') \right )\text{Tr}\left (\tilde{\tau}_3 \tilde{G}_0(l+l'-l'')\tilde{\tau}_3 \tilde{G}_0(l) \right ) \nonumber \\
& -\text{Tr}\left (\tilde{\tau}_3\tilde{G}_0(l'')\tilde{\tau}_3 \tilde{G}_0(l)\tilde{\tau}_3\tilde{G}_0(l+l'-l'')\tilde{\tau}_3\tilde{G}_0(l')\right )\bigg ] \ .
\label{eq:W2}
\end{align}
Taking the functional derivatives and projecting onto the pairing basis gives
\begin{align}
(K^{[2]})_{11} \ &=  -ig_0^2\int_l \!\left [F_0(l)F_0(l+\bar{Q})\ + \ F_0(l)F_0(l+Q) \ + \ G_0(l)G_0(l+\bar{Q}) \right ] 
\label{eq:K11}
\end{align}
and
\begin{align}
    (K^{[2]})_{12} \ &= \ -ig_0^2\int_l \!\left [F_0(l)F_0(l+Q)\right ] \ ,
    \label{eq:K12}
\end{align}
where $Q = k - k'$ and $\bar{ Q} = k + k'$ with $\pm k$ and $\pm k'$ the incoming and outgoing momentum respectively.
We have specialized to the case of zero total four-momentum as is relevant to the gap equation.
Deriving these expressions made use of the fact that both $F_0$ and $G_0$ are invariant under ${\bf l} \to - {\bf l}$ and $F_0$ is additionally invariant under $l_0 \to -l_0$.
Combining these expressions gives
\begin{align}
K^{[2]}_{-} \ &=  -ig^2\int_l\left [F_0(l)F_0(l+\bar{Q}) \ + \ G_0(l)G_0(l+\bar{Q}) \right ] \ ,
\end{align}
where $K^{[2]}_-=K^{[2]}_{11}-K^{[2]}_{12}$.

\subsection{Details on the next-to-leading order computation}
\label{app:NLO}
\noindent
After the energy integrals, the next-to-leading-order contribution in Eq.~\eqref{eq:R2} is
\begin{align}
R^{[2]} \ = \ -\frac{g^3}{8}\int\!\frac{d^3l_0}{(2\pi)^3}\int\!\frac{d^3l_1}{(2\pi)^3}\int\!\frac{d^3l_3}{(2\pi)^3}\,\frac{\Delta^2+\xi_1\xi_2-E_1E_2}{E_0E_1E_2E_3(E_0+E_1+E_2+E_3)} \ ,
\label{eq:R22}
\end{align}
with ${\bf l}_2={\bf l}_0+{\bf l}_1+{\bf l}_3$.
We work in units $M=k_F=1$, so that, at this order, $\mu=E_F=1/2$.
The integrand depends only on the magnitudes $\lmag{i}$.
To reduce the angular integrals, we introduce
\begin{align}
\bar{\bf l} \ = \ {\bf l}_0+{\bf l}_1 \ = \ {\bf l}_2-{\bf l}_3 \ , \qquad p \ \equiv \ |\bar{\bf l}| \ ,
\label{eq:bark}
\end{align}
and integrate over $\bar{\bf l}$, ${\bf l}_1$ and ${\bf l}_3$.
This transformation has unit Jacobian.
At fixed $\bar{\bf l}$, the angular integrals over ${\bf l}_1$ and ${\bf l}_3$ give
\begin{align}
d^3l_1 \ \to \ \frac{2\pi}{p}\,\lmag{0}\,\lmag{1}\,d\lmag{0}\,d\lmag{1} \ , \qquad
d^3l_3 \ \to \ \frac{2\pi}{p}\,\lmag{2}\,\lmag{3}\,d\lmag{2}\,d\lmag{3} \ ,
\end{align}
subject to the triangle constraints
\begin{align}
C_1:\ \big|\lmag{0}-\lmag{1}\big|\leq p\leq \lmag{0}+\lmag{1} \ , \qquad
C_2:\ \big|\lmag{2}-\lmag{3}\big|\leq p\leq \lmag{2}+\lmag{3} \ .
\end{align}
Together with $d^3\bar{l}=4\pi p^2\,dp$, all dependence on $p$ cancels, and the $p$ integral gives the length of the allowed interval,
\begin{align}
{\cal L} \ \equiv \ \text{length}(C_1\cap C_2) \ = \ \max\!\Big[0,\ \min\big(\lmag{0}+\lmag{1},\,\lmag{2}+\lmag{3}\big)-\max\big(\big|\lmag{0}-\lmag{1}\big|,\,\big|\lmag{2}-\lmag{3}\big|\big)\Big] \ .
\end{align}
Changing variables to $\xi_i=(\lmag{i}^2-1)/2$, with $d\xi_i=\lmag{i}\,d\lmag{i}$, gives
\begin{align}
R^{[2]} \ = \ -\frac{g^3}{256\pi^6}\int_{-1/2}^{\infty}\!d\xi_0\,d\xi_1\,d\xi_2\,d\xi_3\ {\cal L}(\xi_0,\xi_1,\xi_2,\xi_3)\,\frac{\Delta^2+\xi_1\xi_2-E_1E_2}{E_0E_1E_2E_3(E_0+E_1+E_2+E_3)} \ ,
\label{eq:IKPi_Simplified_app}
\end{align}
which reproduces Eq.~\eqref{eq:IKPi_Simplified}.

In the fully soft $(S,S,S,S)$ region, all four lines lie within $\Delta$ of the Fermi surface.
We set $\xi_i=\Delta t_i$ and keep the leading terms as $\Delta\to0$ giving $\lmag{i}=1+O(\Delta)$ and ${\cal L}=2+O(\Delta)$.
At leading order, the lower bound of the integral $-1/(2\Delta)\to-\infty$ so that
\begin{align}
R^{[2]}_{S} \ = \ -\frac{g^3\Delta}{128\pi^6}\int_{\mathbb{R}^4}\!d^4t\ \frac{1+t_1t_2-e_1e_2}{e_0e_1e_2e_3(e_0+e_1+e_2+e_3)} \ , \qquad e_i=\sqrt{1+t_i^2} \ .
\label{eq:R2S}
\end{align}
The $t_1t_2$ term is odd and drops out upon integration.
Changing variables to $t_i=\sinh u_i$, so that $dt_i/e_i=du_i$, and using $1/c=\int_0^\infty ds\,e^{-sc}$ to decouple the integrals gives
\begin{align}
R^{[2]}\big \vert_{S} \ &= \ -\frac{g^3\Delta}{128\pi^6}\int_0^\infty\! ds\int_{\mathbb{R}^4}\!d^4u\ e^{-s\sum_i\cosh u_i}\,\big(1-\cosh u_1\cosh u_2\big) \nonumber\\
&= \ -\frac{g^3\Delta}{8\pi^6}\int_0^\infty\! ds\ K_0(s)^2\big[K_0(s)^2-K_1(s)^2\big] \ ,
\end{align}
where we have used the integral representation of the modified Bessel function of the second kind, $\int_{-\infty}^{\infty}du\,e^{-s\cosh u}\cosh(nu)=2K_n(s)$ for $s> 0$.

This integral diverges as $s\to0$ which comes from the $t_i\to \infty$ region of Eq.~\eqref{eq:R2S}. 
This divergence arises because, for $t_i\gg 1$, we are no longer in the soft region and are picking up contributions to the $c_{n,0}$ coefficients that diverge when the $\Delta\to0$ limit is taken.
These pieces can be removed by subtracting the leading $s\to0$ dependence
\begin{align}
    K_0(s)^2 \left (K_0(s)^2 - K_1(s)^2 \right )\ = \ -\frac{\eta(s)^2}{s^2} + O(\log^4{s}) \ , \quad \eta(s) \equiv \gamma_E + \log(s/2) \ ,
    \label{eq:RS_integrand}
\end{align}
where $\gamma_E$ is the Euler constant. 
There is no $1/s$ term which would produce $\Delta\log\Delta$ and its absence confirms that $c_{1,1}$ receives no contribution from $R^{[2]}$.
The soft contribution is therefore the finite part (FP).
To isolate it, define $y=K_0(s)$, $y'=-K_1(s)$, and use the modified Bessel equation $y''+y'/s-y=0$ to rewrite the integrand in Eq.~\eqref{eq:RS_integrand} as
\begin{align}
y^4-y^2y'^2 \ &= \ \frac{4}{3}y^4-\frac{16}{3}s^2y^4+{\cal F}' \ ,  \quad\
{\cal F} \ = \ 2s^3\big(y^2-y'^2\big)^2-\frac{s}{3}y^4+s\,y^2y'^2+2s^2y\,y'^3-\frac{2}{3}s^2y^3y' \ .
\end{align}
${\cal F}'$ is a total derivative and enters only through its boundary values. It vanishes as $s\to\infty$, and
\[
{\cal F}(\epsilon) \ = \ \frac{\eta(\epsilon)^2+2\eta(\epsilon)+2}{\epsilon} \ + \ O(\epsilon\,\eta(\epsilon)^4)  \ .
\]
The boundary term $-{\cal F}(\epsilon)$ is therefore exactly the $s\to0$ divergence of $\int_\epsilon ds\,(-\eta(s)^2/s^2)$, the singular part of the integrand.
The total derivative carries the entire divergence, and
\begin{align}
J_A \ \equiv \ \text{FP}\!\int_0^\infty\! ds\,K_0^2\big(K_0^2-K_1^2\big) \ = \ \int_0^\infty\! ds\,\left ( \frac{4}{3}K_0^4-\frac{16}{3}s^2K_0^4\right ) \ = \ \frac{4}{3}M_{4,0}-\frac{16}{3}M_{4,2} \ ,
\label{eq:j_Adefined}
\end{align}
where moments of the modified Bessel functions have been defined as,
\begin{align}
    M_{n,k} \ \equiv \ \int_0^\infty\! ds\,s^kK_0(s)^n \ .
\end{align}
The exact linear combination of moments in Eq.~\eqref{eq:j_Adefined} is provided in Equation~83 of Ref.~\cite{Bailey:2008ib} and gives,
\begin{align}
J_A \ = \ \pi^2+\frac{\pi^4}{4}\ {}_4F_3\!\left(\tfrac12,\tfrac12,\tfrac12,\tfrac12;\,2,1,1;\,1\right) \ = \ 35.27767455575576\ldots
\end{align}
With $g=2\pi^2\lambda$ and $\Delta=x/2$ in these units, $R^{[2]}\big \vert_S=-\tfrac12\lambda^3x\,J_A$.
Inserting this into the next-to-leading order gap equation $R^{[1]}+R^{[2]}=1$, with $R^{[1]}=-\lambda\log(8/(e^2x))$, shifts $\log x$ by $-R^{[2]}\big \vert_S/\lambda$, so that
\begin{align}
c_{2,1}^{[2]} \ = \ \frac{J_A}{2} \ = \ 17.63883727787788\ldots
\end{align}

\subsection{Details on the next-to-next-to-leading order computation: \texorpdfstring{$c_{1,0}$}{}}
\label{app:NNLO_c10}
\noindent
This appendix corrects the mistakes made in Ref.~\cite{BCF2024} that computed $c_{1,0}$.
We will copy all of the notation from Ref.~\cite{BCF2024} and use (their Eq.~XX) to refer to equations from that paper.
The extra terms relative to the previous work will be highlighted in blue.
The kinetic energy of free particles are defined as $\omega_l = |{\bf l}|^2/(2M)$.

The gap is determined by the pole condition (their Eq.~32),
\begin{align}
    \left (\Gamma_{3a}+\Gamma_{3b}+\Gamma_{3c}+\Gamma_{3d}+\Gamma_{3e} \right )^{-1} \ = \ 0
    \label{eq:gapEq1}
\end{align}
where the $\Gamma_{i}$ correspond to classes of resummed Feynman diagrams.
The first mistake is in $\Gamma_{3b}$ (their Eq.~C2) which incorrectly contains off-shell propagators multiplied by the one-loop static potential $V^{g^2}$.
Correcting this gives,
\begin{align}
    \Gamma_{3b} \ = \ \frac{-V^{(g^2)}}{D^2}\left (1 \ - \ 2 \frac{g\Pi_{pp}^{(g^0)}}{D} \ + \ \textcolor{blue}{2 g\Pi_{pp}^{(g^0)}} \right )
\end{align}
where $D \equiv (1+g\Pi_{pp}^{(g^{-1})})$.
The second mistake was in $\Gamma_{3c}$ (their Eq.~C3) where a term of the form $0/0$ was dropped. 
Correcting this gives,
\begin{align}
\Gamma_{3c}=\frac{\big(V^{(g^2)}\big)^2\,\Pi_{pp}^{(g^{-1})}}{D^3}
\ = \ -\frac{\big(V^{(g^2)}\big)^2}{g\,D^3} \ + \ \textcolor{blue}{\frac{\big(V^{(g^2)}\big)^2}{g\,D^2}}
\end{align}
The final mistake was in a double counting that occurred in the evaluation of ${\cal I}$ in
\begin{align}
 \Gamma_{3e} \ = \ \frac{2g}{D^2}\textcolor{blue}{{\cal I}\left [\Pi_{pp}^{(g^0)}V^{(g^2)} \right ]}   \ .
\end{align}
This will be discussed in more detail when we derive ${\cal I}$ in closed form.
Note that $\Gamma_{3b}$ and $\Gamma_{3e}$ are not independent and the term added to $\Gamma_{3b}$ changes what needs to be subtracted from $\Gamma_{3e}$.
Eq.~\eqref{eq:gapEq1} with the corrected expression gives the corrected version of their Eq.~32,
\begin{align}
    0 \ = \ 1 + g\left (\Pi_{pp}^{(g^{-1})}+\Pi_{pp}^{(g^{0})}+\Pi_{pp}^{(g^{1})} \right ) \ - \ \frac{1}{g}\left (V^{(g^2)}+V^{(g^3)} \right ) \  + \ \textcolor{blue}{\left ( \frac{V^{(g^{2})}}{g}\right )^2} \ - \ \textcolor{blue}{2\Pi_{pp}^{(g^0)}V^{(g^2)}} \ + \  2\textcolor{blue}{{\cal I}\left [\Pi_{pp}^{(g^0)}V^{(g^2)} \right ]}
\end{align}
The first two terms in blue can be evaluated exactly using their Eq.~28 and Eq.~30.
We next determine both $V^{(g^3)}$ and $\textcolor{blue}{{\cal I}\left [\Pi_{pp}^{(g^0)}V^{(g^2)} \right ]}$ in closed form.

\subsubsection{Closed form of \texorpdfstring{$V^{(g^3)}$}{}}
\noindent
We work in units $M=k_F=1$, so that $E=E_F=1/2$, $\lambda=g/(2\pi^2)$, and define $n_l\equiv\theta(1-l)$.
Their Eq.~C11 is written as
\begin{align}
V^{(g^3)} \ = \ g^3\int_0^\pi d\theta\,\sin\theta\left[-\Pi^{(a)}_{ppph}+\Pi^{(b)}_{phph}+\Pi_{ph}^2\right] \ \equiv \ g\lambda^2\left(A+B+C\right) \ ,
\end{align}
with
\begin{align}
A \ = \ -4\pi^4\!\int_0^\pi d\theta\,\sin\theta\,\Pi^{(a)}_{ppph} \ , \qquad
B \ = \ 4\pi^4\!\int_0^\pi d\theta\,\sin\theta\,\Pi^{(b)}_{phph} \ , \qquad
C \ = \ 4\pi^4\!\int_0^\pi d\theta\,\sin\theta\,\Pi_{ph}^2 \ .
\end{align}
Their Monte Carlo evaluation gave $A=0.972041(15)$ and $B=-0.0418477(64)$, while $C=\frac{1}{4}\left(1+\frac{7}{2}\zeta(3)\right)$ was obtained analytically (their Eq.~C11).
We now evaluate $A$ and $B$ in closed form.
Both $\Pi^{(a)}_{ppph}$ and $\Pi^{(b)}_{phph}$ contain two propagators, with momenta ${\bf q}_1$ and ${\bf q}_2$, that carry the same loop energy.
After the energy integrations, the terms in each of their Eqs.~C12 and C17 combine into the divided-difference form
\begin{align}
\int\frac{d^3l'}{(2\pi)^3}\,\frac{h({\bf q}_1)-h({\bf q}_2)}{\omega_{q_1}-\omega_{q_2}} \ ,
\label{eq:divdiff}
\end{align}
where $h({\bf q})$ is a one-loop integral over the remaining loop momentum ${\bf l}$, given in Eq.~\eqref{eq:hB} for $\Pi^{(b)}_{phph}$ and in Eq.~\eqref{eq:hA} for $\Pi^{(a)}_{ppph}$.
For each diagram we choose a polar axis such that, at fixed magnitude of ${\bf l}'$, $h({\bf q}_1)$ depends on the external momenta only through one direction cosine $u$, $h({\bf q}_2)$ only through a second one $v$, and $\omega_{q_1}-\omega_{q_2}\propto u-v$.
The s-wave projection then averages $u$ and $v$ independently, $\int_0^\pi d\theta\,\sin\theta\int d\Omega=2\pi\int_{-1}^{1}du\int_{-1}^{1}dv$, where $d\Omega$ is over the direction of the polar axis.
Writing $h({\bf q}_1)=f(u)/(2\pi)^3$ and $h({\bf q}_2)=f(v)/(2\pi)^3$, the identity
\begin{align}
\int_{-1}^{1}du\int_{-1}^{1}dv\,\frac{f(u)-f(v)}{u-v} \ = \ 2\int_{-1}^{1}du\,f(u)\,L(u) \ , \qquad L(u) \ \equiv \ \log\frac{1+u}{1-u} \ ,
\label{eq:Lident}
\end{align}
which follows from $\int_{-1}^{1}dv/(u-v)=L(u)$, reduces each diagram to a single average over $u$.

\paragraph{The $\Pi_{phph}^{(b)}$ term.}
Define ${\bf Q}={\bf k}+{\bf k}'-{\bf l}'$, so that ${\bf q}_3={\bf l}+{\bf Q}$, ${\bf q}_1={\bf k}-{\bf Q}$, and ${\bf q}_2={\bf k}'-{\bf Q}$.
The three terms of their Eq.~C17 combine into Eq.~\eqref{eq:divdiff} with
\begin{align}
h({\bf q}) \ = \ \int\frac{d^3{\bf l}}{(2\pi)^3}\,\frac{w_{\bf q}({\bf l})}{E-\omega_q+\omega_l-\omega_{|{\bf l}+{\bf Q}|}} \ , \qquad
w_{\bf q} \ = \ \begin{cases} n_l\left(1-n_{|{\bf l}+{\bf Q}|}\right) \ , & q>1 \ , \\ \left(1-n_l\right)n_{|{\bf l}+{\bf Q}|} \ , & q<1 \ . \end{cases}
\label{eq:hB}
\end{align}
Take $\hat{\bf Q}$ as the polar axis, with $r=|{\bf Q}|$, $u=-\hat{\bf k}\cdot\hat{\bf Q}$, and $v=-\hat{\bf k}'\cdot\hat{\bf Q}$.
Then $\omega_{q_1}=\frac{1}{2}(1+r^2+2ru)$, $\omega_{q_1}-\omega_{q_2}=r(u-v)$, and ${\bf q}_1$ is a particle for $u>-r/2$.
After ${\bf l}\to-{\bf l}-{\bf Q}$ in the hole case, both cases involve the same region $n_l(1-n_{|{\bf l}+{\bf Q}|})$: holes at ${\bf l}$ that can be excited to particles at ${\bf l}+{\bf Q}$.
Slicing it perpendicular to $\hat{\bf Q}$ at $l_z=z$ gives the cross-sectional area
\begin{align}
\mathcal{A}_r(z) \ = \ \begin{cases} \pi(1-z^2) \ , & \max(-1,1-r)<z<1 \ , \\ \pi r(2z+r) \ , & -\frac{r}{2}<z<1-r \ , \\ 0 \ , & \text{otherwise} \ , \end{cases}
\end{align}
and $h({\bf q}_1)=H_r(u)/(2\pi)^3$, i.e.\ $f=H_r$ in Eq.~\eqref{eq:Lident}, with
\begin{align}
H_r(u) \ = \ \begin{cases} -\dfrac{1}{r}\displaystyle\int dz\,\frac{\mathcal{A}_r(z)}{z+r+u} \ , & -\frac{r}{2}<u<1 \ , \\[10pt] \dfrac{1}{r}\displaystyle\int dz\,\frac{\mathcal{A}_r(z)}{z-u} \ , & -1<u<-\frac{r}{2} \ . \end{cases}
\label{eq:Hr}
\end{align}
Both denominators are positive on the support of $\mathcal{A}_r$, so no principal value is needed.
Using Eq.~\eqref{eq:Lident},
\begin{align}
B \ = \ \frac{1}{4\pi}\int_0^\infty r\,dr\int_{-1}^{1}du\,H_r(u)\,L(u) \ .
\label{eq:Bred}
\end{align}
For $r>2$ the Fermi balls centered at ${\bf 0}$ and $-{\bf Q}$ do not overlap, $\mathcal{A}_r(z)=\pi(1-z^2)$, and $H_r(u)=\pi J(r+u)/r$ with
\begin{align}
J(c) \ = \ \int_{-1}^{1}dz\,\frac{1-z^2}{z-c} \ = \ -2c+(1-c^2)\log\left|\frac{1-c}{1+c}\right| \ .
\end{align}
Subtracting the $u$-independent $J(r)$ is allowed because $L(u)$ integrates to zero.
With the primitive
\begin{align}
\mathcal{J}(t) \ = \ -\frac{2t^2}{3}+\left(t-\frac{t^3}{3}\right)\log\frac{t-1}{t+1}-\frac{2}{3}\log(t^2-1) \ , \qquad \mathcal{J}'(t) \ = \ J(t) \ ,
\end{align}
the region $r>2$ gives
\begin{align}
B_> \ = \ \frac{1}{4}\int_{-1}^{1}du\,L(u)\left[\mathcal{J}(2)-\mathcal{J}(2+u)\right] \ = \ \frac{20-80\log2+5\pi^2}{36} \ .
\end{align}
For $0<r<2$ the $z$-integrals in Eq.~\eqref{eq:Hr} are elementary,
\begin{align}
\frac{H_r(u)}{\pi} \ &= \ 1-\frac{r}{2}-u+(r+2u)\log\frac{1-r-u}{-u-r/2}+\frac{1-u^2}{r}\log\frac{1-u}{1-r-u} \ , && -1<u<-\frac{r}{2} \ , \nonumber\\
\frac{H_r(u)}{\pi} \ &= \ -\left(1+\frac{r}{2}+u\right)-(r+2u)\log\left(u+\frac{r}{2}\right)-\frac{1-(r+u)^2}{r}\log(1+r+u)+\frac{1-u^2}{r}\log(1+u) \ , && -\frac{r}{2}<u<1 \ ,
\end{align}
and at $r=2$ the second branch reduces to $J(2+u)/2$.
Direct iterated integration gives
\begin{align}
B_< \ = \ \frac{-48+96\log2-5\pi^2}{72} \ ,
\end{align}
and therefore
\begin{align}
B \ = \ \frac{-8-64\log2+5\pi^2}{72}
\end{align}

\paragraph{The $\Pi_{ppph}^{(a)}$ term.}
Here ${\bf q}_1={\bf l}'-{\bf k}'$, ${\bf q}_2={\bf l}'-{\bf k}$, and ${\bf q}_3={\bf l}'-{\bf l}$. The terms of their Eq.~C12 combine into Eq.~\eqref{eq:divdiff} with

\begin{align}
h({\bf q}) \ = \ \int\frac{d^3l}{(2\pi)^3}\,\frac{n_qn_3-n_q(1-n_l)-n_3n_l}{D_{\bf q}({\bf l})}\ , \quad D_{\bf q}({\bf l}) \ =\ E+\omega_q-\omega_l-\omega_{q_3}
\label{eq:hA}
\end{align}

\noindent The integral of $n_3/D_{\bf q}({\bf l})$ is convergent and is evaluated by changing integration variables from $\bf l$ to $\bf q_3$. The measure is decomposed in spherical coordinates with the polar angle chosen to be the angle between ${\bf q_3}$ and ${\bf l'}$. This gives
\begin{align}
    n_q \ \int\frac{d^3l}{(2\pi)^3}\,\frac{n_3}{D_{\bf q}({\bf l})} \ &= \ \frac{n_q}{8\pi^2|{\bf l'}|}\left\{\left(1-\left(\frac{|{\bf l'}|}{2}+b\right)^2\right)\log\left|\frac{1-\frac{|{\bf l'}|}{2}-b}{1+\frac{|{\bf l'}|}{2}+b}\right|+\left(1-\left(\frac{|{\bf l'}|}{2}-b\right)^2\right)\log\left|\frac{1-\frac{|{\bf l'}|}{2}+b}{1+\frac{|{\bf l'}|}{2}-b}\right|-2|{\bf l'}|\right\}\ , 
\end{align}
where $b= \left| {\bf q}-\frac{\bf l'}{2}\right|$.
The integral of $(1-n_l)/D_{\bf q}({\bf l})$ is linearly divergent, but as is outlined in Ref.~\cite{BCF2024}, this divergence can be isolated and subtracted off by expanding the integrand around large $|{\bf l}|$. The divergent piece, evaluated with dimensional regularization in the PDS scheme, is 

\begin{align}
    n_q \ \int\frac{d^3l}{(2\pi)^3}\frac{1-n_l}{2\omega_l}\left(1+\frac{{\bf l'}\cdot{\bf l}}{|{\bf l}|^2}\right) = n_q\left(\frac{\mu_{\text{PDS}}}{4\pi}-\frac{1}{2\pi^2}\right)\ .
\end{align}
What remains of the integral is UV finite and can be evaluated in spherical coordinates by regulating the integral with a cutoff $|{\bf l}|\leq\Lambda$ and then sending $\Lambda\to\infty$. The remainder is 

\begin{align}
    n_q\ \int\frac{d^3l}{(2\pi)^3}\,\frac{1-n_l}{D_{\bf q}({\bf l})} \ \Rightarrow \ -\frac{n_q}{8\pi^2|{\bf l'}|}\left\{\left(1-\left(\frac{|{\bf l'}|}{2}+b\right)^2\right)\log\left|\frac{1-\frac{|{\bf l'}|}{2}-b}{1+\frac{|{\bf l'}|}{2}+b}\right|+\left(1-\left(\frac{|{\bf l'}|}{2}-b\right)^2\right)\log\left|\frac{1-\frac{|{\bf l'}|}{2}+b}{1+\frac{|{\bf l'}|}{2}-b}\right|\right\}\ .
\end{align}
For the integral of $n_3n_l/D_{\bf q}({\bf l})$, the two occupation factors restrict $\bf l$ to the intersection of two unit balls whose centers are separated by a distance $|{\bf l'}|$; therefore, this contribution vanishes for $|{\bf l'}|> 2$. The integral is evaluated in cylindrical coordinates where the longitudinal axis is chosen to point parallel to ${\bf l'}$ from the center of one unit ball to the other. The result is

\begin{align}
    \ \int\frac{d^3l}{(2\pi)^3}\,\frac{n_3n_l}{D_{\bf q}({\bf l})}\ = \ \frac{\theta(2-|{\bf l'}|)}{8\pi^2|{\bf l'}|}\left(2|{\bf l'}|b\log\left|\frac{|{\bf l'}|-2-2b}{|{\bf l'}|-2+2b}\right|+(1-|{\bf q}|^2)\log\left|\frac{(|{\bf l'}|-1)^2-|{\bf q}|^2}{1-|{\bf q}|^2}\right|+|{\bf l'}|^2-2|{\bf l'}|\right).
\end{align}

\noindent Substituting this expression for $h({\bf q})$, Eq.~\eqref{eq:Lident} contains two categories of terms. The PDS subtraction term of $h({\bf q)}$ contributes 

\begin{align}
    A_{PDS} \ &= \ -4\pi^4\int_0^1dv\int_0^1du\int\frac{d^3l'}{(2\pi)^3}\frac{n_{|{\bf l'}-{\bf k'}|}}{\omega_{|{\bf l'}-{\bf k'}|}-\omega_{|{\bf l'}-{\bf k}|}}\left(\frac{\mu}{4\pi}-\frac{1}{2\pi^2}\right)\nonumber  \\
    &=\ -4\pi^4\left(\frac{\mu_{\text{PDS}}}{4\pi}-\frac{1}{2\pi^2}\right)\int_0^1d\theta\ \Pi_{ph}({\bf k}-{\bf k'},0) \nonumber \\
    &=\ \frac{2}{3}+\frac{4}{3} \log2\ ,
\end{align}
where, in the last line, $\mu_{PDS}=0$. For the remaining, UV finite contributions, the measure is decomposed into $d^3l' = |{\bf l}'|^2d|{\bf l}'|d\Omega$. The angular integration $d\Omega$ trivially contributes a factor of $4\pi$, leaving
\begin{align}
    A_{fin} = -4\pi^4\int_{-1}^1du\left(\int_0^\infty \frac{{\bf l}'^2d|{\bf l}'|}{2\pi^2}h_{fin}\left({\bf l'}-\bf{k'}\right)\right)L(u)
\end{align}
\noindent $h_{fin}\left({\bf l'}-\bf{k'}\right)$ contains two sets of terms: those proportional to $n_{q_1}$ and those proportional to $\theta(2-|{\bf l}'|)$. $n_{q_1}$ reduces to $\theta(u)\theta(2u-|{\bf l}'|)$. For these terms, $|{\bf l}'|$ is integrated from $0$ to $2u$, and what remains is an integral of $u$ from $0$ to $1$. For the second set of terms, $|{\bf l}'|$ is integrated from $0$ to $2$, and the integration domain $-1\leq u\leq 1$ is transformed to match that of the first set of terms via
\begin{align}
    \int_{-1}^1du\, H(u)L(u) = \int_{0}^1du\, \left(H(u)-H(-u)\right)L(u)\ , \quad L(-u) = -L(u)
\end{align}
All that remains is an integral on $u$ from $0$ to $1$. Most terms consist of polynomials multiplying either products of logarithms or dilogarithms and can be evaluated by integration by parts. The only nontrivial term is

\begin{align}
    I=\int_0^1du\,(1-u^2)^{3/2}\log\left|\frac{1+\sqrt{1-u^2}}{u}\right|\log\left|\frac{1-u}{1+u}\right| &= 2\int_0^\infty dt\, t\operatorname{sech} t \tanh^4t \log\tanh\frac{t}{2} \quad\quad (u=\operatorname{sech} t) \nonumber \\
    &= -4\int_0^1dx\frac{\left(1-x^2\right)^4}{\left(1+x^2\right)^5}\log x\log\left|\frac{1-x}{1+x}\right| \quad\quad (x=e^{-t}) \nonumber  \\
    &= -4\int_0^{\pi/4}d\theta\cos^42\theta\log\tan\theta\log\tan\left(\frac{\pi}{4}-\theta\right) \quad\quad (x=\tan\theta)
\end{align}

\noindent Since the quantity $\log\tan\theta\log\tan(\pi/4-\theta)$ is symmetric with respect to $\theta\to\pi/4-\theta$, $\cos^42\theta$ can be replaced by $(\cos^42\theta+\cos^4(\pi/2-2\theta))/2 = (\cos^42\theta+\sin^4(2\theta))/2$. 
Then $I$ can be written as
\begin{align}
    I=-\frac{3}{2}I_0-\frac{1}{2}I_8 \ , \quad I_n = \int_0^{\pi/4}d\theta\cos n\theta\log\tan\theta\log\tan\left(\frac{\pi}{4}-\theta\right) \ .
\end{align}
$I_8$ is evaluated by using integration by parts to eliminate the $\log\tan$s. 
The result is $I_8 = (\log2-1)/3$. $I_0$ is more easily evaluated by changing variables back to $x=\tan\theta$.
\begin{align}
    I_0 = \int_0^1dx\frac{1}{1+x^2}\log x\log\left|\frac{1-x}{1+x}\right| \ .
\end{align}
Expanding $1/(1+x^2)$ and $\log\left|\left(1-x\right)/\left(1+x\right)\right|$ in power series yields
\begin{align}
    I_0&=-2\sum_{m,n=0}^\infty\frac{(-1)^n}{2m+1}\int_0^1dx\, x^{2n+2m+1}\log x \nonumber\\
    &=\frac{1}{2}\sum_{m,n=0}^\infty\frac{1}{2m+1}\frac{(-1)^n}{(n+m+1)^2}\nonumber\\
    &=\frac{1}{2}\int_0^1dt\sum_{m,n=0}^\infty t^{2m}\frac{(-1)^n}{(n+m+1)^2}\ .
\end{align}
This sum is re-indexed by replacing $n$ with $r=n+m+1$,
\begin{align}
    I_0 &= \frac{1}{2}\int_0^1dt\sum_{r=1}^{\infty}\left(\sum_{m=0}^{r-1}\left(-t^{2}\right)^m\right)\frac{(-1)^{r+1}}{r^2}\nonumber\\
    &=-\frac{1}{2}\int_0^1dt\sum_{r=1}^{\infty}\frac{\left(1-\left(-t^2\right)^r\right)}{1+t^2}\frac{(-1)^{r}}{r^2} \nonumber\\
    &= -\frac{1}{2}\int_0^1dt\frac{1}{1+t^2}\sum_{r=1}^{\infty}\left(\frac{(-1)^{r}}{r^2}-\frac{t^{2r}}{r^2}\right)\nonumber\\
    &= -\frac{1}{2}\int_0^1dt\frac{1}{1+t^2}\left(\operatorname{Li}_2(-1)-\operatorname{Li}_2(t^2)\right)\ .
\end{align}
The first term is integrated trivially and gives $\pi^3/96$. The dilogarithm in the second term can be eliminated via integration by parts.
\begin{align}
    I_0 &= \frac{\pi^3}{96}+\frac{\pi^3}{48}+\int_0^1 \frac{dt}{t}\arctan t \log(1-t^2) \nonumber \\
    &= \frac{\pi^3}{32}+\Im\left\{\int_0^1\frac{dt}{t}\log(1+it)\left(\log(1+t)+\log(1-t)\right)\right\}
\end{align}
The remaining two terms are products of two logs divided by $t$. Using
\begin{align}
    ab = \left(a^2+b^2-(a-b)^2\right)/2\ ,
\end{align}
each term can be written as a linear combination of squared logarithms. The subsequent integrals are evaluated using integration by parts, producing di- and trilogarithms. This yields

\begin{align}
    I_0 = -\frac{3\pi^3}{32}-\frac{\pi}{8}\log^22+2{\rm G}\log2+4\Im\operatorname{Li}_3\left(\frac{1+i}{2}\right)\,
\end{align}
where ${\rm G}$ is Catalan's constant. Combining the finite contributions gives

\begin{align}
    A_{fin} = -\frac{4}{9}-\frac{7\pi^2}{72}+\frac{22}{9}\log2+\frac{3\pi^3}{16}+\frac{\pi}{4}\log^22-4{\rm G}\log2-8\Im\operatorname{Li}_3\left(\frac{1+i}{2}\right)\ ,
\end{align}
and 
\begin{align}
    A = A_{PDS}+A_{fin} = \frac{2}{9}-\frac{7\pi^2}{72}+\frac{34}{9}\log2+\frac{3\pi^3}{16}+\frac{\pi}{4}\log^22-4{\rm G}\log2-8\Im\operatorname{Li}_3\left(\frac{1+i}{2}\right)\ .
\end{align}

\paragraph{Result.}
Combining the three terms,
\begin{align}
V^{(g^3)} \ = \ g\lambda^2\left[\frac{13}{36}-\frac{\pi^2}{36}+\frac{3\pi^3}{16}+\frac{26}{9}\log2+\frac{\pi}{4}\log^22-4{\rm G}\log2-8\,{\rm Im}\,{\rm Li}_3\!\Big(\frac{1+i}{2}\Big)+\frac{7}{8}\zeta(3)\right] \ = \ 2.2319891\ldots\,g\lambda^2 \ ,
\end{align}
in agreement with their Monte Carlo value $2.231993(16)\,g\lambda^2$ from their Eq.~32.

\subsubsection{Closed form of \texorpdfstring{${\cal I}$}{}}
\noindent
In their Eq.~C21, the counterterm is added back with $\log(\Delta/2E_F)$ replaced by its $O(g^0)$ value.
With the corrected $\Gamma_{3b}$, the Cooper logarithm on either side of $V^{(g^2)}$ is already counted, so only the regular part of the in-medium bubble is added back.
The last line of their Eq.~C21 becomes
\begin{align}
-\frac{M^2k_F^2}{12\pi^4}\left(1+2\log2\right)\left[\textcolor{blue}{-2(\log2-1)}-\frac{\Lambda}{k_F}+\coth^{-1}\left(\frac{\Lambda}{k_F}\right)\right] \ .
\end{align}
We write ${\cal I}={\cal I}_{\rm sub}+2F_0(\log2-1)\lambda^2$, where ${\cal I}_{\rm sub}$ contains no add-back and
\begin{align}
F_0 \ = \ \frac{1+2\log2}{3} \ = \ \frac{V^{(g^2)}}{g\lambda} \ .
\end{align}
${\cal I}_{\rm sub}$ is evaluated using the spectral representation of the particle-hole loop,
\begin{align}
\Pi_{ph}(q) \ = \ \int_0^\infty d\Omega\,\frac{\sigma(\Omega,q)}{\Omega} \ , \qquad \sigma(\Omega,q) \ = \ \frac{s(\Omega,q)}{4\pi^2 q} \ ,
\end{align}
\begin{align}
s(\Omega,q) \ = \ \begin{cases} 2\Omega \ , & q<2 \ , \ \ 0<\Omega<q-\frac{q^2}{2} \ , \\[2pt] 1-\left(\frac{\Omega}{q}-\frac{q}{2}\right)^2 \ , & \left|q-\frac{q^2}{2}\right|<\Omega<q+\frac{q^2}{2} \ . \end{cases}
\end{align}
Each spectral component is a dynamic exchange $2\Omega/(l_0^2-\Omega^2+i\epsilon)$, and the Cooper energy integral is
\begin{align}
\int\frac{dl_0}{2\pi i}\,\frac{1}{l_0-\xi+i\epsilon}\,\frac{1}{-l_0-\xi+i\epsilon}\,\frac{2\Omega}{l_0^2-\Omega^2+i\epsilon} \ = \ \frac{1}{2|\xi|}\left(-\frac{2}{\Omega+|\xi|}\right) \ ,
\end{align}
where $\xi=(p^2-1)/2$.
Compared to the static potential, $1/\Omega\to1/(\Omega+|\xi|)$.
With one leg on the Fermi surface and the other at $|{\bf l}|=p$, the momentum transfer runs over $|1-p|<q<1+p$ with weight $q\,dq/2p$, giving the dynamic s-wave potential
\begin{align}
p\,F(\xi) \ = \ \frac{1}{4}\int_{|1-p|}^{1+p}dq\int d\Omega\,\frac{s(\Omega,q)}{\Omega+|\xi|} \ , \qquad F(0) \ = \ F_0 \ .
\end{align}
The in-medium Cooper loop has weight $p/(2|\xi|)$ on both sides of the Fermi surface, and in $\overline{\rm MS}$ its Fermi-surface part is twice its hole side (their Eq.~B4).
Therefore
\begin{align}
\frac{{\cal I}_{\rm sub}}{\lambda^2} \ = \ \lim_{\delta\to0}\left[\int_{|\xi|>\delta}d\xi\,\frac{p\,F(\xi)}{2|\xi|}-2F_0\int_{-1/2}^{-\delta}d\xi\,\frac{p}{2|\xi|}\right] \ .
\end{align}
At fixed $(q,\Omega)$ the condition $|1-p|<q<1+p$ is $\xi_-<\xi<\xi_+$ with $\xi_\pm=q(q\pm2)/2$, and the $\xi$-integration is elementary,
\begin{align}
\int \frac{d\xi}{2|\xi|(\Omega+|\xi|)} = \frac{\mathrm{sgn}(\xi)}{2\Omega}\log\frac{|\xi|}{\Omega+|\xi|} \ .
\end{align}
For $q<2$ the interval contains $\xi=0$, and its $\log\delta$ cancels against the subtraction, leaving $F_0(2-\log2)$.
This gives
\begin{align}
\frac{{\cal I}_{\rm sub}}{\lambda^2} \ = \ \frac{1}{8}\int_0^\infty dq\int d\Omega\,\frac{s(\Omega,q)}{\Omega}\,\Lambda(\Omega,q)+F_0\left(2-\log2\right) \ ,
\end{align}
\begin{align}
\Lambda(\Omega,q) \ = \ \begin{cases} \log\frac{|\xi_-|}{\Omega+|\xi_-|}+\log\frac{\xi_+}{\Omega+\xi_+}+2\log\Omega \ , & q<2 \ , \\[4pt] \log\frac{\xi_+}{\Omega+\xi_+}-\log\frac{\xi_-}{\Omega+\xi_-} \ , & q>2 \ . \end{cases}
\end{align}
The $\Omega$-integrations produce dilogarithms, and the $q$-integrations over $q<2$ and $q>2$ give
\begin{align}
J_< \ = \ \frac{1}{9}-\frac{22}{9}\log2+\frac{2}{3}\log^22 \ , \qquad
J_> \ = \ \frac{\pi^2+12\log2-9}{18} \ .
\end{align}
The $\log^22$ terms cancel against $F_0(2-\log2)$ and
\begin{align}
\frac{{\cal I}_{\rm sub}}{\lambda^2} \ = \ \frac{\pi^2}{18}+\frac{5}{18}-\frac{7}{9}\log2 \ .
\end{align}
Restoring the regular part of the add-back,
\begin{align}
{\cal I}\left[\Pi_{pp}^{(g^0)}V^{(g^2)}\right] \ = \ \lambda^2\left[\frac{\pi^2}{18}-\frac{7}{18}-\frac{13}{9}\log2+\frac{4}{3}\log^22\right] \ = \ -0.2011861\ldots\,\lambda^2 \ .
\end{align}
A direct numerical evaluation of the unswapped three-dimensional integral agrees with this to $10^{-9}$.
For reference, the same method applied to the add-back of their Eq.~C21 gives $\lambda^2\left[\frac{\pi^2}{18}+\frac{7}{18}-\frac{1}{3}\log2+\frac{4}{9}\log^22\right]=0.9196859\,\lambda^2$, compared to their Monte Carlo value $0.917746(98)\,\lambda^2$ in their Eq.~C23.
This discrepancy is likely due to the uncertainty in Monte Carlo being underestimated due to the integrand having heavy tails in Ref.~\cite{BCF2024}.

\subsubsection{Corrected result}
\noindent
Using their Eq.~28, $V^{(g^2)}/g=\frac{\lambda}{3}\left(1+2\log2\right)$, and their Eq.~30, $g\Pi_{pp}^{(g^0)}=2\lambda\left(\log2-1\right)$, the first two terms in blue in the corrected pole condition are
\begin{align}
\left(\frac{V^{(g^2)}}{g}\right)^2 \ = \ \frac{\lambda^2}{9}\left(1+2\log2\right)^2 \ , \qquad
-2\Pi_{pp}^{(g^0)}V^{(g^2)} \ = \ \frac{4\lambda^2}{3}\left(1+\log2-2\log^22\right) \ .
\end{align}
Combining these with $V^{(g^3)}$ and ${\cal I}$ gives the corrected version of their Eq.~32,
\begin{align}
0 \ = \ 1+\lambda\left[\log\left(\frac{8E_F}{e^2\Delta}\right)-\frac{1}{3}\left(1+2\log2\right)\right]
+\lambda^2 B \ ,
\end{align}
with
\begin{align}
B \ = \ \frac{11}{36}+\frac{5\pi^2}{36}-\frac{3\pi^3}{16}-4\log2+\Big(\frac{4}{9}-\frac{\pi}{4}\Big)\log^22+4G\log2+8\,{\rm Im}\,{\rm Li}_3\!\Big(\frac{1+i}{2}\Big)-\frac{7}{8}\zeta(3) \ = \ -1.0253286\ldots \ .
\end{align}
Including the effective mass $M^*$ and wavefunction renormalization $Z$ of their Eq.~33, their Eq.~34 becomes
\begin{align}
\Delta_{\rm NNLO} \ = \ \frac{8E_F}{e^2(4e)^{1/3}}\left(1+B\,\lambda\right)\exp\left(\frac{\pi}{2k_Fa}\frac{M}{M^*}\frac{1}{Z^2}\right) \ .
\end{align}
Extracting the $O(\lambda)$ coefficient in $\log\Delta_{\rm NNLO}$, using $\frac{M}{M^*}\frac{1}{Z^2}=1+(2+16\log2)\lambda^2/15+O(\lambda^3)$~\cite{landau1980course}, gives
\begin{align}
c_{1,0} \ = \ \frac{79}{180}+\frac{5\pi^2}{36}-\frac{3\pi^3}{16}-\frac{44}{15}\log2+\Big(\frac{4}{9}-\frac{\pi}{4}\Big)\log^22+4{\rm G}\log2+8\,{\rm Im}\,{\rm Li}_3\!\Big(\frac{1+i}{2}\Big)-\frac{7}{8}\zeta(3)  \ .
\end{align}

\subsection{Details on the next-to-next-to leading order computation: soft contribution to the self-energy}
\label{app:NNLO_Sigma}
\noindent
The self-energy comes from a single cut of $W^{[2]}$ in Eq.~\eqref{eq:W2}.
The self-energy in Eq.~\eqref{eq:Sigmadef} is parameterized through
\begin{align}
    \tilde{\Sigma}(k) \ = \ \sigma_o(k)\tilde{\mathds{1}} \ + \ \sigma_e(k)\tilde{\tau}_3 \ + \ \Sigma_1(k)\tilde{\tau}_1 \ .
\end{align}
Projecting the single-cut Feynman diagram onto this basis gives,
\begin{align}
    \sigma_o(k) \ &= \ -\frac{g^2}{4}\int_{l,l'} \left [G_0(l+l'-k)-G_0(\bar{l}+\bar{l}'-\bar{k})\right ]\left [2 F_0(l) F_0(l')+G_0(l') G_0(\bar{l})+G_0(l) G_0(\bar{l}')\right ] \ , \nonumber \\
    \sigma_e(k) \ &= \ \frac{g^2}{4}\int_{l,l'}\left[G_0(l+l'-k)+G_0(\bar l+\bar l'-\bar k)\right]\left[2F_0(l)F_0(l')+G_0(l')G_0(\bar l)+G_0(l)G_0(\bar l')\right] \ , \nonumber \\
\Sigma_1(k)&=\frac{g^2}{2}\int_{l,l'}F_0(l+l'-k)\left [2 F_0(l) F_0(l')+G_0(l') G_0(\bar{l})+G_0(l) G_0(\bar{l}')\right ] \ ,
\end{align}
where, for example, $\bar{l} = (-l_0,{\bf l})$.
After doing the energy integrals this becomes,
\begin{align}
    \sigma_o(k)&=g^2\!\int\!\frac{d^3l_1}{(2\pi)^3}\frac{d^3l_2}{(2\pi)^3}\frac{k_0(\bar{\Delta}^2+\xi_{1
    }\xi_{2}-E_{1}E_{2})}{4E_{1}E_{2}\,(E_T^2-k_0^2-i\epsilon)} \ , \nonumber \\
\sigma_e(k)&=g^2\!\int\!\frac{d^3l_1}{(2\pi)^3}\frac{d^3l_2}{(2\pi)^3}\frac{\xi_3E_T(\bar{\Delta}^2+\xi_{1}\xi_{2}-E_{1}E_{2})}{4E_{1}E_{2}E_3\,(E_T^2-k_0^2-i\epsilon)} \ ,\nonumber \\
\Sigma_1(k)&=-g^2\bar{\Delta}\!\int\!\frac{d^3l_1}{(2\pi)^3}\frac{d^3l_2}{(2\pi)^3}\frac{ E_T(\bar{\Delta}^2+\xi_{1}\xi_{2}-E_{1}E_{2})}{4E_{1}E_{2}E_3\,(E_T^2-k_0^2-i\epsilon)},
\label{eq:selfEnergyints}
\end{align}
with $E_T=E_1+E_{2}+E_3$ and ${\bf l}_1={\bf l}$, ${\bf l}_2={\bf l}'$ and ${\bf l}_3={\bf l}_1+{\bf l}_2-{\bf k}$.
We are using the shorthand $E_i = E_{l_i}$ and $\xi_i = \xi_{l_i}$ and work in units $M=k_F=1$, so that $\mu=E_F=1/2$.

The strategy for evaluating these integrals closely follows Appendix~\ref{app:NLO}.
The integrand depends only on the magnitudes $\lmag{i}$.
To reduce the angular integrals, we introduce
\begin{align}
\bar{\bf l} \ = \ {\bf l}_1-{\bf l}_3 \ = \ {\bf k}-{\bf l}_2 \ , \qquad p \ \equiv \ |\bar{\bf l}| \ ,
\label{eq:bark_sigma}
\end{align}
and integrate over $\bar{\bf l}$ and ${\bf l}_1$ instead of ${\bf l}_1$ and ${\bf l}_2$.
This transformation has unit Jacobian.
At fixed $\bar{\bf l}$, the angular integral over ${\bf l}_1$, with polar axis along $\bar{\bf l}$, trades $\cos\theta_1$ for $\lmag{3}$, while the angular integral over $\bar{\bf l}$, with polar axis along ${\bf k}$, trades $\cos\theta$ for $\lmag{2}$:
\begin{align}
d^3l_1 \ \to \ \frac{2\pi}{p}\,\lmag{1}\,\lmag{3}\,d\lmag{1}\,d\lmag{3} \ , \qquad
d^3\bar l \ \to \ \frac{2\pi p}{|{\bf k}|}\,\lmag{2}\,dp\,d\lmag{2} \ ,
\end{align}
subject to the triangle constraints
\begin{align}
C_1:\ \big|\lmag{1}-\lmag{3}\big|\leq p\leq \lmag{1}+\lmag{3} \ , \qquad
C_2:\ \big||{\bf k}|-\lmag{2}\big|\leq p\leq |{\bf k}|+\lmag{2} \ .
\end{align}
The factors of $p$ cancel, and the $p$ integral gives the length ${\cal L}\equiv\text{length}(C_1\cap C_2)$.
Changing variables to $\xi_i=(\lmag{i}^2-1)/2$, the measures in Eq.~\eqref{eq:selfEnergyints} become
\begin{align}
\int\!\frac{d^3l_1}{(2\pi)^3}\frac{d^3l_2}{(2\pi)^3} \ \to \ \frac{1}{16\pi^4|{\bf k}|}\int_{-1/2}^{\infty}\!d\xi_1\,d\xi_2\,d\xi_3\ {\cal L} \ .
\end{align}

The $O(\bar{\Delta}^2)$ contribution that effects $c_{2,1}$ comes from the soft region $\{k_0,\xi_1,\xi_2,\xi_3\}\sim\bar{\Delta}$.
When inserted and integrated over in $\int_k\Pi_-^{[2]}(k)$, we also have the requirement that $\xi_k\sim\bar{\Delta}$ and therefore $\lmag{1},\lmag{2},\lmag{3},|{\bf k}|\to1$ to lowest order.
The intervals $C_1$ and $C_2$ both become $[0,2]$ and ${\cal L}=2+O(\bar{\Delta})$.
To isolate the soft region, we rescale $\xi_i = \bar{\Delta} t_i$ and $k_0=\bar{\Delta} \bar{k}_0$ giving
\begin{align}
\sigma_o(k)\big|_S&=g^2\bar{\Delta}^2 \frac{1}{8\pi^4}\!\int_{\mathbb{R}^3}\!d^3t\frac{\bar{k}_0(1+t_1t_2-e_{1}e_{2})}{4e_{1}e_{2}\,(e_T^2-\bar{k}_0^2-i\epsilon)} \ , \nonumber \\
\sigma_e(k)\big|_S&=g^2\bar{\Delta}^2\frac{1}{8\pi^4}\!\int_{\mathbb{R}^3}\!d^3t\frac{t_3e_T(1+t_{1}t_{2}-e_{1}e_{2})}{4e_{1}e_{2}e_3\,(e_T^2-\bar{k}_0^2-i\epsilon)} \ ,\nonumber \\
\Sigma_1(k)\big|_S&=-g^2\bar{\Delta}^2\frac{1}{8\pi^4}\!\int_{\mathbb{R}^3}\!d^3t\frac{ e_T(1+t_{1}t_{2}-e_{1}e_{2})}{4e_{1}e_{2}e_3\,(e_T^2-\bar{k}_0^2-i\epsilon)} \ , 
\label{eq:selfEnergyints2}
\end{align}
where $e_i = \sqrt{1+t_i^2}$ and $e_T=e_1+e_2+e_3$.
The bounds of integration at leading order are $\int_{-1/(2\bar{\Delta})}^{\infty} \to \int_{-\infty}^{\infty}$.
The odd terms in the integrand drop out giving $\sigma_e\big|_S = 0$ and
\begin{align}
\sigma_o(k)\big|_S&=g^2\bar{\Delta}^2 \frac{1}{8\pi^4}\!\int_{\mathbb{R}^3}\!d^3t\frac{\bar{k}_0(1-e_{1}e_{2})}{4e_{1}e_{2}\,(e_T^2-\bar{k}_0^2-i\epsilon)} \ , \nonumber \\
\Sigma_1(k)\big|_S&=-g^2\bar{\Delta}^2\frac{1}{8\pi^4}\!\int_{\mathbb{R}^3}\!d^3t\frac{ e_T(1-e_{1}e_{2})}{4e_{1}e_{2}e_3\,(e_T^2-\bar{k}_0^2-i\epsilon)} \ .
\label{eq:selfEnergyints3}
\end{align}

To treat the pole in $\bar{k}_0$, we put these self-energy terms into the integral over $\Pi_-^{[2]}$,
\begin{align}
    \int_k\Pi_-^{[2]}(k) \ = \ \int_k\frac{i\big(2k_0\,\sigma_o(k)-2\bar{\Delta}\,\Sigma_1(k)\big)}{(k_0^2-E_k^2+i\epsilon)^2}\ .
\end{align}
This expression has poles at $k_0 = (E_k-i\epsilon),(-E_k+i\epsilon),(E_T-i\epsilon),(-E_T+i\epsilon)$.
These lie in the second and fourth quadrant of the complex plane and can be avoided by Wick rotating counterclockwise with $k_0= i\nu$.
The $i\epsilon$ prescription can now be dropped giving,
\begin{align}
    \int_k\Pi_-^{[2]}(k)\Big|_S \ &= \ \int\frac{d^3 k}{(2\pi)^3}\int\frac{d\nu}{2\pi}\frac{-1}{{(\nu^2+E_k^2)^2}}\big [2i\nu \sigma_o(i\nu)\big|_S-2\bar{\Delta}\,\Sigma_1(i\nu)\big|_S\big] \nonumber \\
    &=\frac{1}{4\pi^3}\int_{-1/2}^{\infty} d\xi_k|{\bf k}|\int d\nu\frac{-1}{{(\nu^2+E_k^2)^2}}\big [2i\nu \sigma_o(i\nu)\big|_S-2\bar{\Delta}\,\Sigma_1(i\nu)\big|_S\big] \nonumber \\
    &=\frac{1}{4\pi^3\bar{\Delta}}\int_{-\infty}^{\infty} dt_0\int d\bar{\nu}\frac{-1}{{(\bar{\nu}^2+e_0^2)^2}}\big [2i\bar{\nu} \sigma_o(i\bar{\nu})\big|_S-2\,\Sigma_1(i\bar{\nu})\big|_S\big] 
    \label{eq:Pi2reduce}
\end{align}
where in the second line we changed variables $d^3 k \to (4\pi) d\xi_k|{\bf k}|$ using the fact that the integrand only depends on $|{\bf k}|$. 
In the third line we took the soft limit where $|{\bf k}|\to 1$ and defined $\xi_k=\bar{\Delta} t_0$, $e_0 = \sqrt{1 +t_0^2}$ and $\nu = \bar{\Delta} \bar{\nu}$.

The self-energies in Eq.~\eqref{eq:selfEnergyints3} are evaluated at $\bar k_0=i\bar\nu$.
The $[.]$ in Eq.~\eqref{eq:Pi2reduce} becomes
\begin{align}
\left [2i\bar\nu\,\sigma_o(i\bar\nu)-2\,\Sigma_1(i\bar\nu)\right ]  \ = \ -\frac{g^2\bar{\Delta}^2}{8\pi^4}\int_{\mathbb{R}^3}\!d^3t\;\frac{1-e_1e_2}{4e_1e_2}\left[\frac{2\bar\nu^2}{ e_T^2+\bar\nu^2}-\frac{2e_T}{e_3\big[e_T^2+\bar\nu^2\big]}\right] \ ,
\end{align}
giving
\begin{align}
    \int_k\Pi_-^{[2]}(k)\Big|_S \ = \ \frac{g^2\bar{\Delta}}{32\pi^7}\int_{-\infty}^{\infty} dt_0\int d\bar{\nu}\frac{1}{{(\bar{\nu}^2+e_0^2)^2}}\int_{\mathbb{R}^3}\!d^3t\;\frac{1-e_1e_2}{4e_1e_2}\left[\frac{2\bar\nu^2}{ e_T^2+\bar\nu^2}-\frac{2e_T}{e_3\big[e_T^2+\bar\nu^2\big]}\right] \ .
\end{align}
The outer $t_0$ integral is straightforward
\begin{align}
\int_{-\infty}^{\infty}\!dt_0\,\frac{1}{(\bar\nu^2+e_0^2)^2} \ = \ \frac{\pi}{2(\bar\nu^2+1)^{3/2}} \ .
\end{align}
The remaining $t_i$-integrals are decoupled using
\begin{align}
\frac{\bar\nu}{ e_T^2+\bar\nu^2}=\int_0^\infty\!ds\,\sin(\bar\nu s)\,e^{-se_T} \ , \qquad
\frac{e_T}{e_T^2+\bar\nu^2}=\int_0^\infty\!ds\,\cos(\bar\nu s)\,e^{-se_T} \ .
\end{align}
There is cancellation by changing variables to $t_i=\sinh u_i$ giving
\begin{align}
\int_{\mathbb{R}^3}\!d^3t\;\frac{1-e_1e_2}{4e_1e_2}\,e^{-se_T}=2K_1(s)(K_0(s)^2-K_1(s)^2) \ , \qquad
\int_{\mathbb{R}^3}\!d^3t\;\frac{1-e_1e_2}{4e_1e_2e_3}\,e^{-se_T}=2K_0(s)(K_0(s)^2-K_1(s)^2) \ ,
\end{align}
where $K_n$ are modified Bessel functions of the second kind. 
Exchanging the order of integration and using $g=2\pi^2\lambda$ gives,
\begin{align}
\int_k\Pi_-^{[2]}(k)\Big|_S \ = \ \frac{\lambda^2\bar{\Delta}}{4\pi^2}\int_0^\infty\!ds\,\left (K_0(s)^2-K_1(s)^2\right )\int_{-\infty}^{\infty}\!d\bar\nu\;\frac{\bar\nu\sin(\bar\nu s)\,K_1(s)-\cos(\bar\nu s)\,K_0(s)}{(\bar\nu^2+1)^{3/2}} \ .
\end{align}
The $\bar\nu$ integrals follow from Basset's integral for $K_1$, and for $K_0$ (see Section 10.32 of Ref.~\cite{NIST:DLMF}) after an integration by parts using $\bar\nu/(\bar\nu^2+1)^{3/2}=-\partial_{\bar\nu}(\bar\nu^2+1)^{-1/2}$,
\begin{align}
\int_{-\infty}^{\infty}\!d\bar\nu\,\frac{\cos(\bar\nu s)}{(\bar\nu^2+1)^{3/2}}=2sK_1(s) \ , \qquad
\int_{-\infty}^{\infty}\!d\bar\nu\,\frac{\bar\nu\sin(\bar\nu s)}{(\bar\nu^2+1)^{3/2}}=s\!\int_{-\infty}^{\infty}\!d\bar\nu\,\frac{\cos(\bar\nu s)}{(\bar\nu^2+1)^{1/2}}=2sK_0(s) \ .
\end{align}
Putting everything together gives the soft contribution to the gap equation
\begin{align}
\int_k\Pi_-^{[2]}(k)\Big|_S \ = \ \frac{\lambda^2\bar{\Delta}}{4\pi^2}\int_0^\infty\!ds\,(K_0(s)^2-K_1(s)^2)\Big[2sK_0(s)K_1(s)-2sK_1(s)K_0(s)\Big] \ = \ 0 \ .
\end{align}
Thus the self-energy does not contribute directly to $c_{2,1}$.

As discussed in Sec.~\ref{sec:NNLO}, the gap extracted from QMC calculations and measured in experiments is the fermionic excitation gap $\Delta$ that is the lowest positive energy pole in the Nambu-Gor'kov propagator.
This is distinct from the gap $\bar{\Delta}$ that appears in the leading propagator, and at lowest order they are related via Eq.~\eqref{eq:Delta_Deltaqp}.
Unlike for the pole condition, the shift depends on the difference $\left ( \sigma_o(\bar{\Delta})\big|_S-\Sigma_1(\bar{\Delta})\big|_S\right )$, which does not vanish.
The soft self-energies in Eq.~\eqref{eq:selfEnergyints3} therefore need to be evaluated at $\bar k_0=1$.
This point lies below the three-quasiparticle threshold since $e_T\geq3$ and the $i\epsilon$ can be safely dropped giving,
\begin{align}
\sigma_o(\bar{\Delta})\big|_S-\Sigma_1(\bar{\Delta})\big|_S \ = \ \frac{g^2\bar{\Delta}^2}{8\pi^4}\int_{\mathbb{R}^3}\!d^3t\;\frac{1-e_1e_2}{4e_1e_2}\left[\frac{1}{e_T^2-1}+\frac{e_T}{e_3(e_T^2-1)}\right] \ .
\end{align}
The $t_i$-integrals are decoupled as before, now using
\begin{align}
\frac{1}{e_T^2-1}=\int_0^\infty\!ds\,\sinh s\,e^{-se_T} \ , \qquad
\frac{e_T}{e_T^2-1}=\int_0^\infty\!ds\,\cosh s\,e^{-se_T} \ ,
\end{align}
which converge since $e_T-1\geq2$.
The change of variables $t_i=\sinh u_i$ gives
\begin{align}
\sigma_o(\bar{\Delta})\big|_S-\Sigma_1(\bar{\Delta})\big|_S \ = \ \lambda^2\bar{\Delta}^2\int_0^\infty\!ds\,\big[\cosh s\,K_0(s)+\sinh s\,K_1(s)\big]\big(K_0(s)^2-K_1(s)^2\big) \ .
\label{eq:Jos_int}
\end{align}

The integral converges at large $s$ but diverge as $s\to0$:
\begin{align}
(\cosh s\,K_0+\sinh s\,K_1)(K_0^2-K_1^2) \ = \ \frac{\eta(s)-1}{s^2}+O(\eta^3) \ , \quad \eta(s) \equiv \gamma_E + \log(s/2)\ .
\end{align}
Similar to $R^{[2]}\big \vert_S$ in Appendix~\ref{app:NLO}, this divergence comes from $t_i\to\infty$, where the lines are far from the Fermi surface and the region is no longer soft. Multiplied by the overall $\bar{\Delta}^2$, the $1/s^2$ terms are independent of $\bar{\Delta}$ up to logarithms.
They are the low-energy end of hard-region contributions, which belong to the $O(\bar{\Delta}^0)$ sector and shift $c_{1,0}$ through the wavefunction renormalization. 
A $1/s$ term would instead give $\bar{\Delta}^2\log\bar{\Delta}$, i.e. a contribution to $c_{1,1}$.
There is no $1/s$ term, so the soft contribution is the finite part.
Defining, $y=K_0(s)$, $y'=-K_1(s)$, $C=\cosh s$ and $S=\sinh s$, the integrand can be rewritten as,
\begin{align}
&C\,y\big(y^2-y'^2\big)-S\,y'\big(y^2-y'^2\big) \ = \ \frac{23}{12}\,C\,y^3-\frac{2}{3}\,s\,S\,y^3+{\cal G}' \ , \nonumber\\
&{\cal G} \ = \ -\frac{1}{3}Sy^3-\frac{1}{2}sSy'^3+sCyy'^2+\frac{1}{4}sSy^2y'-\frac{7}{12}sCy^3+\frac{1}{2}s^2Cy'^3-\frac{1}{2}s^2Syy'^2-\frac{1}{2}s^2Cy^2y'+\frac{1}{2}s^2Sy^3 \ .
\end{align}
${\cal G}$ vanishes as $s\to\infty$ and ${\cal G}(\epsilon)=-\eta(\epsilon)/\epsilon+o(1)$, so the boundary term $-{\cal G}(\epsilon)$ is exactly the divergence of $\int_\epsilon ds\,(\eta-1)/s^2$, and
\begin{align}
J_{os} \ \equiv \ \text{FP}\!\int_0^\infty\!ds\,\big[\cosh s\,K_0+\sinh s\,K_1\big]\big(K_0^2-K_1^2\big) \ = \ \frac{23}{12}\int_0^\infty\!ds\,\cosh s\,K_0^3-\frac{2}{3}\int_0^\infty\!ds\,s\sinh s\,K_0^3
 .
\label{eq:Jos}
\end{align}
These moments are not among those evaluated in Ref.~\cite{Bailey:2008ib} but the following form was identified with PSLQ~\cite{Ferguson1999AnalysisOP} and confirmed to 500 digits, 
\[
J_{os}\ = \ \sqrt{2}\left (\frac{47}{1536}[\Gamma(1/8)\Gamma(3/8)]^2-\frac{\pi^4}{[\Gamma(1/8)\Gamma(3/8)]^2}\right )
\]
This gives,
\begin{align}
\Delta \ = \ \bar{\Delta}+\lambda^2\bar{\Delta}^2\,J_{os} \ = \  \bar{\Delta}+\lambda^2\bar{\Delta}^2\sqrt{2}\left (\frac{47}{1536}[\Gamma(1/8)\Gamma(3/8)]^2-\frac{\pi^4}{[\Gamma(1/8)\Gamma(3/8)]^2}\right ) \ ,
\end{align}
which determines the shift to $c_{2,1}$ in Eq.~\eqref{eq:c21qp}.

\subsection{Static limit of the pairing potential}
\noindent
\label{app:contactph}
\noindent
In this appendix, we derive the $s$-wave irreducible pairing potential $\hat{K}^{[2]}$ in the static limit up to $O(\Delta^2\log^2{\Delta})$.
The static approximation is not used to compute the $c_{2,1}$ coefficient important to this work, but may be useful in other contexts.
The relevant integrals for the $K_{11}$ and $K_{12}$ matrix elements are given in Eq.~\eqref{eq:K11} and Eq.~\eqref{eq:K12} respectively.
We break this into two integrals that are evaluated separately,
\begin{align}
    I_{FF} \ &= \ -i \int_l F_0(l)F_0(l+q) \nonumber \\
    I_{GG} \ &=\ -i \int_l G_0(l)G_0(l+q) \ . 
    \label{eq:I_FFI_GG}
\end{align}
In the static approximation, the external legs are put on the Fermi surface and therefore, with our external kinematics, $q= (0,{\bf q})$ where ${\bf q} = {\bf k} - {\bf k}'$ is the momentum transfer.
Because the external momentum are at the Fermi surface, we have $0\leq|{\bf q}|\leq 2k_F$.
The choice of $q$ instead of $\bar {q}$ does not matter as they are equivalent when projected onto the $s$-wave.
Throughout this appendix, we use units with $k_F=M=1$ and take $\mu = 1/2$ as pairing-induced corrections are higher order than we will be working.

\subsubsection{\texorpdfstring{$I_{FF}$}{}}
\noindent
After doing the $l_0$ integral, the $I_{FF}$ term in Eq.~\eqref{eq:I_FFI_GG} is,
\begin{align}
I_{FF}(\Delta,{\bf q}) \ &= \ \int \frac{d^3 l}{(2\pi)^3} \frac{\Delta^2}{2(E_{{\bf l}}+E_{{\bf l}+{\bf q}})E_{{\bf l}}E_{{\bf l}+{\bf q}}} \nonumber \\
&= \ \frac{1}{4\pi^2 q}\int_{{\cal D}} d\xi_1 d\xi_2 \frac{\Delta^2}{2\left (E_1+E_2\right )E_1E_2} 
\end{align}
where we have changed variables to $\xi_1 = l^2/2 - 1/2$ and $\xi_2 = |{\bf l} + {\bf q}|^2/2 - 1/2$.
In these variables, $E_{1,2}=\sqrt{\Delta^2+\xi_{1,2}^2}$ and the integration domain ${\cal D}$ is 
\[
\xi_1 \in[-1/2,\infty] \quad  \text{and} \quad\xi_2 \in[y_-,y_+] \quad  \text{where} \quad y_{\pm} = \left (\xi_1 \pm |{\bf q}|\sqrt{2\xi_1+1}+|{\bf q}|^2/2 \right )
\]
First,  perform partial fractions and then integrate over $\xi_2$ giving,
\begin{align}
I_{FF}[B] \ &= \ \frac{1}{4\pi^2 q}\int_{{\cal D}} d\xi_1 d\xi_2 \frac{\Delta^2}{2E_1^2}\left (\frac{1}{E_2}-\frac{1}{E_1 +E_2} \right ) \nonumber \\
&= \ \frac{1}{4\pi^2 q}\int_{-1/2}^{\infty} d\xi_1 \frac{\Delta^2}{2(\xi_1^2+\Delta^2)}B(\xi_1,\Delta)
\label{eq:Iff}
\end{align}
where 
\begin{align}
B(\xi_1,\Delta) \ = \ \left (\sinh^{-1}\frac{y_+}{\Delta}-\sinh^{-1}\frac{y_-}{\Delta} \right ) - \left (\int_{y_-}^{y+} d\xi_2 \frac{1}{E_1+E_2}\right ) \ .
\label{eq:B}
\end{align}
The second integral can be expressed in terms of logs but is more convenient to leave it symbolic.
The integrals branch depending on the sign of $y_{\pm}$.
The cases are given in Table~\ref{tab:signs}, where we have defined $A_{\pm}=(|{\bf q}|^2/2 \pm |{\bf q}|)$. Note that $|{\bf q}|\in [0,2]$ and will be integrated over when the irreducible potential is projected onto the s-wave.
Throughout this section we will assume that $\{|{\bf q}|,A_{\pm}\}\sim O(\Delta^0)$.
The correction from the region of the s-wave projection that has $\{|{\bf q}|,A_{\pm}\}\sim O(\Delta)$ will be discussed at the end of the appendix.
\begin{figure}[t]
\centering
\setlength{\tabcolsep}{4pt} 
\renewcommand{\arraystretch}{1.5}

\begin{tabular}{|c|c|c|c|}
    \hline
    region & band & signs of $(y_{-},y_+)$ & sign of $\xi_1$\\
    \hline
    1 & $\xi_1\in(A_-,A_+)$ & $(-,+)$ & $\xi_1$ crosses 0\\
    \hline
    2 & $\xi_1>A_+$ & $(+,+)$ & $+$\\
    \hline
    3 & $\xi_1<A_-$ & $(-,-)$ if $|{\bf q}|<1$; $(+,+)$ if $|{\bf q}|>1$ & $-$\\
    \hline
\end{tabular}
\hfill
\begin{tabular}{|c|c|}
    \hline
    Boundary & zero for \\
    \hline
    $y_-(A_+)$ & $0\leq |{\bf q}|\leq2$ \\
    \hline
    $y_-(A_-)$ & $|{\bf q}|>1$ \\
    \hline
    $y_+(A_+)$ & never \\
    \hline
    $y_+(A_-)$ & $|{\bf q}| <1$ \\
    \hline
\end{tabular}
\caption{Left:The region of integration over $\xi_1$ organized by the sign of ($y_-,y_+$). Right: The combinations of $y_{\pm}(A_{\pm'})$ that vanish.}
\label{tab:signs}
\end{figure}

The goal is to extract the coefficients of the $\{\Delta,\Delta^2\log{\Delta},\Delta^2\}$ terms in Eq.~\eqref{eq:Iff}.
To begin with, consider the naive $\Delta\to0$ limit of Eq.~\eqref{eq:B},
\begin{align}
B_{\text{out}} \ = \ \left (\sgn(y_+)\log\frac{2|y_+|}{\Delta} - \sgn(y_-)\log\frac{2|y_-|}{\Delta} \right ) \ - \ \left (\sgn(y_+)\log{\left ( 1+\frac{|y_+|}{|\xi_1|}\right )}-\sgn(y_-)\log{\left (1+\frac{|y_-|}{|\xi_1|} \right )} \right ) \ .
\label{eq:Bout}
\end{align}
Where we have used $\sinh^{-1}(x) = \sgn{(x)}\log{(2|x|)} + O(1/x^2)$ and performed the integral with $E_i \to |\xi_i|$.
This is naive because both $y_{\pm}$ and $\xi_2$ can pass through zero making this expansion invalid.
Our strategy will be to compute,
\begin{align}
I_{FF} \ = \ I_{FF}[B_{\text{out}}] \ + \ I_{FF}[B-B_{\text{out}}] \ . 
\label{eq:IFF_decomp}
\end{align}
Splitting the integral in this way makes it easier to isolate the $\Delta$ dependence.

\paragraph{\texorpdfstring{$I_{FF}[B_{\text{out}}]$}{}}
\noindent
First, we compute the $I_{FF}[B_{\text{out}}]$ component of $I_{FF}$ in Eq.~\eqref{eq:IFF_decomp}.
This is done by breaking the integral into the three regions that are enumerated in Table~\ref{tab:signs}.
In region 1 we have,
\begin{align}
B_{\text{out}}^{(1)} \ = \ \log \left (\frac{ 4\xi_1^2}{\Delta^2} \right )  \ + \ R(\xi_1) \quad , \quad R(\xi_1)=\log{\left ( \frac{-y_+ y_-}{(|\xi_1| - y_-)(|\xi_1| + y_+)} \right )} \ .
\end{align}
The first term integrates to,
\begin{align}
&\frac{1}{4\pi^2 |{\bf q}|}\int_{A_-}^{A_+}d\xi_1 \frac{\Delta^2}{2(\xi_1^2+\Delta^2)}\log{\left ( \frac{4\xi_1^2}{\Delta^2} \right )} \nonumber \\
& = \ \frac{\Delta}{4\pi^2 |{\bf q}|}\int_{A_-/\Delta}^{A_+/\Delta}dt \frac{1}{2(1+t^2)}\log{\left ( 4 t^2 \right )} \nonumber \\
& =  \frac{\Delta}{4\pi^2 |{\bf q}|}\left [\pi \log{2}  + \Delta\log{(\Delta)}\left (\frac{1}{A_+} - \frac{1}{A_-}\right )  +  \Delta\left (\frac{1}{A_-}(1 + \log{(-2 A_-)})-\frac{1}{A_+}(1 + \log{(2 A_+))} \right ) \right ]  +  O(\Delta^3)
\end{align}
where in the second line we changed variables to  $t = \xi_1/\Delta$.

In the second term, all $\Delta$ dependence is carried by the $\Delta^2/(\xi_1^2 + \Delta^2)$ prefactor.
It is straightforward to show that any function that is smooth at $\xi_1=0$ cannot generate a $\log\Delta$.
Therefore, we split $R(\xi_1)$ into a component that is smooth and kinked at $\xi_1=0$.
The latter is extracted by expanding around the origin
\begin{align}
R(\xi_1) \ &= \ \frac{1}{2}\left (\frac{dR}{d\xi_1}\bigg |_{0^+} - \frac{dR}{d\xi_1}\bigg |_{0^-} \right )|\xi_1| \ + \ \frac{1}{2}\left (\frac{dR}{d\xi_1}\bigg |_{0^+} + \frac{dR}{d\xi_1}\bigg |_{0^-} \right )\xi_1 \ + \ O(\xi_1^2) \nonumber \\ 
& = \ -\left (\frac{1}{A_+} -\frac{1}{A_-}\right )|\xi_1| \ + \ O(\xi_1^2) \ . 
\end{align}
The $\Delta^2 \log \Delta$ piece is extracted as,
\begin{align}
&\frac{1}{4\pi^2 |{\bf q}|}\int_{A_-}^{A_+} d\xi_1 \frac{-\Delta^2}{2(\xi^2 + \Delta^2)}\left (\frac{1}{A_+} -\frac{1}{A_-}\right )|\xi_1| \nonumber \\
&=\frac{\Delta^2}{4\pi^2 |{\bf q}|}\left (\frac{1}{A_+}-\frac{1}{A_-} \right )\left (\log{\Delta} - \frac{1}{2}\log{(-A_+ A_-)} \right )
\end{align}
The remaining $\Delta^2$ component can be extracted by subtracting the kink and taking $E_1\to\xi_1$  inside the integrand,
\begin{align}
&\frac{1}{4\pi^2 |{\bf q}|}\int_{A_-}^{A_+} d\xi_1 \frac{\Delta^2}{2\xi_1^2}\left (R(\xi_1)+ \left (\frac{1}{A_+} -\frac{1}{A_-}\right )|\xi_1|\right ) \nonumber \\
&\equiv \frac{\Delta^2}{4\pi^2 |{\bf q}|} a^{(1)}(|{\bf q}|)
\end{align}
The integral $a^{(1)}(|{\bf q}|)$ can be done analytically and has different expressions for $|{\bf q}|<1$ and $|{\bf q}|>1$,
\begin{align}
&a^{(1)}(|{\bf |{\bf q}|}|<1) \ = \
\frac{1}{4}\left (\frac{1}{A_+}-\frac{1}{A_-} \right ) \bigg[5 \log \left(4-|{\bf q}|^2\right)+\log \left(36-25 |{\bf q}|^2\right)+\sqrt{4-|{\bf q}|^2} \log \left(\frac{5 \sqrt{4-|{\bf q}|^2}+8}{5 \sqrt{4-|{\bf q}|^2}-8}\right) \nonumber \\
& + 7 |{\bf q}| \tanh ^{-1}\left(\frac{|{\bf q}|}{2}\right)-|{\bf q}| \tanh ^{-1}\left(\frac{5 |{\bf q}|}{6}\right)-4-16 \log (2)\bigg]\ , \nonumber \\[5pt]
&a^{(1)}(|{\bf q}|>1) \ = \ \frac{1}{8}\left (\frac{1}{A_+}-\frac{1}{A_-} \right )\bigg [ 6 \sqrt{4-|{\bf q}|^2} \log \left(\frac{\sqrt{4-|{\bf q}|^2}+2}{|{\bf q}|}\right)-2 \sqrt{4-|{\bf q}|^2} \tanh ^{-1}\left(\frac{(3 |{\bf q}|+2) \sqrt{4-|{\bf q}|^2}}{4 |{\bf q}|^2+6 |{\bf q}|+4}\right) \nonumber \\
&+8 \log (|{\bf q}|)+(7 |{\bf q}|+10) \log (|{\bf q}|+2)+(2-3 |{\bf q}|) \log ((2-|{\bf q}|) (3 |{\bf q}|-2))+(2-|{\bf q}|) \log (5 |{\bf q}|+6)-8-32 \log (2)\bigg ]
\end{align}
Summing together we get the naive region 1 contribution of,
\begin{align}
&I_{FF}[B_{\text{out}}^{(1)}] \ = \  \frac{\Delta}{4\pi^2 |{\bf q}|}\bigg  [\pi \log{2} \ + \ 2\Delta\log{(\Delta)}\left (\frac{1}{A_+} - \frac{1}{A_-}\right )  \nonumber \\
&+ \ \Delta\left (\frac{1}{A_-}(1 + \log{(-2 A_-)})-\frac{1}{A_+}(1 + \log{(2 A_+))}  -\frac{1}{2}\left (\frac{1}{A_+}-\frac{1}{A_-} \right )\log{(-A_+ A_-)}   +   a^{(1)}(|{\bf q}|)\right )\bigg ]
\end{align}

In regions 2, both $y_{\pm}$ are positive and
from Eq.~\eqref{eq:B} we have
\begin{align}
B_{\text{out}}^{(2)} \ = \ \log{\left (\frac{y_+}{y_-} \frac{\xi_1 + y_-}{\xi_1 + y_+}\right )}  \ .
\end{align}
This does not cross $\xi_1=0$ and does not generate a $\log{\Delta}$ piece. Therefore we can take $E_1\to\xi_1$ in the integrand giving,
\begin{align}
&\frac{1}{4\pi^2 |{\bf q}|}\int_{A_+}^{\infty} d\xi_1 \frac{\Delta^2}{2\xi_1^2}\log{\left (\frac{y_+}{y_-} \frac{\xi_1 + y_-}{\xi_1 + y_+}\right )} \nonumber \\
& \equiv \frac{\Delta^2}{4\pi^2 |{\bf q}|}a^{(2)}(|{\bf q}|)
\end{align}
where
\begin{align}
a^{(2)}(|{\bf q}|) \ &= \ \frac{1}{8}\left (\frac{1}{A_+}-\frac{1}{A_-}\right )\bigg (4 \sqrt{4-|{\bf q}|^2} \tanh ^{-1}\left(\frac{2 \sqrt{2-|{\bf q}|} (|{\bf q}|+1)}{(|{\bf q}|+2)^{3/2}}\right)+8 \log (|{\bf q}|) \nonumber \\
&\ -(|{\bf q}|+2) \log (|{\bf q}|+2) +  (|{\bf q}|-2) \log (5 |{\bf q}|+6)\bigg )
\end{align}

In regions 3, both $y_{\pm}$ have the same sign, positive for $|{\bf q}|>1$ and negative for $|{\bf q}|<1$.
For $|{\bf q}|>1$ the integral is,
\begin{align}
&\frac{1}{4\pi^2 |{\bf q}|}\int_{-1/2}^{A_-} d\xi_1 \frac{\Delta^2}{2\xi_1^2}\log{\left (\frac{y_+}{y_-} \frac{y_- -\xi_1}{y_+ -\xi_1}\right )} \nonumber \\
& \equiv \frac{\Delta^2}{4\pi^2 |{\bf q}|}a^{(3,+,+)}(|{\bf q}|)
\end{align}
where
\begin{align}
a^{(3,+,+)}(|{\bf q}|) \ = \ \frac{1}{8}\left (\frac{1}{A_+}-\frac{1}{A_-}\right )\left ( 3 (2-|{\bf q}|) \log (2-|{\bf q}|)+(3 |{\bf q}|-2) \log (3 |{\bf q}|-2) \right )
\end{align}
For $|{\bf q}|<1$ we have,
\begin{align}
&\frac{1}{4\pi^2 |{\bf q}|}\int_{-1/2}^{A_-} d\xi_1 \frac{\Delta^2}{2\xi_1^2}\log{\left (\frac{y_-}{y_+} \frac{y_+ + \xi_1 }{y_- +\xi_1}\right )} \nonumber \\
& \equiv \frac{\Delta^2}{4\pi^2 |{\bf q}|}a^{(3,-,-)}(|{\bf q}|)
\end{align}
where
\begin{align}
a^{(3,-,-)}(|{\bf q}|) \ &= \ \frac{1}{8}\left (\frac{1}{A_+}-\frac{1}{A_-}\right )\bigg (4 \sqrt{4-|{\bf q}|^2} \left(\tanh ^{-1}\left(\frac{2-3 |{\bf q}|}{\sqrt{4-|{\bf q}|^2}}\right)+\tanh ^{-1}\left(\sqrt{\frac{4}{|{\bf q}|+2}-1}\right)\right)\nonumber \\
& + \ (|{\bf q}|-2) \log (2-|{\bf q}|)+8 \log (|{\bf q}|)-(|{\bf q}|+2) \log (6-5 |{\bf q}|)\bigg )
\end{align}
Adding the contributions from all three regions we get,
\begin{align}
&I_{FF}[B_{\text{out}}] \ = \ \frac{\Delta}{4\pi^2 |{\bf q}|}\bigg  [\pi \log{2} + \Delta\left (\frac{1}{A_+} - \frac{1}{A_-}\right )\left ( 2\log{(\Delta)}-\log{\left (|{\bf q}|^2(2-\frac{|{\bf q}|^2}{2}) \right )} -1 +\frac{|{\bf q}|}{2} \tanh^{-1}(|{\bf q}|/2)\right )   \nonumber \\
&+ \ \Delta \left ( a^{(1)}(|{\bf q}|) +a^{(2)}(|{\bf q}|) + \theta(|{\bf q}|-1)a^{(3,+,+)}(|{\bf q}|)+ \theta(1-|{\bf q}|)a^{(3,-,-)}(|{\bf q}|)\right )\bigg ] \nonumber \\
&=\frac{\Delta}{4\pi^2 |{\bf q}|}\Bigg  [\pi \log{2} + \frac{8\Delta}{|{\bf q}|(4-|{\bf q}|^2)}\left ( \log{(\Delta^2)}  + 2|{\bf q}|\tanh^{-1}\left (\frac{|{\bf q}|}{2} \right )+\sqrt{4-|{\bf q}|^2}\tanh^{-1}\left (\frac{\sqrt{4-|{\bf q}|^2}}{2}\right ) -2-\log{8} \right )\Bigg ] \ . 
\label{eq:IFF_Bout}
\end{align}
There is a tremendous amount of cancellation that occurs when combining the $O(\Delta^2)$ terms.

\paragraph{\texorpdfstring{$I_{FF}[(B-B_{\text{out}})]$}{}}
\noindent
Next we compute all terms up to $O(\Delta^2)$ in $I_{FF}[(B-B_{\text{out}})]$, see Eq.~\eqref{eq:B} and Eq.~\eqref{eq:Bout}.
First, consider 
\[
\sinh^{-1}(y_\pm/\Delta) - \sgn(y_{\pm})\log{(2|y_\pm|/\Delta)}
\]
In the region of the $\xi_1$ integral where $y_{\pm} > \Delta$ we can expand $\sinh^{-1}(x) = \sgn(x)(\log{(2|x|)} + 1/(4x^2))$.
The correction to $B_{\text{out}}$ in this region is $O(\Delta^2)$ and beyond working order.
In the region of the integral where $y_{\pm} \sim \Delta$ we can linearize the integral as,
\begin{align}
&\int_{A_{\pm} - \delta}^{A_{\pm} + \delta} d\xi_1 \frac{\Delta^2}{2(\xi_1^2 + \Delta^2)}\left \{\sinh^{-1}(y_\mp/\Delta) - \sgn(y_{\mp})\log{(2|y_\mp|/\Delta)}\right \} \nonumber \\
&\int_{A_{\pm} - \delta}^{A_{\pm} + \delta} d\xi_1 \frac{\Delta^2}{A_{\pm}^2 }\left \{- \sgn(y_{\mp})\log{(2|y_\mp|/\Delta)} + O\left (\frac{y_{\mp}}{\Delta} \right )\right \} = \Delta^3 f(|{\bf q}|) 
\end{align}
where we used $y_{+}(A_{-})=0$ and $y_-(A_{+}) = 0$ for $|{\bf q}|<1$ and $y_-(A_-)=0$ for $|{\bf q}|>1$, and that the perturbation $\delta \propto\Delta$. This difference is beyond working order.
Therefore, up to $O(\Delta^2)$, there is no error by taking $\sinh^{-1}(y_\pm/\Delta) \to\sgn(y_{\pm})\log{(2|y_\pm|/\Delta)}$.

Next, consider the remaining term
\begin{align}
\int_{y_-}^{y_+}d\xi_2 \left ( \frac{1}{|\xi_1| + |\xi_2|} -\frac{1}{E_2+E_1}\right) \ = \ \int_{y_-}^{y_+}d\xi_2 \frac{(E_1-|\xi_1| ) + (E_2-|\xi_2|)}{(E_1+E_2)(|\xi_1| + |\xi_2|)}
\end{align}
In regions 2 and 3 we have $\xi_1 \sim O(1)$ and it is straightforward to show that the integral over $\xi_2$ is $\leq|\Delta/\xi_1^2|$ which is beyond working order.
There is a contribution from region 1,
\begin{align}
&\int_{y_-}^{y_+}d\xi_2 \left ( \frac{1}{|\xi_1| + |\xi_2|} -\frac{1}{E_2+E_1}\right) = \left (\int_{-\infty}^{\infty}du - \int_{-\infty}^{y_-/\Delta}du- \int_{y_+/\Delta}^{\infty}du\right ) \left (\frac{1}{|u| + |t|} - \frac{1}{\sqrt{1+u^2}+\sqrt{1+t^2}}\right) \nonumber \\
&= \ \left [\frac{2\sqrt{1+t^2}}{|t|}\sinh^{-1}(|t|) -\log(4t^2)\right ] +\Delta\left (\frac{1}{A_+}-\frac{1}{A_-} \right )(|t|-\sqrt{t^2+1})+ O(\Delta^2)
\end{align}
where in the first line we have changed variables to $u = \xi_2/\Delta$ and $t = \xi_1/\Delta$.
In the second line, the integrand of the terms $ (\int_{-\infty}^{y_-/\Delta}du- \int_{y_+/\Delta}^{\infty}du)$ has been expanded for large $u$ prior to integration. 
We have also used $y_{\pm}(\Delta t) = A_{\pm}+O(\Delta)$.
This gives,
\begin{align}
&I_{FF}[(B-B_{\text{out}})] =\nonumber \\
&\ \frac{\Delta}{4 \pi^2 |{\bf q}|}\int_{A_-/\Delta}^{A_+/\Delta}dt \frac{1}{2(1+t^2)}\left ( \left [\frac{2\sqrt{1+t^2}}{|t|}\sinh^{-1}(|t|)- \log(4t^2)  \right ] + \Delta\left (\frac{1}{A_+} - \frac{1}{A_-} \right )(|t|-\sqrt{t^2 +1})\right ) 
\end{align}
The term in $[.]$ is evaluated by $\int_{-\infty}^{\infty}dt[.] = \frac{\pi^2}{2} -\pi\log{2}$ and noting that residual term is beyond working order
\begin{align}
&\mp\int_{\pm\infty}^{A_\pm/\Delta}dt \frac{1}{2(1+t^2)}\left [\frac{2\sqrt{1+t^2}}{|t|}\sinh^{-1}(|t|) - \log(4t^2) \right ]\nonumber \\
& = \mp\int_{\pm\infty}^{A_\pm/\Delta}dt \frac{1}{2t^2}\left (\frac{1+\log{4t^2}}{2t^2}\right ) \sim O(\Delta^3\log{\Delta}) \ .
\end{align}
The other term is an elementary integral, and the result is
\begin{align}
I_{FF}[(B-B_{\text{out}})] =\frac{\Delta}{4 \pi^2 |{\bf q}|}\left ( \left [\frac{\pi^2}{2} -\pi\log{2}\right ]  -\Delta\left (\frac{1}{A_+} - \frac{1}{A_-} \right )\log{2} + O(\Delta^2)\right ) \ .
\end{align}
Combining with Eq.~\eqref{eq:IFF_Bout} gives
\begin{align}
\boxed{I_{FF} = \frac{1}{4\pi^2 |{\bf q}|}\left  [\frac{\pi^2}{2}\Delta + \frac{16\Delta^2}{|{\bf q}|(4-|{\bf q}|^2)}\left ( \log{(\Delta)}  + |{\bf q}|\tanh^{-1}\left (\frac{|{\bf q}|}{2} \right )+\frac{\sqrt{4-|{\bf q}|^2}}{2}\tanh^{-1}\left (\frac{\sqrt{4-|{\bf q}|^2}}{2}\right ) -1-\log{4} \right )\right ]}   
\end{align}
%

\subsubsection{\texorpdfstring{$I_{GG}$}{}}
\label{app:phGG}
\noindent
After making the change of variables, the normal part of the p-h bubble is,
\begin{align}
I_{GG}(\Delta,|{\bf q}|) \ = \ \frac{1}{4\pi^2 |{\bf q}|}\int_{{\cal D}}d\xi_1 d\xi_2 \frac{\xi_1 \xi_2 - E_1 E_2}{2(E_1 +E_2)E_1 E_2} \ . 
\end{align}
after partial fractions we get,
\begin{align}
I_{GG} \ &= \ \frac{1}{4\pi^2 |{\bf q}|}\int_{{\cal D}}  d\xi_1 d\xi_2 \left [ \frac{\xi_1\xi_2}{2 E_1^2}\left ( \frac{1}{E_2} - \frac{1}{E_1 +E_2}\right ) -\frac{1}{2(E_1 +E_2)}\right ] \nonumber \\
&= \ \frac{1}{4\pi^2 |{\bf q}|} \int_{-1/2}^{\infty}d\xi_1 \left ( \frac{\xi_1}{2E_1}\log{\left (\frac{E_1 + E_2(y_+)}{E_1+E_2(y_-)} \right )} - \int_{y_-}^{y_+}d\xi_2\frac{1}{2(E_1 + E_2)}\right )
\end{align}
For our purposes, we can subtract the integral with $\Delta=0$ (the standard Lindhard function),
\begin{align}
\delta I_{GG} \ = \ &\frac{1}{4\pi^2 |{\bf q}|}\int_{-1/2}^{\infty}d\xi_1 \Bigg [ \frac{\xi_1}{2E_1}\log{\left (\frac{E_1 + E_2(y_+)}{E_1+E_2(y_-)} \right )}-\frac{\sgn{(\xi_1)}}{2}\log{\frac{|\xi_1|+|y_+|}{|\xi_1|+|y_-|}}  \nonumber \\
&- \int_{y_-}^{y_+}\left (\frac{1}{2(E_1+E_2)}-\frac{1}{2(|\xi_1| + |\xi_2|)}\right ) \Bigg ]
\label{eq:deltaIGG}
\end{align}
%

\paragraph{Region  1}
\noindent
The goal is to compute,
\begin{align}
\delta I_{GG}^{(1)} \ = \ &\frac{1}{4\pi^2 |{\bf q}|}\int_{A_-}^{A_+}d\xi_1 \Bigg [ \frac{\xi_1}{2E_1}\log{\left (\frac{E_1 + E_2(y_+)}{E_1+E_2(y_-)} \right )}-\frac{\sgn{(\xi_1)}}{2}\log{\frac{|\xi_1|+|y_+|}{|\xi_1|+|y_-|}} \nonumber \\
&- \int_{y_-}^{y_+}\left (\frac{1}{2(E_1+E_2)}-\frac{1}{2(|\xi_1| + |\xi_2|)}\right ) \Bigg ]
\label{eq:deltaIGG1}
\end{align}
First, consider the term in the second line of Eq.~\eqref{eq:deltaIGG1},
\begin{align}
    &\frac{1}{4\pi^2 |{\bf q}|}\int_{A_-}^{A_+}d\xi_1\frac{1}{2}\int_{y_-}^{y_+}\left (\frac{1}{|\xi_1| + |\xi_2|} - \frac{1}{E_1 + E_2}\right )\nonumber \\
    &= \frac{1}{4\pi^2 |{\bf q}|}\int_{A_-}^{A_+}d\xi_1\frac{1}{2}\left (\int_{-\infty}^{\infty}d\xi_2-\int_{y_+}^{\infty}d\xi_2-\int_{-\infty}^{y_-}d\xi_2\right )\left (\frac{1}{|\xi_1| + |\xi_2|} - \frac{1}{E_1 + E_2}\right ) \nonumber \\
    &= \frac{1}{4\pi^2 |{\bf q}|}\int_{A_-}^{A_+}d\xi_1\left [\frac{\sqrt{\Delta^2+\xi_1^2}}{|\xi_1|}\sinh^{-1}(|\xi_1|/\Delta) -\log(2|\xi_1|/\Delta)\right ]\nonumber \\
    &-\frac{1}{2}\left (\int_{y_+}^{\infty}d\xi_2+\int_{-\infty}^{y_-}d\xi_2\right )\left (\frac{1}{|\xi_1| + |\xi_2|} - \frac{1}{E_1 + E_2}\right )
\end{align}
The term in $[.]$ can be directly evaluated
giving,
\begin{align}
&\frac{1}{4\pi^2 |{\bf q}|}\int_{A_-}^{A_+}d\xi_1\left [\frac{\sqrt{\Delta^2+\xi_1^2}}{|\xi_1|}\sinh^{-1}(|\xi_1|/\Delta) -\log(2|\xi_1|/\Delta)\right ] \nonumber \\
&= \frac{\Delta}{4\pi^2 |{\bf q}|}\left [\frac{\pi^2}{2} +\frac{\Delta}{2}\log{(\Delta)}\left (\frac{1}{A_+}-\frac{1}{A_-} \right ) + \Delta\left ( \frac{3+\log{(4A_-^2)}}{4A_-} - \frac{3+\log{(4A_+^2)}}{4A_+}\right )\right]
\label{eq:Igg11}
\end{align}

For the $\left (\int_{y_+}^{\infty}d\xi_2+\int_{-\infty}^{y_-}d\xi_2\right )$ terms, it is straightforward to show that $|\xi_2|>\Delta$ up to corrections that are $O(\Delta^3)$.
By performing partial fractions, and expanding to $O(\Delta^2)$ using $|\xi_2|>\Delta$ the integrand becomes,
\begin{align}
\left (\frac{1}{|\xi_1| + |\xi_2|} - \frac{1}{E_1 + E_2}\right ) \ = \ \frac{E_1 - |\xi_1|}{(|\xi_1| + |\xi_2|)(E_1+|\xi_2)} + \frac{\Delta^2}{2 |\xi_2|(E_1 + |\xi_2|)^2}
\end{align}
which integrates to
\begin{align}
\frac{1}{2}\left (\int_{y_+}^{\infty}d\xi_2+\int_{-\infty}^{y_-}d\xi_2\right )\frac{E_1 - |\xi_1|}{(|\xi_1| + |\xi_2|)(E_1+|\xi_2)} + \frac{\Delta^2}{2 |\xi_2|(E_1 + |\xi_2|)^2}\  = \ T(\xi_1,y_+)+T(\xi_1,-y_-)
\end{align}
where
\begin{align}
T(\xi_1,y) \ &= \ \frac{1}{2}\log \left (\frac{E_1 + y}{|\xi_1| + y}\right ) + \frac{\Delta^2}{4}\left [\frac{\log{\left (1+E_1/y \right )}}{E_1^2} -\frac{1}{E_1(E_1+y)}\right ] \nonumber \\
&= \ \frac{1}{2}\log \left (\frac{E_1 + y}{|\xi_1| + y}\right ) + \frac{\Delta^2}{4}\left [\frac{\log{\left (1+|\xi_1|/y \right )}}{\xi_1^2} -\frac{1}{|\xi_1|(|\xi_1|+y)}\right ] + O(\Delta^3)
\end{align}
The second line has taken $E_1\to |\xi_1|$ in the second term which is accurate up to $O(\Delta^3)$ even in the integration region $|\xi_1| \lesssim \Delta$.
The remaining integral over $\xi_1$,adding the term in the first line of Eq.~\eqref{eq:deltaIGG}, is
\begin{align}
\frac{1}{4\pi^2 |{\bf q}|}\int_{A_-}^{A_+}d\xi_1  \left (\frac{\xi_1}{2E_1}\log{\left (\frac{E_1 + E_2(y_+)}{E_1+E_2(y_-)} \right )}-\frac{\sgn{(\xi_1)}}{2}\log{\frac{|\xi_1|+|y_+|}{|\xi_1|+|y_-|}} -T(\xi_1,y_+) - T(\xi_1,-y_-)\right )
\end{align}
The strategy for evaluating this integral will be to first manipulate it into terms with and without an explicit $\Delta^2$ prefactor.

Putting everything together, 
\begin{align}
&\frac{1}{4\pi^2 |{\bf q}|} \int_0^{\pm A_\pm}d\xi_1 \bigg\{ \left [\pm\frac{\xi_1}{2E_1}\log{\left (\frac{E_1 + y_+}{E_1-y_-} \right )}\mp\frac{1}{2}\log{\frac{\xi_1+y_+}{\xi_1-y_-}} -\frac{1}{2}\log \left (\frac{E_1 + y_+}{\xi_1 + y_+}\right ) -\frac{1}{2}\log \left (\frac{E_1 - y_-}{\xi_1 - y_-}\right )\right ] \nonumber \\
&\pm\frac{\xi_1}{2E_1}\log{\left (\frac{E_1 + E_2(y_+)}{E_1+E_2(y_-)}\frac{E_1 - y_-}{E_1+y_+}\right )}-\frac{\Delta^2}{4}\left ( \frac{\log{\left (1+\xi_1/y_+ \right )}}{\xi_1^2} -\frac{1}{\xi_1(\xi_1+y_+)} + \frac{\log{\left (1-\xi_1/y_- \right )}}{\xi_1^2} -\frac{1}{\xi_1(\xi_1-y_-)}\right ) \bigg \}
\end{align}
where the integral has been split into two pieces $(\int_0^{A_+} + \int_{A_-}^0)$ and in the second integral $\xi_1\to-\xi_1$ is used.
In the second integral, all the $y_{\pm}$ need to have $\xi_1\to-\xi_1$ inside them.
The term in $[.]$ becomes
\begin{align}
& \left [\pm\frac{\xi_1}{2E_1}\log{\left (\frac{E_1 + y_+}{E_1-y_-} \right )}+\log{(\xi_1\mp y_\mp)}-\frac{1}{2}\log{((E_1 + y_+)(E_1 - y_-))} \right ]  \nonumber \\
& = \left [\left ( \frac{\pm\xi_1-E_1}{2E_1} \right )\log{\left (E_1 + y_+ \right )}-\left ( \frac{\pm\xi_1 + E_1}{2E_1} \right )\log{\left (E_1 - y_- \right )}+\log{(\xi_1\mp y_\mp)} \right ] \nonumber \\
& = \left [\frac{-\Delta^2}{2E_1(\xi_1+E_1)}\log{\left (E_1 \pm y_{\pm} \right )}+\left ( \frac{\Delta^2}{2E_1(\xi_1+E_1)} - 1 \right )\log{\left (E_1 \mp y_{\mp} \right )}+\log{(\xi_1\mp y_\mp)} \right ]
\end{align}
In the third line we have used the identity $E_1-\xi_1 = \Delta^2/(E_1 + \xi_1)$.
There are only two terms without explicit $\Delta^2$ dependence. 
We will integrate these terms first, and then deal with the $\Delta^2$ pieces.
We have,
\begin{align}
&\frac{1}{4\pi^2 |{\bf q}|}\int_0^{\pm A_{\pm}}d\xi_1\log{\left (\frac{\xi_1\mp y_\mp}{E_1 \mp y_{\mp}} \right )} = \frac{1}{4\pi^2 |{\bf q}|}\int_0^{\pm A_{\pm}}d\xi_1\left \{ \left [\log{\left (\frac{\xi_1\mp y_\mp}{E_1 \mp y_{\mp}}  \right )} - \log{\left (\frac{\xi_1\mp A_\mp}{E_1 \mp A_\mp}  \right )}\right ] + \log{\left (\frac{\xi_1\mp A_\mp}{E_1 \mp A_\mp} \right )}\right \} \nonumber \\
&= \frac{1}{4\pi^2 |{\bf q}|}\int_0^{\pm A_{\pm}}d\xi_1\left \{ \left [\frac{\Delta^2}{2}\frac{\mp(y_{\mp}-A_{\mp})}{\xi_1(\xi_1\mp A_{\mp})(\xi_1\mp y_{\mp})} + O(\Delta^3)\right ] + \log{\left (\frac{\xi_1\mp A_\mp}{E_1 \mp A_\mp} \right )} \right \} \nonumber \\
&= \frac{\Delta^2}{4\pi^2 |{\bf q}|}\Bigg \{ \left [\frac{4 \log (| 1-|{\bf q}|| +1)-2 |{\bf q}| \log (2 | 1-|{\bf q}|| +|{\bf q}|)+(2-|{\bf q}|) \log (2-|{\bf q}|)+3 (|{\bf q}|+2) \log (|{\bf q}|+2)-16 \log (2)}{|{\bf q}| \left(4-|{\bf q}|^2\right)}\right ] \nonumber \\
& + \frac{1}{2}\left (\frac{1}{A_+} - \frac{1}{A_-} \right )\left (\log{\left (\frac{\Delta(A_+ - A_-)}{-2A_+A_-} \right )} -\frac{1}{2} \right )\Bigg \}
\label{eq:Igg12}
\end{align}
In the second line we have Taylor expanded the quantity in $[.]$ to $O(\Delta^2)$. This Taylor expansion is valid even in the regime $\xi_1\sim\Delta$ because of the subtraction that was performed in the first line.
In the third line, the quantity in $[.]$ is a standard integral. The remaining term is evaluated using the identity,
\begin{align}
\int_0^A d\xi_1 \log{\left (\frac{\xi_1 + c}{\sqrt{\xi_1^2 + \Delta^2}+c} \right )} \ = \ \frac{\Delta^2}{2c}\left (\log{\left (\frac{\Delta(A+c)}{2A c} \right )}-\frac{1}{2} \right) + O(\Delta^4)
\label{eq:FeynamTrickIntegral}
\end{align}
which is derived by using Feynman's trick of first applying $\frac{d}{d\Delta^2}$ to undo the log, then integrating with respect to $\xi_1$, then dropping all terms that will be higher than $O(\Delta^2)$ and then integrating back $\Delta^2$.

Next are the remaining terms,
\begin{align}
&\frac{1}{4\pi^2 |{\bf q}|} \int_0^{\pm A_\pm}d\xi_1 
\Bigg \{ \pm\frac{\xi_1}{2E_1}\log{\left (\frac{E_1 + E_2(y_+)}{E_1+E_2(y_-)}\frac{E_1 - y_-}{E_1+y_+}\right )} \nonumber \\
&-\frac{\Delta^2}{4}\left ( \frac{\log{\left (1+\xi_1/y_+ \right )}}{\xi_1^2} -\frac{1}{\xi_1(\xi_1+y_+)} + \frac{\log{\left (1-\xi_1/y_- \right )}}{\xi_1^2} -\frac{1}{\xi_1(\xi_1-y_-)} -\frac{2}{E_1(\xi_1+E_1)}\log{\left (\frac{E_1 \mp y_{\mp}}{E_1\pm y_{\pm}} \right )}\right ) \Bigg \} \nonumber \\
&=\frac{1}{4\pi^2 |{\bf q}|} \int_0^{\pm A_\pm}d\xi_1 \Bigg \{ \pm\frac{\xi_1}{2E_1}\log{\left (\frac{E_1 + E_2(y_+)}{E_1+E_2(y_-)}\frac{E_1 - y_-}{E_1+y_+}\right )} \pm\frac{E_1-\xi_1}{2E_1}\log{\left ( \frac{E_1-y_-}{E_1+y_+}\right )}\nonumber \\
&-\frac{\Delta^2}{4}\left ( \frac{\log{\left (1+\xi_1/y_+ \right )}}{\xi_1^2} -\frac{1}{\xi_1(\xi_1+y_+)} + \frac{\log{\left (1-\xi_1/y_- \right )}}{\xi_1^2} -\frac{1}{\xi_1(\xi_1-y_-)} \right ) \Bigg \}
\end{align}
The second line has used $\Delta^2 = (E_1+\xi_1)(E_1-\xi_1)$.
First the terms with the $\Delta^2/4$ out front.
Integrating by parts gives the identity,
\begin{align}
\int_0^A d\xi_1\left (\frac{\log{\left (1+\xi_1/y \right )}}{\xi_1^2} -\frac{1}{\xi_1(\xi_1+y)} \right ) \ = \ -\frac{1}{\xi_1}\log{\left (1+\xi_1/y\right )}\bigg \vert_0^{A}-\int_0^{A}d\xi_1\frac{y'}{y(\xi_1+y)}
\label{eq:ibpidentity}
\end{align}
Defining
\[
c_{\pm \pm'}\  = \ \int_0^{\pm A_{\pm}}\left (\frac{\log{\left (1\pm'\xi_1/y_{\pm'} \right )}}{\xi_1^2} -\frac{1}{\xi_1(\xi_1\pm'y_{\pm'})} \right ) \ .
\]
There are two cases that have to be treated separately; $y(A)=0$ and $y(A)\neq 0$.
The former causes the explicit log in Eq.~\eqref{eq:ibpidentity} to diverge which is canceled by a log in the integral. This is made manifest by rewriting,
\[
\frac{y'}{y(\xi_1 + y)} \ = \ \frac{1}{A}\frac{d}{d\xi_1}\log{y} + \frac{A-\xi_1-y}{A(\xi_1+y)}\frac{y'}{y}
\]
It is straightforward to show that in this case,
\begin{align}
\int_0^A d\xi_1\left (\frac{\log{\left (1+\xi_1/y \right )}}{\xi_1^2} -\frac{1}{\xi_1(\xi_1+y)} \right ) \ = \ -\frac{1}{A}\log{(A/y(0))} + \frac{1}{y(0)} - \int_0^A d\xi_1 \frac{A-\xi_1-y}{A(\xi_1 + y)}\frac{y'}{y}
\label{eq:ibpidentity2}
\end{align}
Putting together, the four integrals are,
\begin{align}
c_{++} \ &= \ \frac{1}{A_+}\left (1-\log{\left (1+\frac{A_+}{y_+(A_+)} \right )} \right )-\int_0^{A_+}d\xi_1\frac{y_+'}{y_+(\xi_1+y_+)} \ , \nonumber \\
c_{+-}\ &= \ -\frac{1}{A_+}\log{(-A_+/A_-)} - \frac{1}{A_-} - \int_0^{A_+} d\xi_1\frac{A_+-\xi_1+y_-}{A_+(\xi_1 - y_-)}\frac{y_-'}{y_-} \nonumber \\
c_{-+}(|{\bf q}|<1) \ &= \ \frac{1}{A_-}\log{(-A_-/A_+)} + \frac{1}{A_+} + \int_0^{-A_-} d\xi_1 \frac{A_- +\xi_1+y_+[-\xi_1]}{A_-(\xi_1 + y_+[-\xi_1])}\frac{y_+'[-\xi_1]}{y_+[-\xi_1]} \nonumber \\
c_{--}(|{\bf q}|<1) \ &= \ \frac{1}{A_-}\left (\log{\left (1+\frac{A_-}{y_-(A_-)}\right )} -1\right ) + \int_0^{-A_-} d\xi_1\frac{y_-'[-\xi_1]}{y_-[-\xi_1](\xi_1 - y_-[-\xi_1])}  \nonumber \\
c_{-+}(|{\bf q}|>1) \ &= \ \frac{1}{A_-}\log{\left (1-\frac{A_-}{y_+(A_-)} \right )} + \frac{1}{A_+} + \int_0^{-A_-} d\xi_1 \frac{y'_+[-\xi_1]}{y_+[-\xi_1](\xi_1+y_+[-\xi_1])} \nonumber \\
c_{--}(|{\bf q}|>1) \ &= \ -\frac{1}{A_-} + \int_0^{-A_-} d\xi_1 \frac{A_- +\xi_1-y_-[-\xi_1]}{A_-(\xi_1 - y_-[-\xi_1])}\frac{y_-'[-\xi_1]}{y_-[-\xi_1]}
\end{align}
where $y' = dy/d\xi_1$. 
The expression splits at $|{\bf q}|=1$ because $y_-(A_+)=0$ for $0<|{\bf q}|<2$ but there is an additional zero, $y_-(A_-)=0$ for $|{\bf q}|>1$.
The remaining integrals can be solved in closed form giving a summed total of,
\begin{align}
    & \sum_{\pm \pm'}c_{\pm \pm'}  \nonumber \\ 
    & = \frac{1}{|{\bf q}| \left(|{\bf q}|^2-4\right)}\Bigg [2 \sqrt{4-|{\bf q}|^2} \left(\log \left(\frac{4 \sqrt{4-|{\bf q}|^2}-3 |{\bf q}|+10}{6-5 |{\bf q}|}\right)\nonumber -2 \tanh ^{-1}\left(\frac{3 |{\bf q}|+2}{\sqrt{4-|{\bf q}|^2}}\right)  +2 \coth ^{-1}\left(\sqrt{\frac{4}{|{\bf q}|+2}-1}\right)\right)\nonumber \\
    &+4 \left(\log \left(\frac{1}{256} \left(25 |{\bf q}|^4-136 |{\bf q}|^2+144\right)\right)+2 |{\bf q}| \tanh ^{-1}\left(\frac{|{\bf q}|}{2}\right)-4\right) \Bigg ] \quad , \quad |{\bf q}|<1 \nonumber \\
    &=-\frac{2}{|{\bf q}| \left(|{\bf q}|^2-4\right)}\Bigg [ -2 \sqrt{4-|{\bf q}|^2} \left(\tanh ^{-1}\left(\frac{-3 |{\bf q}|-2}{\sqrt{4-|{\bf q}|^2}}\right)+2 \tanh ^{-1}\left(\sqrt{\frac{4}{|{\bf q}|+2}-1}\right)+\coth ^{-1}\left(\sqrt{\frac{4}{|{\bf q}|+2}-1}\right)\right) \nonumber \\
    &+2 \log (|{\bf q}|+1)-2 (|{\bf q}|+2) \log (|{\bf q}|+2)+|{\bf q}| \log ((8-3 |{\bf q}|) |{\bf q}|-4)-2 \log ((5 |{\bf q}|+6) ((8-3 |{\bf q}|) |{\bf q}|-4))+4 \coth ^{-1}(2 |{\bf q}|+3) \nonumber \\
    &+8+\log (65536) \Bigg ] \quad , \quad |{\bf q}|>1 \ .
\end{align}

The last term for region 1 is
\begin{align}
&\frac{1}{4\pi^2 |{\bf q}|} \int_0^{\pm A_\pm}d\xi_1 \Bigg \{ \pm\frac{\xi_1}{2E_1}\log{\left (\frac{E_1 + E_2(y_+)}{E_1+E_2(y_-)}\frac{E_1 - y_-}{E_1+y_+}\right )} \pm\frac{E_1-\xi_1}{2E_1}\log{\left ( \frac{E_1-y_-}{E_1+y_+}\right )} \Bigg \} \nonumber \\
&=\frac{1}{4\pi^2 |{\bf q}|} \int_0^{\pm A_\pm}d\xi_1 \Bigg \{ \pm\frac{1}{2}\log{\left (\frac{\xi_1 + E_2(y_+)}{\xi_1+E_2(y_-)}\frac{\xi_1 - y_-}{\xi_1+y_+}\right )} \pm\frac{E_1-\xi_1}{2E_1}\log{\left ( \frac{\xi_1-y_-}{\xi_1+y_+}\right )} \Bigg \} \ .
\end{align}
In the second line we replaced $E_1\to\xi_1$ in the first term as the difference is $O(\Delta^3)$.
The same can be done in the second term because the error is canceled when summing both the $\pm$ contributions.
First, the second term. This we solve by subtracting and adding back expression as $\xi_1\to 0$,
\begin{align}
&\pm\Bigg \{ \frac{E_1-\xi_1}{2E_1}\left [\log{\left ( \frac{\xi_1-y_-}{\xi_1+y_+}\right )}-\log{\left (\frac{-A_-}{A_+} \right )} + \frac{1}{A_+}\left (E_1\pm\xi_1(1+|{\bf q}|) \right ) +\frac{1}{A_-}\left(E_1\mp\xi_1(1-|{\bf q}|) \right )  + \frac{E_1-\xi_1}{2E_1}\right ] \nonumber \\
&+\frac{E_1-\xi_1}{2E_1}\left ( \log{\left (\frac{-A_-}{A_+} \right )} - \frac{1}{A_+}\left (E_1\pm\xi_1(1+|{\bf q}|) \right ) -\frac{1}{A_-}\left(E_1\mp\xi_1(1-|{\bf q}|) \right )  - \frac{E_1-\xi_1}{2E_1}\right )\Bigg \} \nonumber \\
&=\pm\Bigg \{ \frac{\Delta^2}{4\xi_1^2}\left [\log{\left ( \frac{\xi_1-y_-}{\xi_1+y_+}\right )}-\log{\left (\frac{-A_-}{A_+} \right )}  +\xi_1\left (\frac{\pm2}{|{\bf q}|}-\frac{2}{2\mp |{\bf q}|} \right )\right ] \nonumber \\
&+\frac{E_1-\xi_1}{2E_1}\left ( \log{\left (\frac{-A_-}{A_+} \right )} - \frac{1}{A_+}\left (E_1\pm\xi_1(1+|{\bf q}|) \right ) -\frac{1}{A_-}\left(E_1\mp\xi_1(1-|{\bf q}|) \right )  - \frac{E_1-\xi_1}{2E_1}\right )\Bigg \}
\end{align}
The term in  $[.]$ has been Taylor expanded in the third line as the subtracted term makes the error $O(\Delta^3)$ for $\xi_1<\Delta$.
The integral over the expression in $[.]$ can be solved with the substitution, $\xi_1 = \pm \frac{s^2-1}{2}$.
The integral over the expression in the fourth line can be solved directly.
The result is
\begin{align}
S_2 \ = \ \frac{\pm1}{4\pi^2 |{\bf q}|} \int_0^{\pm A_\pm}d\xi_1  \frac{E_1-\xi_1}{2E_1}\log{\left ( \frac{\xi_1-y_-}{\xi_1+y_+}\right )} 
\end{align}
with
\begin{align}
&S_2(|{\bf q}|<1) \ = \ \frac{1}{4\pi^2 |{\bf q}|} \frac{\Delta^2}{|{\bf q}| \left(|{\bf q}|^2-4\right)}\Bigg[ (|{\bf q}|^2-2)\log \left(\Delta^2\right ) -|{\bf q}|^2\log{\left (\frac{16|{\bf q}|^2}{4-|{\bf q}|^2} \right )}+5|{\bf q}| \tanh ^{-1}\left(\frac{|{\bf q}|}{2}\right)+|{\bf q}|\tanh ^{-1}\left(\frac{5 |{\bf q}|}{6}\right) 
\nonumber \\
&+\log \left(\frac{256 |{\bf q}|^4}{25 |{\bf q}|^4-136 |{\bf q}|^2+144}\right)+2-|{\bf q}|^2 \nonumber \\
&+2 \sqrt{4-|{\bf q}|^2} \left(\tanh ^{-1}\left(\frac{3 |{\bf q}|+2}{\sqrt{4-|{\bf q}|^2}}\right)+\tanh ^{-1}\left(\frac{2-3 |{\bf q}|}{\sqrt{4-|{\bf q}|^2}}\right)-\tanh ^{-1}\left(\sqrt{\frac{4}{|{\bf q}|+2}-1}\right)-\coth ^{-1}\left(\sqrt{\frac{4}{|{\bf q}|+2}-1}\right)\right)\Bigg ]
\end{align}
and
\begin{align}
&S_2(|{\bf q}|>1) \ = \ \frac{1}{4\pi^2 |{\bf q}|}\frac{\Delta^2}{2 |{\bf q}| \left(|{\bf q}|^2-4\right)} \Bigg[-2 \Bigg( (2-|{\bf q}|^2 )\log (\Delta^2)+|{\bf q}|^2+2 \sqrt{4-|{\bf q}|^2} \bigg(\tanh ^{-1}\left(\frac{-3 |{\bf q}|-2}{\sqrt{4-|{\bf q}|^2}}\right)\nonumber \\
&+2 \tanh ^{-1}\left(\sqrt{\frac{4}{|{\bf q}|+2}-1}\right)+\coth ^{-1}\left(\sqrt{\frac{4}{|{\bf q}|+2}-1}\right)\bigg)-\log (-256 (|{\bf q}|-2))+\log (5 |{\bf q}|+6)-2\Bigg)  \nonumber \\
&+|{\bf q}| \bigg(4 |{\bf q}| \left(\log \left(\frac{1}{4}\right)+\tanh ^{-1}\left(\frac{|{\bf q}|}{2}\right)\right)-3 \log (2-|{\bf q}|)+2 \log (|{\bf q}|+2)+\log (5 |{\bf q}|+6)\bigg)+(6 |{\bf q}|-4) \coth ^{-1}\left(\frac{2 |{\bf q}|}{2-|{\bf q}|}\right)\Bigg]
\end{align}

The last piece to region 1 is,
\begin{align}
\frac{\pm1}{8\pi^2 |{\bf q}|} \int_0^{\pm A_\pm}d\xi_1 \left [ \log{\left (\frac{\xi_1 + E_2(y_+)}{\xi_1+y_+}\right )}-\log{\left (\frac{\xi_1+E_2(y_-)}{\xi_1 - y_-} \right ) }  \right ]\ .
\end{align}
In the $\int_0^{A_+}$, the first term can be expanded in $y_+>\Delta$, and the second term can be solved using Feynman trick.
The result is,
\begin{align}
&U_+ \ = \ \frac{1}{8\pi^2 |{\bf q}|} \int_0^{A_+}d\xi_1 \left [ \log{\left (\frac{\xi_1 + E_2(y_+)}{\xi_1+y_+}\right )}-\log{\left (\frac{\xi_1+E_2(y_-)}{\xi_1 - y_-} \right ) }  \right ] \nonumber \\
& =\frac{1}{4\pi^2 |{\bf q}|}\frac{\Delta^2}{4|{\bf q}|(|{\bf q}|^2-4)}\Bigg [2 |{\bf q}| \left(\log \left(\frac{|{\bf q}| (|{\bf q}|+2)^2 (5 |{\bf q}|+6)}{4 \Delta (|{\bf q}|+1)^2}\right)+|{\bf q}| \log \left(\frac{\Delta (2-|{\bf q}|) (|{\bf q}|+1)}{4 |{\bf q}| (|{\bf q}|+2)}\right)\right)-4 \log \left(\Delta \left(\frac{1}{|{\bf q}|}+1\right)\right)-|{\bf q}|^2\nonumber \\
&+4 \sqrt{4-|{\bf q}|^2} \left(\tanh ^{-1}\left(\frac{3 |{\bf q}|+2}{\sqrt{4-|{\bf q}|^2}}\right)-\coth ^{-1}\left(\sqrt{\frac{4}{|{\bf q}|+2}-1}\right)\right)+|{\bf q}|-4 (|{\bf q}|+2) \tanh ^{-1}\left(\frac{1}{3} (1-2 |{\bf q}|)\right)\nonumber \\
&+8 \coth ^{-1}(2 |{\bf q}|+3)+2+\log (256)\Bigg ]
\end{align}
The $\int_0^{-A_-}$ splits into branches because the zeros split for at $|{\bf q}|=1$ as shown in the right table of Table~\ref{tab:signs}.
For $|{\bf q}|<1$, the second term can be expanded in $y_->\Delta$ and the first term can be solved using Feynman's trick.
We have,
\begin{align}
&U_-\ = \ \frac{-1}{8\pi^2 |{\bf q}|} \int_0^{-A_-}d\xi_1 \left [ \log{\left (\frac{\xi_1 + E_2(y_+[-\xi_1])}{\xi_1+y_+[-\xi_1]}\right )}-\log{\left (\frac{\xi_1+E_2(y_-[-\xi_1])}{\xi_1 - y_-[-\xi_1]} \right ) }  \right ]
\end{align}
and with $|{\bf q}|<1$,
\begin{align}
&U_-(|{\bf q}|<1) \ = \ \frac{1}{4\pi^2 |{\bf q}|}\frac{\Delta^2}{4 |{\bf q}| \left(4-|{\bf q}|^2\right)} \Bigg [ 2\left(|{\bf q}|^2+|{\bf q}|-2\right) \log \left(\frac{(2-|{\bf q}|) |{\bf q}|}{\Delta (1-|{\bf q}|)}\right)+|{\bf q}|^2-4 \log \left(\frac{4 (|{\bf q}|-1)}{|{\bf q}|^2-4}\right)\nonumber \\
&-2 \sqrt{4-|{\bf q}|^2} \log \left(\frac{6-5 |{\bf q}|}{4 \sqrt{4-|{\bf q}|^2}-3 |{\bf q}|+10}\right)+|{\bf q}|-2 |{\bf q}| \left(|{\bf q}| \log \left(\frac{|{\bf q}|+2}{4}\right)+\log \left(\frac{4 (|{\bf q}|+2)}{(|{\bf q}|-2) (5 |{\bf q}|-6)}\right)\right)\nonumber \\
&-8 \coth ^{-1}(3-2 |{\bf q}|)-2\Bigg ]
\end{align}
and for $|{\bf q}|>1$ we have
\begin{align}
&U_-(|{\bf q}|>1) \ = \ \frac{1}{4\pi^2 |{\bf q}|}\frac{\Delta^2}{4 |{\bf q}| \left(4-|{\bf q}|^2\right)} \Bigg[2\left(|{\bf q}|^2+|{\bf q}|-2\right) \log \left(\frac{(2-|{\bf q}|) |{\bf q}|}{\Delta (|{\bf q}|-1)}\right)+|{\bf q}|^2-2 |{\bf q}|^2 \log \left(\frac{|{\bf q}|+2}{4}\right)\nonumber \\
&+8 \sqrt{4-|{\bf q}|^2} \tanh ^{-1}\left(\sqrt{\frac{4}{|{\bf q}|+2}-1}\right)
+|{\bf q}|-2 |{\bf q}| \log (4 (|{\bf q}|+2))+4 |{\bf q}| \log (3 |{\bf q}|-2)+4 \log (|{\bf q}|+2)-2-8 \log (2)\Bigg]
\end{align}

Combining all the terms in region 1 gives,
\begin{align}
\delta I_{GG}^{(1)} \ = \ \frac{\Delta}{8|{\bf q}|}
+\frac{14-3|{\bf q}|^2}{4\pi^2 |{\bf q}|^2(4-|{\bf q}|^2)}\,\Delta^2\log\!\left(\frac{\Delta}{4}\right)
+\frac{\Delta^2}{8\pi^2 |{\bf q}|^2(|{\bf q}|^2-4)}\,\Big[R_c(|{\bf q}|)+R_b(|{\bf q}|)\Big]
\end{align}
with
\begin{align}
R_c(|{\bf q}|) \ = \ &30-3|{\bf q}|^2+(12-2|{\bf q}|^2)\log |{\bf q}|+(|{\bf q}|+2)(|{\bf q}|-1)\log|1-|{\bf q}||+(|{\bf q}|-2)(|{\bf q}|+1)\log\!\big[(|{\bf q}|+1)(|{\bf q}|+2)^2\big]\nonumber\\
&+2(|{\bf q}|-2)\log(5|{\bf q}|+6)
+4\sqrt{4-|{\bf q}|^2}\left(2\tanh^{-1}\!\frac{3|{\bf q}|+2}{\sqrt{4-|{\bf q}|^2}}-3\tanh^{-1}\!\sqrt{\frac{2-|{\bf q}|}{|{\bf q}|+2}}\right)
\end{align}
and
\begin{align}
R_b(|{\bf q}|<1) \ = \ &4(4-|{\bf q}|^2)\log |{\bf q}|+2(|{\bf q}|+2)(|{\bf q}|-1)\log(2-|{\bf q}|)-2(|{\bf q}|+2)\log(6-5|{\bf q}|)\nonumber\\
&+2\sqrt{4-|{\bf q}|^2}\left(\log\!\frac{6-5|{\bf q}|}{4\sqrt{4-|{\bf q}|^2}-3|{\bf q}|+10}+2\tanh^{-1}\!\frac{2-3|{\bf q}|}{\sqrt{4-|{\bf q}|^2}}\right)
\end{align}
and
\begin{align}
R_b(|{\bf q}|>1) \ = \ &-2(|{\bf q}|^2-4)\log(2-|{\bf q}|)-12\sqrt{4-|{\bf q}|^2}\,\tanh^{-1}\!\sqrt{\frac{2-|{\bf q}|}{|{\bf q}|+2}}
\end{align}
%

\paragraph{Regions 2 and 3}
\noindent
Next we compute Eq.~\eqref{eq:deltaIGG} in region 2 where Table~\ref{tab:signs} give $(y_+,y_-)=(+,+)$ and $\xi_1\sim O(1)$.
The integral is,
\begin{align}
\delta I_{GG}^{(2)} \ = \ &\frac{1}{4\pi^2 |{\bf q}|}\int_{A_+}^{\infty}d\xi_1 \Bigg [ \frac{\xi_1}{2E_1}\log{\left (\frac{E_1 + E_2(y_+)}{E_1+E_2(y_-)} \right )}-\frac{1}{2}\log{\frac{\xi_1+y_+}{\xi_1+y_-}} - \int_{y_-}^{y_+}d\xi_2\left (\frac{1}{2(E_1+E_2)}-\frac{1}{2(\xi_1 + \xi_2)}\right ) \Bigg ]\ .
\label{eq:deltaIGG2}
\end{align}
Expanding in $\xi_1>\Delta$ and dropping all terms that are beyond $O(\Delta^2)$ gives an integrand,
\begin{align}
&\Bigg [ \frac{1}{2}\log{\left (\frac{\xi_1+y_-}{\xi_1+E_2(y_-)} \right )}-\frac{\Delta^2}{4\xi_1^2}\log{\left (\frac{\xi_1+y_+}{\xi_1+y_-}\right )} +\frac{\Delta^2}{4}\left (\frac{1}{\xi_1y_+}-\frac{1}{\xi_1(\xi_1+y_-)} \right )  +\frac{\Delta^2}{4}\int_{y_-}^{y_+}d\xi_2\frac{1}{\xi_1 \xi_2(\xi_1 + \xi_2)}\Bigg ]\ .
\end{align}
This expansion replaced $E_2(x)\to x$ if the integrated difference is $O(\Delta^3)$.
Every term is an elementary integral except for the first term.
This term is similar to terms we saw in region 1, and can be evaluated by using Feynman's trick.
The result is 
\begin{align}
&\delta I_{GG}^{(2)} \ = \ \frac{1}{4\pi^2 |{\bf q}|}\,\frac{\Delta^2}{4|{\bf q}|(|{\bf q}|^2-4)}
\Bigg[\,2(|{\bf q}|-2)(|{\bf q}|+1)\log\!\left(\frac{\Delta}{4(|{\bf q}|+2)^2}\right)
+2(|{\bf q}|-2)(3|{\bf q}|+7)\log |{\bf q}|-(|{\bf q}|-2)(|{\bf q}|+1)\nonumber \\
& -2(|{\bf q}|-2)(|{\bf q}|+1)\log(|{\bf q}|+1)-4(|{\bf q}|-2)\log(5|{\bf q}|+6)
 -16\sqrt{4-|{\bf q}|^2}\,\tanh^{-1}\!\left(\frac{2(|{\bf q}|+1)\sqrt{2-|{\bf q}|}}{(|{\bf q}|+2)^{3/2}}\right)\Bigg]
\end{align}

In region 3, the sign of $(y_+,y_-)$ depends on $|{\bf q}|$. 
With $|{\bf q}|<1$ we have
\begin{align}
\delta I_{GG}^{(3)}(|{\bf q}|<1) \ = \ &\frac{1}{4\pi^2 |{\bf q}|}\int_{-1/2}^{A_-}d\xi_1 \Bigg [ \frac{\xi_1}{2E_1}\log{\left (\frac{E_1 + E_2(y_+)}{E_1+E_2(y_-)} \right )}+\frac{1}{2}\log{\frac{-\xi_1-y_+}{-\xi_1-y_-}}  \nonumber \\
&- \int_{y_-}^{y_+}\left (\frac{1}{2(E_1+E_2)}-\frac{1}{2(-\xi_1 -\xi_2)}\right ) \Bigg ]
\end{align}
Expanding in $\xi_1>\Delta$ and dropping all terms that are beyond $O(\Delta^2)$ gives an integrand,
\begin{align}
&\Bigg [ \frac{1}{2}\log{\left (\frac{\xi_1+y_+}{\xi_1-E_2(y_+)} \right )}+\frac{\Delta^2}{4\xi_1^2}\log{\left (\frac{\xi_1+y_+}{\xi_1+y_-}\right )} +\frac{\Delta^2}{4}\left (\frac{1}{\xi_1 y_-} -\frac{1}{\xi_1(\xi_1 + y_+)} \right ) -\frac{\Delta^2}{4}\int_{y_-}^{y_+}d\xi_2\frac{1}{\xi_1 \xi_2(\xi_1 + \xi_2)} \Bigg ]
\end{align}
This is evaluated the same way as the region 2 integral.
The result is,
\begin{align}
&\delta I_{GG}^{(3)}(|{\bf q}|<1) \ = \ \frac{1}{4\pi^2 |{\bf q}|}\frac{\Delta^2}{4|{\bf q}|(|{\bf q}|^2-4)}
\Bigg[
2(|{\bf q}|+2)(|{\bf q}|-1)
\log\!\left(
\frac{\Delta}{4(1-|{\bf q}|)(2-|{\bf q}|)^2}
\right) +2(|{\bf q}|+2)(3|{\bf q}|-7)\log{|{\bf q}|}\nonumber \\
&+4(|{\bf q}|+2)\log(6-5|{\bf q}|)
-(|{\bf q}|+2)(|{\bf q}|-1)
-16\sqrt{4-|{\bf q}|^2}\,
\tanh^{-1}\!\left(
\frac{2(1-|{\bf q}|)\sqrt{4-|{\bf q}|^2}}{(2-|{\bf q}|)^2}
\right)
\Bigg]
\end{align}
For $|{\bf q}|>1$ we  have
\begin{align}
\delta I_{GG}^{(3)}(|{\bf q}|>1) \ = \ &\frac{1}{4\pi^2 |{\bf q}|}\int_{-1/2}^{A_-}d\xi_1 \Bigg [ \frac{\xi_1}{2E_1}\log{\left (\frac{E_1 + E_2(y_+)}{E_1+E_2(y_-)} \right )}+\frac{1}{2}\log{\frac{-\xi_1+y_+}{-\xi_1+y_-}}  \nonumber \\
&- \int_{y_-}^{y_+}\left (\frac{1}{2(E_1+E_2)}-\frac{1}{2(-\xi_1 + \xi_2)}\right ) \Bigg ]
\end{align}
Expanding in $\xi_1>\Delta$ and dropping all terms that are beyond $O(\Delta^2)$ gives an integrand,
\begin{align}
&\Bigg [ \frac{1}{2}\log{\left (\frac{E_2(y_-)-\xi_1}{y_- - \xi_1} \right )}+\frac{\Delta^2}{4\xi_1^2}\log{\left (\frac{y_+-\xi_1}{y_- -\xi_1}\right )} +\frac{\Delta^2}{4}\left (\frac{1}{\xi_1 y_+} +\frac{1}{\xi_1(\xi_1 - y_-)} \right ) -\frac{\Delta^2}{4}\int_{y_-}^{y_+}d\xi_2\frac{1}{\xi_1 \xi_2(\xi_2 - \xi_1)} \Bigg ]
\end{align}
The result is,
\begin{align}
&\delta I_{GG}^{(3)}(|{\bf q}|>1) = \frac{1}{4\pi^2 |{\bf q}|}\frac{\Delta^2}{4 |{\bf q}| \left(|{\bf q}|^2-4\right)} \bigg[-2 \left(|{\bf q}|^2+|{\bf q}|-2\right) \log \left(\frac{(|{\bf q}|-1) |{\bf q}|}{\Delta}\right)+4 \left(|{\bf q}|^2-4\right) \log (2-|{\bf q}|)\nonumber \\
&-|{\bf q}| (|{\bf q}|+1) (1+\log (16))+2+\log (256)\bigg]
\end{align}

Adding up all the terms for $\delta I_{GG}$ results in a remarkable simplification,
\begin{align}
\boxed{\delta I_{GG} = \frac{1}{4\pi^2 |{\bf q}|}\left  [\frac{\pi^2}{2}\Delta + \frac{4}{|{\bf q}|}\Delta^2\log{\Delta}+\left (\frac{8}{|{\bf q}|\sqrt{4-|{\bf q}|^2}}\log{\left (\frac{2+\sqrt{4-|{\bf q}|^2}}{|{\bf q}|}\right )} -\frac{8}{|{\bf q}|(4-|{\bf q}|^2)}-\frac{2+8\log{2}}{|{\bf q}|} \right )\Delta^2\right ]}   
\end{align}
%

\subsubsection{\texorpdfstring{$s$}{}-wave projection}
The potential can be projected onto the s-wave by integrating over all relative momentum.
This is given by 
\begin{align}
    K_{11}^{[2]} \ &= \frac{g^2}{2}\int_0^2 d|{\bf q}|\, |{\bf q}|\left [  I _{GG}(|{\bf q}|) \ + \ 2I_{FF}(|{\bf q}|) \right ] \nonumber \\
    K_{12}^{[2]} \ &= \ \frac{g^2}{2}\int_0^2 d|{\bf q}|\, |{\bf q}|\left [ I_{FF}(|{\bf q}|) \right ] \ ,
\end{align}
where we used the invariance of the integral under $|\bar{{\bf q}}| \to |{\bf q}|$ to combine terms.
The $I_{GG}(|{\bf q}|)$ and $I_{FF}(|{\bf q}|)$ integrals were computed analytically up to $O(\Delta^2)$ in the previous sections.
The result after s-wave projection (with $M=k_F=1$) is,
\begin{align}
&K_{11}^{[2]} \ =  \ \frac{g^2}{2}\Bigg \{-\frac{1}{3\pi^2}(1+2\log{2}) +\frac{3\Delta}{4} \ + \ \frac{\Delta^2}{2\pi^2}\int_0^2d|{\bf q}| \frac{1}{|{\bf q}|(4-|{\bf q}|^2)} \nonumber \\
&\Bigg [ 2 \left(12-|{\bf q}|^2\right)\log (\Delta)+|{\bf q}|^2+4 \left(|{\bf q}|^2-12\right) \log (2)+12 \sqrt{4-|{\bf q}|^2} \tanh ^{-1}\left(\frac{\sqrt{4-|{\bf q}|^2}}{2}\right)+16 |{\bf q}| \tanh ^{-1}\left(\frac{|{\bf q}|}{2}\right)-24 \Bigg ]\Bigg \} \ , \nonumber \\[4pt]
&K_{12}^{[2]} \ =\ \frac{g^2}{2} \Bigg \{\frac{\Delta}{4} \ + \ \frac{4\Delta^2}{\pi^2}\int_0^2d|{\bf q}| \frac{1}{|{\bf q}|(4-|{\bf q}|^2)}\left [\log{\Delta} + |{\bf q}|\tanh^{-1}\left (\frac{|{\bf q}|}{2} \right) + \frac{\sqrt{4-|{\bf q}|^2}}{2}\tanh^{-1}\left ( \frac{\sqrt{4-|{\bf q}|^2}}{2}\right ) - 1 - \log{4} \right ] \Bigg \}\ ,
\label{eq:swaveIntNoReg}
\end{align}
where we have performed the integral over the $O(\Delta^0)$ and $O(\Delta)$ terms.
There is a problem, the remaining integral of the $\Delta^2$ pieces diverges near the boundaries.
This is because the derivation of $I_{GG}$ and $I_{FF}$ above assumed that $|{\bf q}|>\Delta$ and $2-|{\bf q}|>\Delta$ when expanding the integrals to $O(\Delta^2)$.
This assumption breaks down at the boundaries of the integral over $|{\bf q}|$, and our integrand is incorrect in that region.
Fortunately, this does not impact the $\Delta^2\log^2\Delta$ and $\Delta^2\log{\Delta}$ coefficients of $\hat{K}$.
This can be shown by putting a regulator into the integral $\int_0^2 \to \int_{\kappa_0 \Delta}^{2-\kappa_1\Delta}$.
The coefficients of the $\Delta^2\log^2\Delta$ and $\Delta^2\log{\Delta}$ coefficients are independent of $\kappa_{0,1}$ and therefore can be reliably extracted.
The result is $K^{[2]}$ expanded up to $O(\Delta^2 \log{\Delta})$,
\begin{align}
&K_{11}^{[2]} \ = \ 
\frac{g^2}{2}\left [-\frac{1}{3\pi^2}(1+2\log{2}) +\frac{3\Delta}{4} -\frac{2}{\pi^2}\Delta^2 \log^2{\Delta} \ + \ \frac{17+12\log 2 }{4\pi^2}\Delta^2\log\Delta  \ + \ O(\Delta^2)\right ] \nonumber \\
&K_{12}^{[2]} \ = \ \frac{g^2}{2} \left [\frac{\Delta}{4} \ - \ \frac{3}{4\pi^2}\Delta^2\log^2\Delta \ + \ \frac{6+4\log{2}}{4\pi^2}\Delta^2 \log{\Delta} \ + \  O(\Delta^2)\right ] \ .
\end{align}

\end{document}